\documentclass[12pt]{article}

\usepackage{times}
\usepackage{helvet}

\usepackage{graphicx}
\usepackage{amsfonts}  
\usepackage{amsmath}   
\usepackage{amssymb}   
\usepackage[dvipsnames,usenames]{xcolor}

\newenvironment{sciabstract}{%
\begin{quote} \bf}
{\end{quote}}

\title{Thermal dependence of the hydrated proton and optimal proton transfer}

\author
{F\'elix Mouhat,$^{1}$ Matteo Peria,$^{2}$\\
 Tommaso Morresi,$^{3}$ Rodolphe Vuilleumier,$^{4}$ \\ 
 Antonino Marco Saitta,$^{2}$ and Michele Casula$^{2\ast}$\\
\\
\normalsize{$^{1}$Saint Gobain Research Paris,}\\
\normalsize{39, Quai Lucien Lefranc, 93300 Aubervilliers, France}\\
\normalsize{$^{2}$IMPMC, Sorbonne Universit\'e, CNRS, MNHN, UMR 7590,}\\
\normalsize{4 Place Jussieu, 75252 Paris, France}\\
\normalsize{$^{3}$ ECT*-Fondazione Bruno Kessler$^{*}$,}\\
\normalsize{286 Strada delle Tabarelle, 38123, Trento, Italy}\\
\normalsize{$^{4}$\'Ecole normale sup\'erieure, PSL Research
  University,}\\
\normalsize{Sorbonne Universit\'e, CNRS, D\'epartement de Chimie, PASTEUR,}\\
\normalsize{24 Rue Lhomond, 75005 Paris, France}\\
\\
\normalsize{$^\ast$To whom correspondence should be addressed; E-mail: michele.casula$@$upmc.fr.}
}

\date{}

\begin{document} 



\baselineskip24pt


\maketitle


\begin{sciabstract}
Water is a key ingredient for life and plays a central
role as solvent in many biochemical reactions. 
However, 
the intrinsically
quantum nature of the hydrogen nucleus, revealing itself in a large
variety of physical manifestations, including proton 
transfer,
gives rise to unexpected phenomena whose description is still elusive.
Here we study, by an unprecedented combination of
state-of-the-art
quantum Monte Carlo methods and
path-integral molecular dynamics, the structure and hydrogen-bond
dynamics of the protonated water hexamer, the fundamental unit for the
hydrated proton. We report a remarkably low thermal expansion of the hydrogen bond from zero temperature up to 300 K, owing to the presence of short-Zundel 
configurations, 
characterised by proton delocalisation and
favoured by the synergy of nuclear quantum effects and thermal
activation.
The hydrogen bond strength
progressively weakens above 300 K, when localised Eigen-like configurations become relevant. Our analysis, supported by the instanton statistics of shuttling protons, reveals that the near-room-temperature range from 250 K to 300 K is a ``sweet spot'' for proton transfer,
and thus for many phenomena depending on it,
including life.
\end{sciabstract}

\section{Introduction}
\label{intro}

For more than 200 years and the seminal work of von
Grotthus, 
the properties of the
hydrated proton H$^+_{(\textrm{aq})}$ have intrigued the scientific
community \cite{200years_after,CUKIERMAN2006876}. Despite significant
advances, the exact role of the solvated proton in proton transfer
(PT) reactions in chemical and biological systems is not fully
elucidated yet. The 
textbook picture is that the hydrated proton exists
as classical hydronium cation H$_{3}$O$^+$, but it looks more
appropriately described as a delocalised 
charge defect
shared by multiple molecules. The spread of this charge defect blurs
the identity of the excess proton among two limiting structures,
namely the Zundel\cite{Zundel1968} and the Eigen\cite{Eigen1964}
ions. Indeed, the hydrated proton Infrared (IR) spectrum displays a
combination of a few discrete absorption bands on top of an absorption continuum, broadly extended over the entire spectrum. Neither the
symmetrically solvated hydronium H$_{9}$O$_4^+$ (Eigen) ion, nor 
the equally shared proton in the H$_{5}$O$_2^+$ (Zundel) ion, can individually rationalise 
this characteristic IR fingerprint. Models involving fast inter-conversions between these two ionic species are also shown to fail\cite{Vuilleumier1999}.

\begin{figure}
\begin{tabular}{c|c|c}
\includegraphics[width=0.33\linewidth]{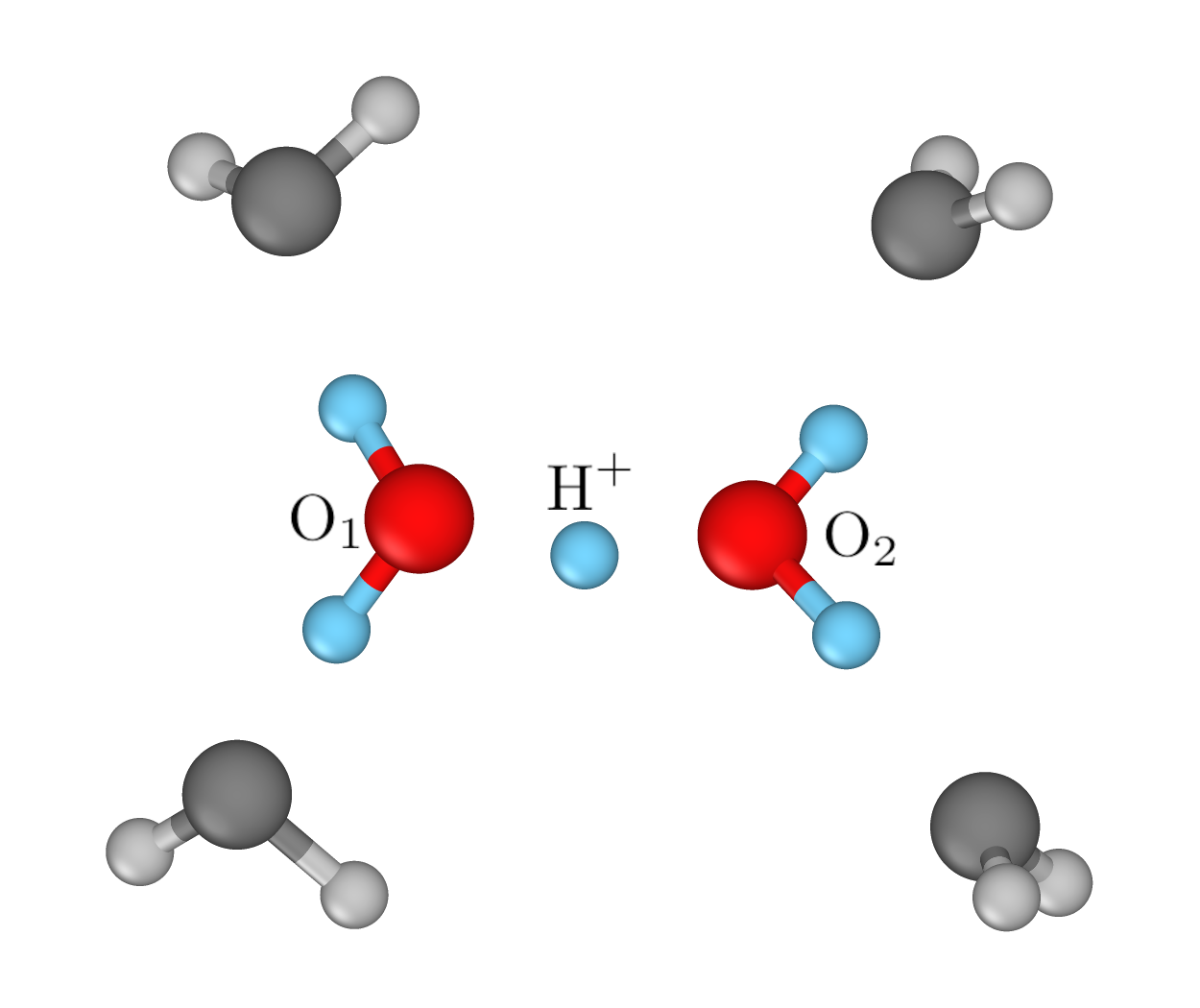}  &
\includegraphics[width=0.33\linewidth]{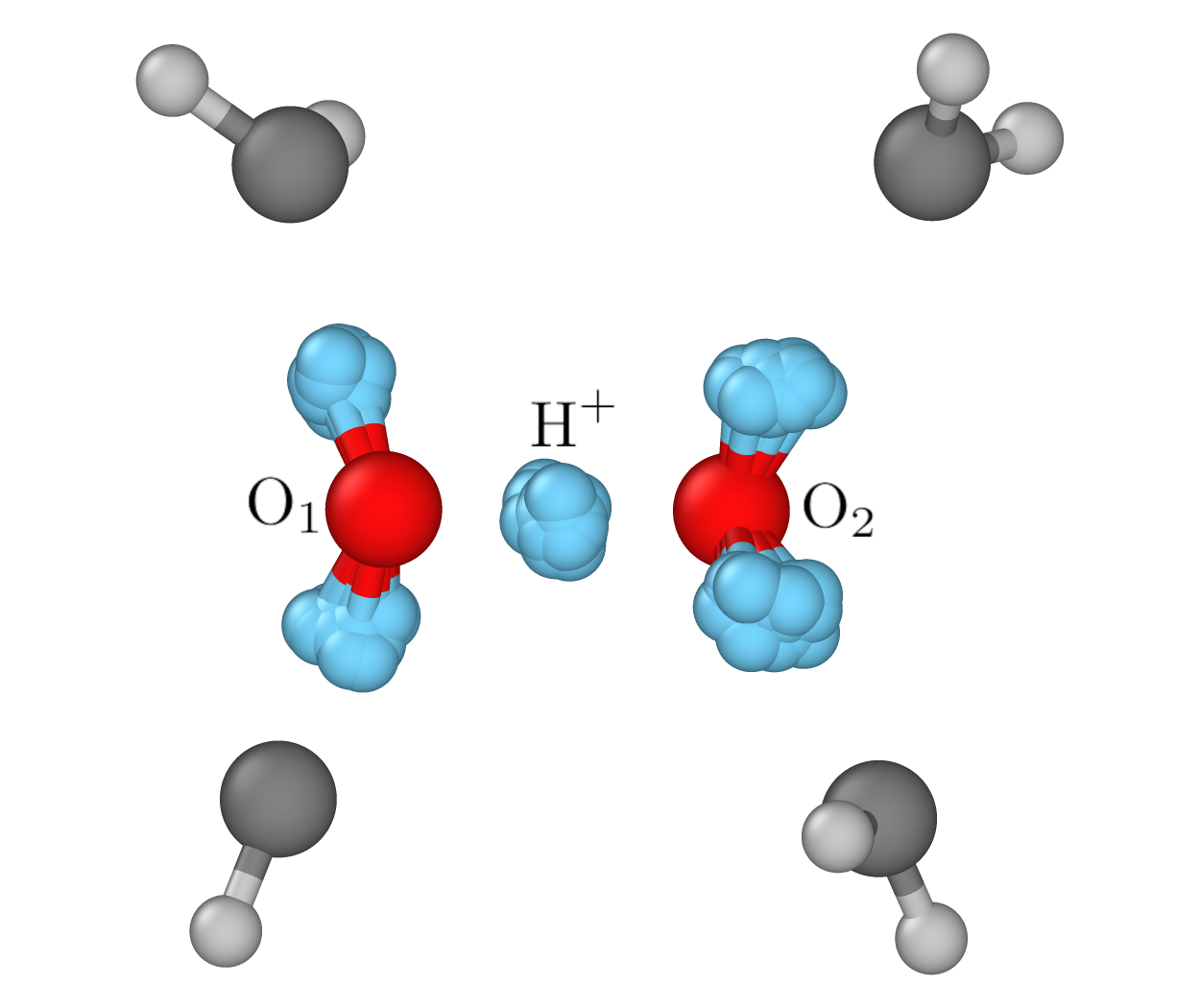} &
\includegraphics[width=0.33\linewidth]{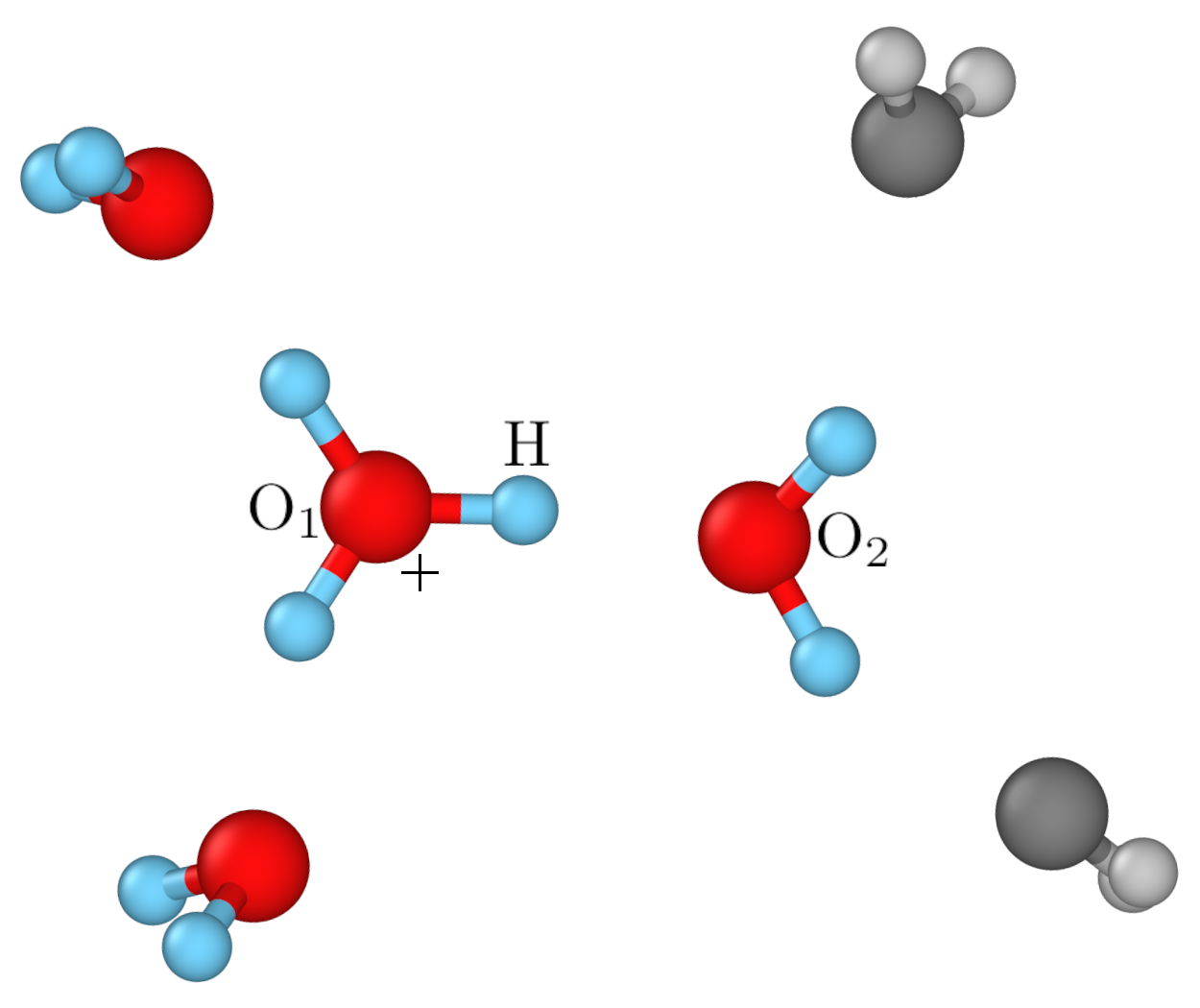} \\
short Zundel  & elongated Zundel & distorted Eigen
\end{tabular}
\caption{\label{fig:config} Different regimes of the protonated
  water hexamer H$_{13}$O$_{6}^+$. Left panel: short-Zundel configuration
  with a Zundel center (H$_{5}$O$_{2}^+$) in colors and its
  first solvation shell (4 H$_{2}$O) in gray shades. Central panel: elongated Zundel with the quantum nature of hydrogen atoms highlighted by the full representation of its imaginary-time positions in a PI configuration. Right panel: distorted-Eigen configuration with an Eigen cation (H$_{9}$O$_{4}^+$) in colors accompanied by two solvating water molecules (2 H$_{2}$O) in gray shades. The O$_1$, O$_2$ and H$^+$ labels are used throughout the paper to refer to the corresponding atoms, as indicated here.}  
\end{figure}

To deal with these issues, Stoyanov \emph{et al.}\cite{Stoyanov2010}
have introduced the stable 
H$_{13}$O$_6^+$ species, the protonated water hexamer, which is Zundel-type in the sense that the
excess proton is equally shared between two water molecules. 
The core of the cluster is characterised by 
a central oxygen-oxygen
distance $\text{d}_{\text{\scriptsize O}_1\text{\scriptsize O}_2}$ that, however, is more elongated than in the Zundel cation\cite{Bell1975}.
On the other hand, recent Molecular Dynamics (MD)
simulations suggest the existence of a distorted, nonsymmetric
Eigen-type cation, remaining at the heart of a dynamical 
charge defect spanning multiple water molecules \cite{Knight2012}. Thus, the
protonated water hexamer represents the smallest protonated water
cluster for which both of these characteristic binding motifs
coexist\cite{Jiang2000,Mizuse2011}. It is the simplest one that can provide a meaningful description of the hydrated proton in water.
The main protonated hexamer
configurations are represented in
Fig.~\ref{fig:config}, for both Zundel- and Eigen-like forms. 

The protonated hexamer Potential Energy Surface (PES) has been
partially explored by IR spectroscopy
\cite{Wang2003,zundel_vibrational,Heine2013} and electronic structure
calculations performed within Density Functional Theory (DFT), 
M\o{}ller-Plesset (MP2), and Coupled Cluster (CC) approaches, also supplemented by Machine Learning techniques,
\cite{Wei1994,zundel_vibrational,Jiang2000,Wang2003,heindel2018,schran2021} confirming that
the two structures introduced above are the lowest energy
isomers. 
The quantum nature of the proton however induces a delocalised structure on this PES.

In this work, we apply
MD simulations fully retaining the nuclear quantum nature of the atoms. In particular,
the quantum proton, described within the Feynman path integral (PI) approach,
evolves in a very accurate 
PES estimated by means
of Quantum Monte Carlo (QMC). This stochastic technique introduces an
intrinsic noise,
which affects
the forces driving the ion dynamics 
and, consequently, the simulation temperature. Relying on the generalised
fluctuation-dissipation theorem, a Langevin-based approach has been
developed to address this issue for classical \cite{Attaccalite2008}
and quantum \cite{Mouhat2017} ions. It allows one to sample microscopic
configurations in the canonical ensemble 
with an unprecedented
level of quantum accuracy. Details of the method are provided in the Methods Section and in 
Secs.~S1 and S2 of the Supplementary Information (SI).

We find that the hydrogen bond (H-bond) mediated by the hydrated
proton shows a remarkably low thermal expansion from zero temperature up to $300$ K, with a nearly temperature-independent length that becomes \emph{shorter} than the classical-ion counterpart in the [$200$-$350$] K temperature range. 
A non-trivial behaviour of the H-bond has also been found in H-rich crystals and ferroelectric materials, such as the potassium dihydrogen phosphate (KDP)\cite{koval2002ferroelectricity}, first detected by Ubbelohde in 1939 upon isotopic substitution\cite{robertson1939structure}.
In the latter case, the lighter the hydrogen, the shorter the H-bond. This was interpreted as a quantum manifestation of 
proton
delocalisation, strengthening the H-bond.   
In the present 
situation,
the strength of the H-bond results 
from a non-trivial cooperation
of nuclear quantum effects (NQEs) and thermal activation, as we will show in this work.
Indeed, NQEs strongly affect the vibrational levels of the proton shuttling mode bridging the central $\text{O}_1$ and $\text{O}_2$ oxygen atoms. These levels are then thermally occupied according to the $\text{d}_{\text{\scriptsize O}_1\text{\scriptsize O}_2}$ distance of a given configuration.
We can thus distinguish three regimes (see Fig.~\ref{fig:config}): (i)
``short-Zundel" configurations with the shortest
$\text{d}_{\text{\scriptsize O}_1\text{\scriptsize O}_2}$, where the
shuttling proton feels a quadratic potential and it is perfectly
shared between the two central water molecules; (ii)
``elongated-Zundel'' configurations for intermediate
$\text{d}_{\text{\scriptsize O}_1\text{\scriptsize O}_2}$, comprising
the equilibrium distance, where a potential energy barrier starts to
develop in between $\text{O}_1$ and $\text{O}_2$ and the proton is
delocalised only due to NQEs; (iii) ``distorted-Eigen" configurations
at even larger $\text{d}_{\text{\scriptsize O}_1\text{\scriptsize
    O}_2}$, where the central barrier is large enough that the
hydrated proton is localised on one of the two flanking water molecules,
forming an Eigen-like complex. We will see that the occurrence of
short-Zundel configurations is key to understand
the H-bond thermal robustness
and to enhance the proton transfer dynamics. Despite being energetically disfavoured by the short $\text{d}_{\text{\scriptsize O}_1\text{\scriptsize O}_2}$ distances at the classical level, these configurations are populated thanks to the synergistic action of NQEs and temperature, yielding a sweet spot for proton transfer in the [$250$-$300$] K temperature range.

\section{Results and discussion}
\label{results}

\paragraph*{Thermal expansion of the H-bond}

To rationalise our main outcome, we first study the zero-temperature
classical geometry and PES of the protonated water hexamer, and compare it with the Zundel cation. While the latter system misses a large part of water solvation effects, the former includes the full contribution of the first and second shells of the solvated proton. The zero-temperature results are reported in Sec.~S3.1 of the SI. We simply highlight here that the Variational Monte Carlo
(VMC) equilibrium O$_1$-O$_2$ distance is found to be
$\text{d}_{\text{\scriptsize O}_1\text{\scriptsize O}_2} = \text{d}_\text{min} = 2.3930(5)$ \AA, in 
good agreement with MP2 calculations, the most widely used post Hartree-Fock theory to study water clusters (see Sec.~S1.2 of the SI for a more extended comparison between VMC and MP2). VMC has a milder scale with the system size than MP2, allowing one to perform extensive calculations of the protonated hexamer.
At variance with the Zundel cation\cite{JONES1989283,Stoyanov2006}, the protonated
hexamer equilibrium geometry is asymmetric, implying that the global minimum is split into a double well, separated by an energy barrier between the two central water molecules. This barrier vanishes at $\text{d}_{\text{\scriptsize O}_1\text{\scriptsize O}_2} = \text{d}_\text{symm} \simeq 2.38$ \AA, a distance that separates the short Zundel 
below from the elongated-Zundel configurations above. The height of the barrier is less than $100$ K (in $k_B$ units) at $\text{d}_{\text{min}}$, rapidly increasing as a function of $\text{d}_{\text{\scriptsize O}_1\text{\scriptsize O}_2}$.
We therefore expect several consequences on the hydrated proton distribution and on its mobility at finite temperature, once the NQEs are taken into account.

To understand how 
the dynamics of the hydrated
proton evolves with temperature, 
QMC-driven \emph{ab initio} MD simulations are
relevant. Such calculations are carried out for both classical and
quantum nuclei of the H$_{13}$O$_{6}^+$ ion, within the temperature
interval T $\in [50-350]$ K, thanks to the methodological developments
detailed in Ref.\cite{Mouhat2017} and in the Methods Section. At these conditions, the clusters are stable during the simulated time frame ($\approx 30$ ps), allowing us to access the thermal properties of the hydrated proton and the $\text{O}_1\text{H}^+\text{O}_2$ bond over an extended temperature range.

\begin{figure}[!htp]
\centering 
\includegraphics[width=1.0\textwidth]{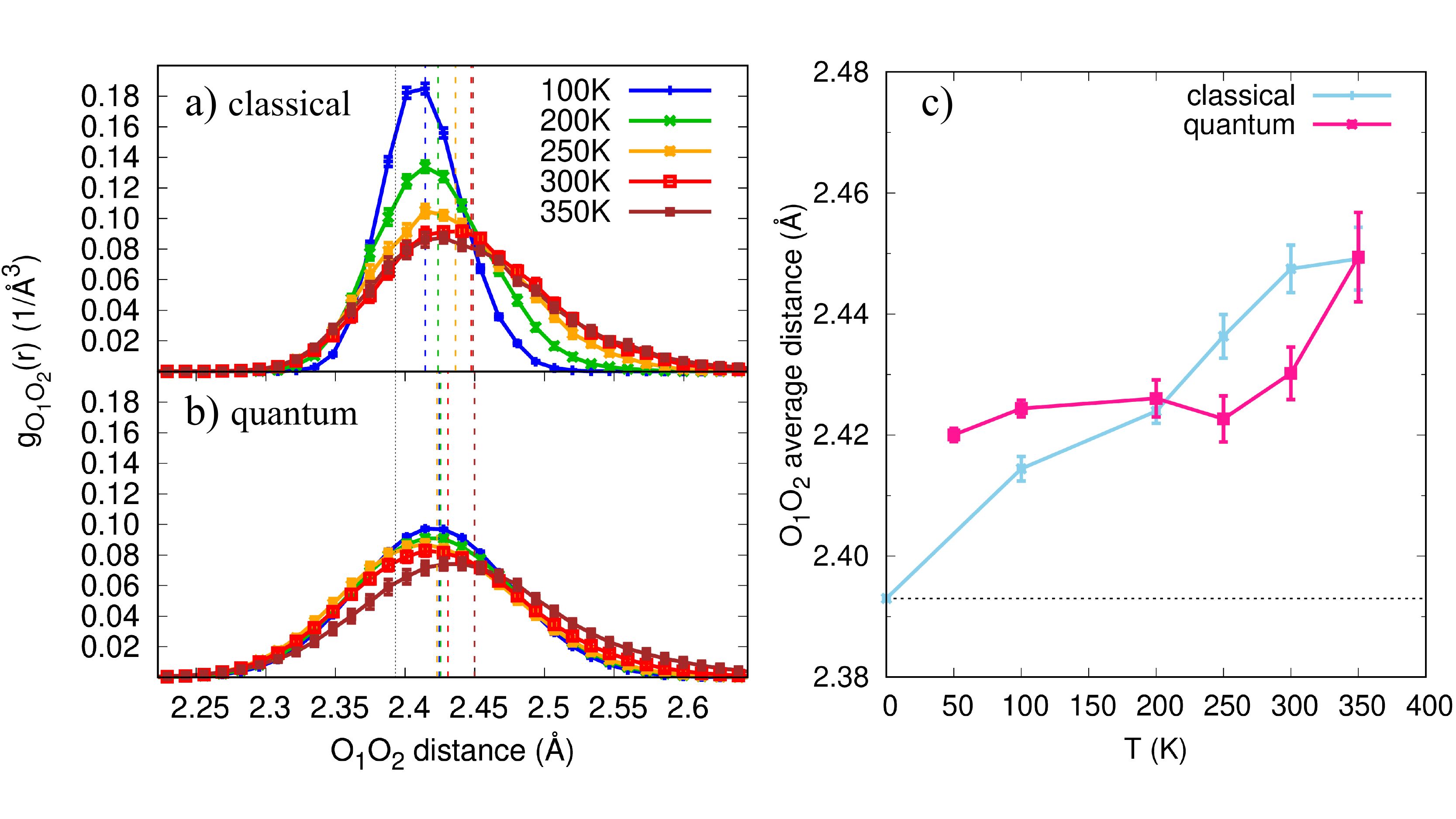}
\caption {\label{figure:radial} Classical (panel a)) and quantum (panel b)) oxygen-oxygen g$_{\text{\scriptsize O}_1\text{\scriptsize O}_2}$ pair correlation functions of the H$_{13}$O$_{6}^+$ ion as a function of temperature. The dashed vertical lines indicate the average  $\langle \textrm{d}_{\text{\scriptsize O}_1\text{\scriptsize O}_2} \rangle $ distance for each
simulation, at the corresponding temperature. The dotted vertical line is located at the classical equilibrium geometry. Panel c) shows the T-dependence of the $\langle \textrm{d}_{\text{\scriptsize O}_1\text{\scriptsize O}_2} \rangle $ average distance. The classical equilibrium geometry is represented by a short-dashed horizontal black line. At 250 K and 300 K the oxygen-oxygen distance is \emph{shortened} by NQEs with respect to the classical counterpart.}  
\end{figure}

From our QMC-MD simulations, we extract the normalised Pair
Correlation Functions (PCFs) g$_{\text{\scriptsize
    O}_1\text{\scriptsize O}_2}$ for the two oxygen atoms O$_1$ and
O$_2$ of the
cluster
core (Fig.~\ref{figure:radial}).
The expected broadening of the PCFs due to nuclear quantisation
is significant over the whole temperature range (Fig.~\ref{figure:radial}(b)). Only at temperatures as high as 350 K, the classical g$_{\text{\scriptsize O}_1\text{\scriptsize O}_2}$ (Fig.~\ref{figure:radial}(a)) starts resembling the quantum distribution.
This implies that
the NQEs cannot be neglected for temperatures up to this value,
above ambient conditions. We also notice that, when comparing to the Zundel
ion results\cite{Suzuki2013}, 
the peak position is shifted up by 
at least
$\sim 0.01$ \AA.
Thus, it appears that the 
H$_{13}$O$_{6}^+$ cluster
frequently
adopts elongated-Zundel configurations\cite{Markovitch2008,kreuer2004transport,asthagiri2005ab} at the lowest temperatures considered here.
This is at variance with the protonated water dimer, 
where the hydrated proton lives in a single minimum symmetrically located between the two water molecules.

Focusing
our attention to $\langle \textrm{d}_{\text{\scriptsize O}_1\text{\scriptsize O}_2} \rangle $ (Fig.~\ref{figure:radial}(c)), its classical and quantum behaviours are remarkably different as
a function of temperature. On the one hand, the classical
$\text{d}_{\text{\scriptsize O}_1\text{\scriptsize O}_2}$ keeps increasing with
temperature, as more energy is given to the intermolecular vibration
modes. On the other hand, the quantum
$\textrm{d}_{\text{\scriptsize O}_1\text{\scriptsize O}_2}$ displays a
nearly
flat behaviour
with the cluster temperature, up to 300 K. This
very low thermal expansion
extended over a wide temperature range leads to a temperature regime where $\textrm{d}_{\text{\scriptsize O}_1\text{\scriptsize O}_2}$ for the quantum system become shorter than the classical values at the same temperatures. This is clearly seen in Fig.~\ref{figure:radial}(c). We will come back to this point later.

Finally, 
as the temperature further increases, 
the NQEs reduction weakens the central H-bond strength. Consequently, $\text{d}_{\text{\scriptsize O}_1\text{\scriptsize O}_2}$ spreads out, due to stochastic fluctuations of the core and the solvent, and a more classical regime is reached, when the averaged $\text{d}_{\text{\scriptsize O}_1\text{\scriptsize O}_2}$ values for classical and quantum nuclei meet again. The 
PCF distributions display longer
tails, with more configurations covering 
regions with $\textrm{d}_{\text{\scriptsize O}_1\text{\scriptsize O}_2} \in
[2.5-2.7]$ \AA, and the peak position rapidly shifts to larger values.
Configurations with such a large $\langle \textrm{d}_{\text{\scriptsize O}_1\text{\scriptsize O}_2} \rangle$ 
are of distorted-Eigen type\cite{Markovitch2008,thecomputer}.

\paragraph*{A cooperative thermal-quantum species: the short-Zundel ion}

\begin{figure}[!htp]
\centering 
\includegraphics[width=1.0\textwidth]{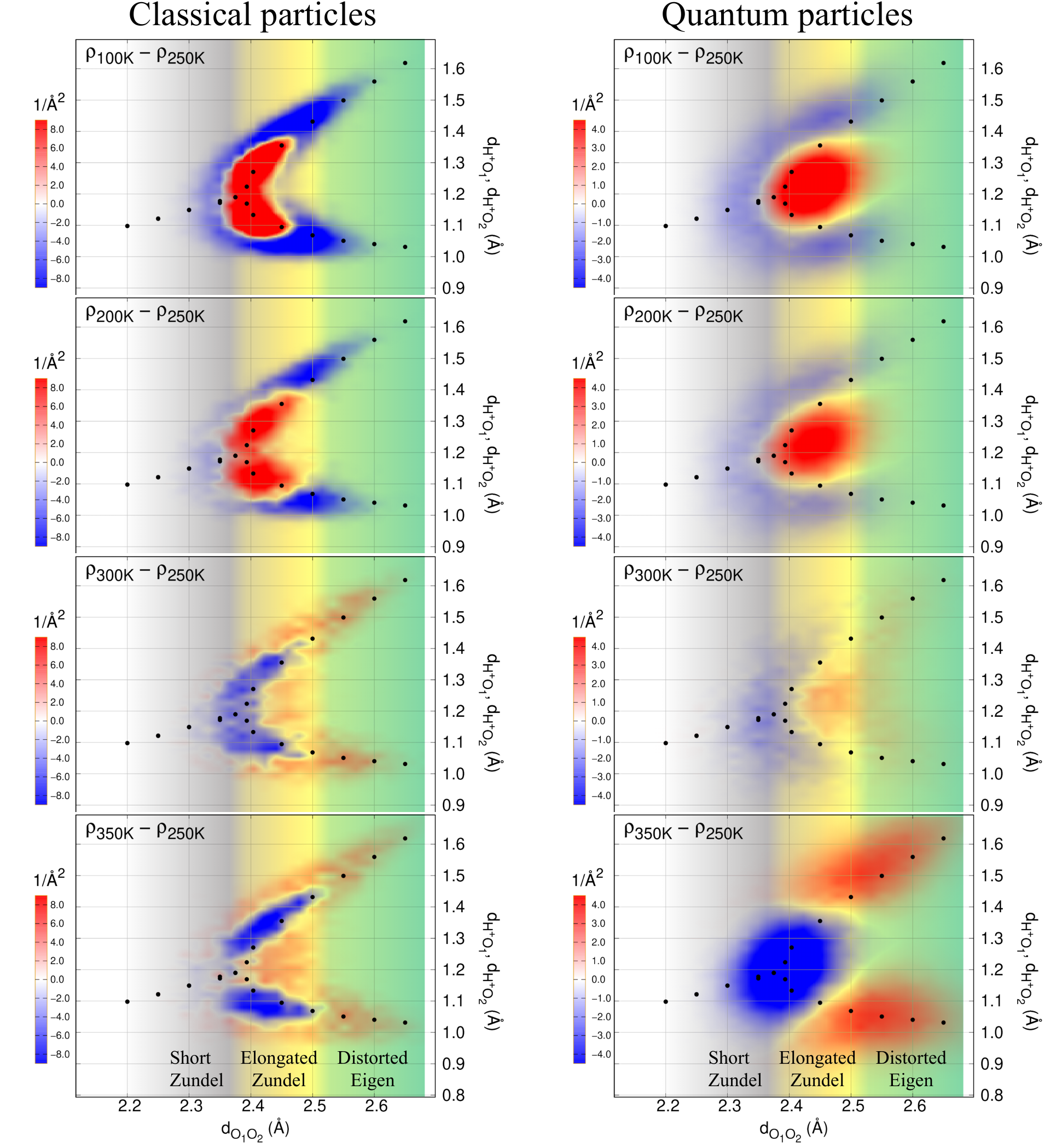}
\caption {\label{figure:2Dgrrnucvst} Difference between bidimensional
  oxygen-oxygen/oxygen-proton distributions $\rho_{2D}$ obtained by QMC-driven
  LD simulations for classical (left panels) and quantum (right
  panels) particles, computed at different temperatures. The bidimensional distribution computed at 250 K is taken as reference.
  Positive (negative) regions are in red (blue) color. The
  black filled circles correspond to the zero-temperature equilibrium
  geometries of the H$_{13}$O$_{6}^+$ ion at a fixed $\textrm{d}_{\text{\scriptsize O}_1\text{\scriptsize O}_2}$ distance. The coloured background highlights the three different regimes explained in the paper: the short Zundel (gray), the elongated Zundel (yellow), and the distorted Eigen (green) species.}  
\end{figure}

To refine our structural analysis, we compute the bidimensional distribution function $\rho_{2D}$, which correlates the oxygen-oxygen (O$_1$O$_2$) and the oxygen-proton (O$_{1/2}$H$^+$) distances, and study its temperature dependence $\rho_{2D}=\rho_{2D}(T)$. In Fig.~\ref{figure:2Dgrrnucvst}, we show the contour plot of the temperature-driven $\rho_{2D}$ variation (see also Sec.~S3.2 of the SI).
By taking $\rho_{2D}(250 \,\, \text{K})$ as reference, four temperature variations are explored: $100$ K, $200$ K, room temperature (RT), and 
$350$ K (from the top to the bottom of Fig.~\ref{figure:2Dgrrnucvst}).

In the \emph{classical} protonated hexamer
(Fig.~\ref{figure:2Dgrrnucvst}, left column), rising the temperature from $250$ K 
up to $350$ K tends to stretch $\langle \text{d}_{\text{\scriptsize O}_1\text{\scriptsize O}_2} \rangle$,
by promoting configurations from the elongated Zundel (blue central distribution with
d$_{\text{\scriptsize O}_1\text{\scriptsize O}_2} \in [2.38,2.5]$ \AA\ in Fig.~\ref{figure:2Dgrrnucvst}) to an Eigen-like arrangement 
with larger d$_{\text{\scriptsize O}_1\text{\scriptsize O}_2}$ and a proton much more localised on one of the two central oxygen atoms (red wings).
The situation is reversed at lower temperatures (100 K and 200 K) if compared to the 250 K reference, with positive (red) variations in the elongated Zundel and negative (blue) variations in the wings. Thus, for classical nuclei, there is a progressive depletion of the elongated Zundel and a corresponding population of the distorted-Eigen wings upon temperature rise. Short-Zundel configurations, highlighted in Fig.~\ref{figure:2Dgrrnucvst} by a gray background, seem to play a very marginal role in the temperature-driven density distribution shift.

The scenario is strikingly different with \emph{quantum} nuclei (right column), particularly at the lowest temperatures (100 K and 200 K). In this regime, distorted-Eigen configurations are barely populated or depleted, and the density shift upon rising temperature takes place between the elongated-Zundel region
and the short-Zundel sector. 
The latter is significantly more populated at 250 K than at lower temperatures at the expense of the elongated Zundel, which instead loses density with respect to the classical counterpart at the same temperature.

In the higher-temperature limit, at 350K, NQEs are less relevant and, by consequence, the classical and quantum variations have a qualitatively similar behaviour. In both classical and quantum case, we notice the presence of red wings at large
oxygen-oxygen distances (d$_{\text{\scriptsize O}_1\text{\scriptsize O}_2} \in [2.5,2.7]$ \AA),
which are the signature of thermally activated Eigen-like states, with a
strongly localised proton. This is related to
less frequent
elongated-Zundel configurations, indicated by
the depleted distribution for d$_{\text{\scriptsize O}_1\text{\scriptsize O}_2} < 2.5$ \AA,
confirming that the distorted-Eigen configurations are indeed 
promoted by
high
temperature. For quantum nuclei, the corresponding depletion goes well below the elongated-Zundel region, by touching also short-Zundel configurations, down to d$_{\text{\scriptsize O}_1\text{\scriptsize O}_2} \sim 2.3$ \AA, at variance with the classical case, where the short-Zundel configurations are not involved.

To interpret these results,
we first construct an accurate effective potential by projecting the full PES, computed during QMC-driven classical MD calculations, onto the degrees of freedom mostly relevant to understand the dynamics of the hydrated proton.  
These are
the d$_{\text{\scriptsize O}_1\text{\scriptsize O}_2}$ distance and the proton sharing coordinate $\delta$, referenced to the midpoint of the $\text{O}_1\text{H}^+\text{O}_2$ complex: $\delta \equiv \tilde{d}_{\text{\scriptsize O}_{1/2}\text{\scriptsize H}^+} -\text{d}_{\text{\scriptsize O}_1\text{\scriptsize O}_2}/2$, with $\tilde{d}_{\text{\scriptsize O}_{1/2}\text{\scriptsize H}^+}$ the $\text{O}_{1/2}$-$\text{H}^+$ distance projected onto the O$_1$O$_2$ direction. The resulting two-dimensional (2D) potential is $V_\textrm{2D}=V_\textrm{2D}(\text{d}_{\text{\scriptsize O}_1\text{\scriptsize O}_2},\delta)$. We refer the reader to Secs.~S4 and S5 of the SI for technical details about the PES projection. We highlight that the potential $V_\textrm{2D}$ is derived here at VMC quality. We also notice that 
$\delta$ 
is the vibrational coordinate
of the proton shuttling mode, while d$_{\text{\scriptsize O}_1\text{\scriptsize O}_2}$ is related to the stretching mode of the two water molecules in the cluster core.

\begin{figure}
    \centering
\includegraphics[width=1.0\textwidth]{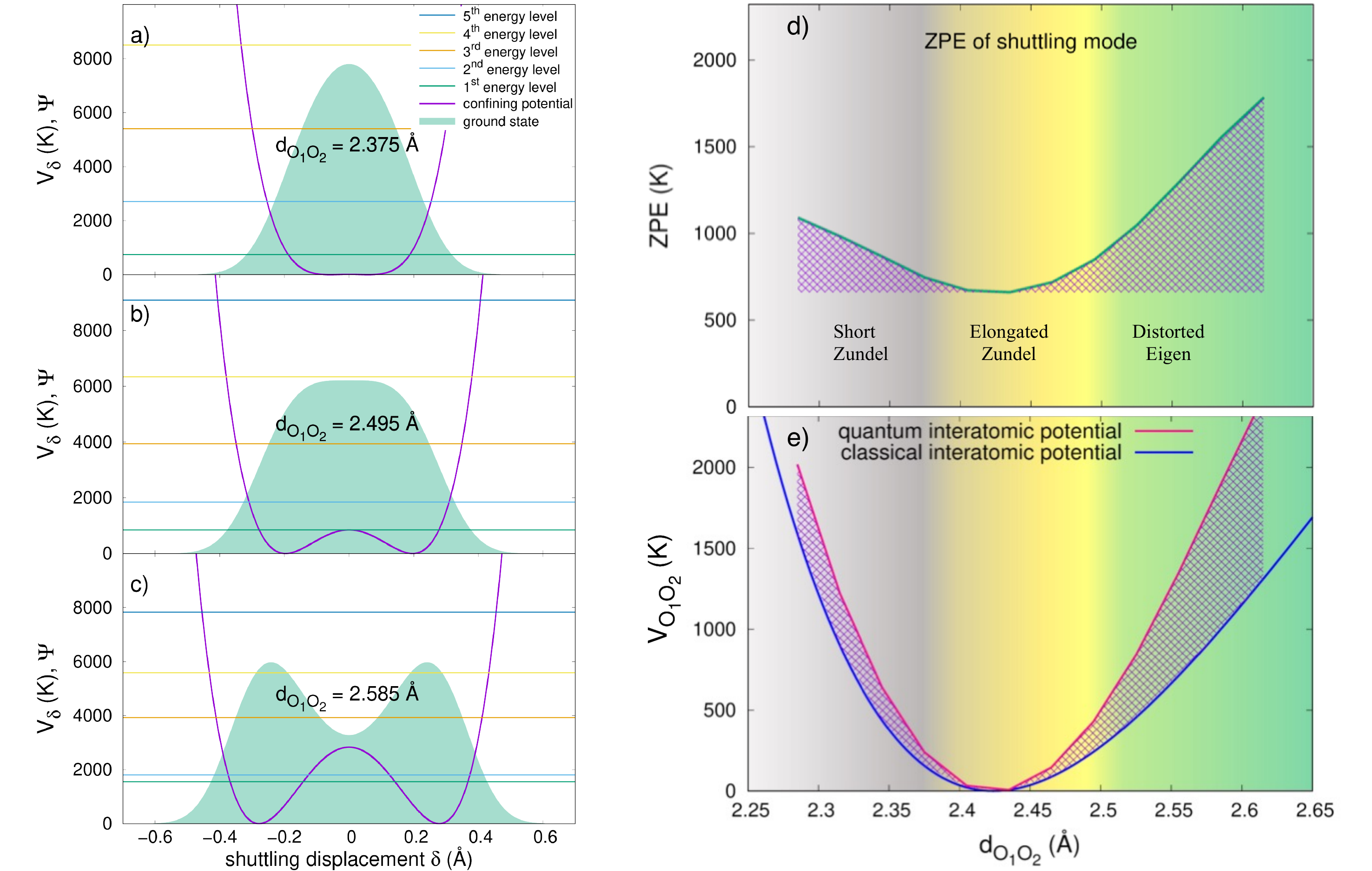}
\caption{NQEs on the shuttling mode, and their impact on the $\textrm{O}_1$-$\textrm{O}_2$ interatomic potential $V_{\text{\scriptsize O}_1\text{\scriptsize O}_2}$. We quantize the proton shuttling mode 
$\delta$, defined as the displacement 
along the segment connecting the two oxygen atoms in the core of the cluster from its mid-point position.
We study the ground-state wave function and the first 5 eigenvalues for the 
confining potential $V_\delta$, 
as a function of $\textrm{d}_{\text{\scriptsize O}_1\text{\scriptsize O}_2}$. 
Panels a), b) and c)
report the
ground state wave function and the lowest 5 energy levels for $\textrm{d}_{\text{\scriptsize O}_1\text{\scriptsize O}_2}$= 2.375, 2.495 and 2.585 \AA, respectively. In panel d), the variation of the zero-point (ground-state) energy (ZPE) as a function of $\textrm{d}_{\text{\scriptsize O}_1\text{\scriptsize O}_2}$ is explicitly plotted. While the ZPE dependence is very flat in the elongated-Zundel region (depicted by the yellow shaded area), the ZPE increases in both short-Zundel (gray shaded area) and distorted-Eigen (green shaded area) regions, with a much steeper slope in the latter. In panel e), the ZPE is added to the classical interatomic potential $V_{\text{\scriptsize O}_1\text{\scriptsize O}_2}$ (solid blue line) to yield the quantum-corrected effective interatomic potential (solid dark-pink line) between the two inner oxygen atoms.
}
    \label{figure:ZPE}
\end{figure}

Given $V_\textrm{2D}(\text{d}_{\text{\scriptsize O}_1\text{\scriptsize O}_2},\delta)$, we then proceed to quantize the variable $\delta$. 
Indeed, while $\text{d}_{\text{\scriptsize O}_1\text{\scriptsize O}_2}$ can be taken as classical, for it is related to the motion of heavier oxygen atoms of mass $m_\text{O}$, the $\delta$ coordinate must be quantised, owing to the light mass ($m_\text{H}$) of the hydrated proton. At the leading order in $2m_\text{H}/(m_\text{O}+m_\text{H})$, we separate the stretching mode from the shuttling one, by invoking a Born-Oppenheimer type of approximation for the two species. 
We finally solve quantum-mechanically the Hamiltonian of a proton in the potential $\text{V}_\delta \equiv V_\textrm{2D}(\alpha,\delta)|_{\alpha=\text{d}_{\text{\scriptsize O}_1\text{\scriptsize O}_2}}$ at fixed $\text{d}_{\text{\scriptsize O}_1\text{\scriptsize O}_2}$ value. In Fig.~\ref{figure:ZPE}(a-c) we plot the ground state distribution and eigenvalues obtained for three distances, i.e. at $\textrm{d}_{\text{\scriptsize O}_1\text{\scriptsize O}_2}$=2.375 \AA, in the short-Zundel region close to the boundary between the short and the elongated Zundel, at $\textrm{d}_{\text{\scriptsize O}_1\text{\scriptsize O}_2}$= 2.495 \AA, in the elongated-Zundel region close to the frontier between the elongated Zundel and the distorted Eigen, and finally at $\textrm{d}_{\text{\scriptsize O}_1\text{\scriptsize O}_2}$= 2.585 \AA, deep into the distorted Eigen regime.

One can notice three different quantum behaviours of the vibrational shuttling mode, that provide a more quantitative ground to the three-regime distinction made at the beginning.
In the short Zundel, $\text{V}_\delta$ is indeed a quadratic potential with a single minimum at the core center, which widens as $\textrm{d}_{\text{\scriptsize O}_1\text{\scriptsize O}_2}$ gets close to $\text{d}_\text{symm} \simeq 2.38 \text{\AA}$, a distance where it becomes quartic because its curvature changes sign. The ground state energy, i.e. the zero point energy (ZPE) of the shuttling mode, decreases as the potential widens, as reported in Fig.~\ref{figure:ZPE}(d). In the elongated Zundel, a central barrier starts to develop, with a ground-state proton distribution that stays uni-modal thanks to a ZPE larger than its height, till $\textrm{d}_{\text{\scriptsize O}_1\text{\scriptsize O}_2} \simeq 2.5$ \AA, where the ZPE equals the barrier height. In this regime, for $\textrm{d}_{\text{\scriptsize O}_1\text{\scriptsize O}_2} \in [\text{d}_\text{symm},2.5 \text{\AA}]$, the ZPE is particularly small, due to the quartic nature of $\text{V}_\delta$, and weakly $\textrm{d}_{\text{\scriptsize O}_1\text{\scriptsize O}_2}$-dependent, as shown in Fig.~\ref{figure:ZPE}(d). Finally, for $\textrm{d}_{\text{\scriptsize O}_1\text{\scriptsize O}_2} > 2.5$ \AA, we enter the distorted-Eigen regime, with an even larger central barrier ($> 1000$ K), such that the quantum proton is instantaneously localised in one of the two wells, and its
distribution is then bimodal. The ZPE starts to rise again as $\textrm{d}_{\text{\scriptsize O}_1\text{\scriptsize O}_2}$ is stretched, with a slope steeper - in absolute value - than the 
ZPE decrease in the short Zundel,
because it is now set by the much deeper lateral minima of the double-well potential. This can be seen again in Fig.~\ref{figure:ZPE}(d). 

We can now correct the classical $\text{O}_1$-$\text{O}_2$ potential, defined as $V_{\text{\scriptsize O}_1\text{\scriptsize O}_2} \equiv V_\textrm{2D}(\text{d}_{\text{\scriptsize O}_1\text{\scriptsize O}_2},\delta)|_{\delta=\delta_\text{min}}$, where $\delta_\text{min}$ is the $V_\textrm{2D}$ minimum at fixed  $\text{d}_{\text{\scriptsize O}_1\text{\scriptsize O}_2}$ value, by adding the ZPE $\forall \text{d}_{\text{\scriptsize O}_1\text{\scriptsize O}_2}$, obtained from the quantisation of the shuttling mode $\delta$. The resulting potential is plotted in Fig.~\ref{figure:ZPE}(e). Amazingly, the anharmonic classical $V_{\text{\scriptsize O}_1\text{\scriptsize O}_2}$ potential becomes 
harmonic after ZPE-correction. It is a consequence of
the much larger ZPE in the distorted-Eigen configurations than in the
short Zundel, which compensates for the underlying
$V_{\text{\scriptsize O}_1\text{\scriptsize O}_2}$ anharmonicity. This
rationalises two main features. On the one hand, it explains the
very low thermal expansion
of $\langle \textrm{d}_{\text{\scriptsize O}_1\text{\scriptsize O}_2} \rangle$, being the 
average position in a harmonic potential
temperature-independent. On the other hand, it proves that NQEs enhance the occurrence of short-Zundel configurations upon heating, while the distorted Eigen is penalised by its large ZPE with respect to the classical counterpart. 

Above RT, 
the distorted Eigen configurations will eventually become dominant again. 
This can be understood within this framework as well.
Indeed, thermal excitations are energetically more available in the distorted Eigen, where  
 the spacing between the ZPE and the first-excited state shrinks, and higher excited states are piled up more densely than in the short and elongated Zundel (see Fig.~\ref{figure:ZPE}(a-c)).

\paragraph*{Optimal proton transfer from instantons statistics}

The 
analysis made 
so far
highlights the paramount importance of the NQEs to set
the non-trivial temperature behaviour of the H$_{13}$O$_6^+$ cluster. At this stage, direct information about the excess proton dynamics along the QMC-PIMD trajectory is necessary to estimate more quantitatively its impact on the PT processes occurring in the system. 

One way to achieve this goal is by analysing the statistics of
selected transition-state (TS) configurations, defined by means of 
instanton theory. Within the PI
formalism, the instanton path seamlessly connects the reactants and products minima, along the minimal action trajectory, periodic in the quantum imaginary time $\tau = \beta \hbar$\cite{Richardson2009}. It provides a generalisation of the TS theory for anharmonic quantum systems\cite{Vanshten1982}, and it has been very recently applied in a QMC framework \cite{Jiang2017,Mazzola2017}, by efficiently recovering the proper scaling of ground-state tunneling rates. TS configurations are therefore identified as those where each half of the instanton path is located on either side of the central O$_1$O$_2$ midpoint, sampled during the QMC-PIMD dynamics.

\begin{figure}
    \centering
\includegraphics[width=1.0\textwidth]{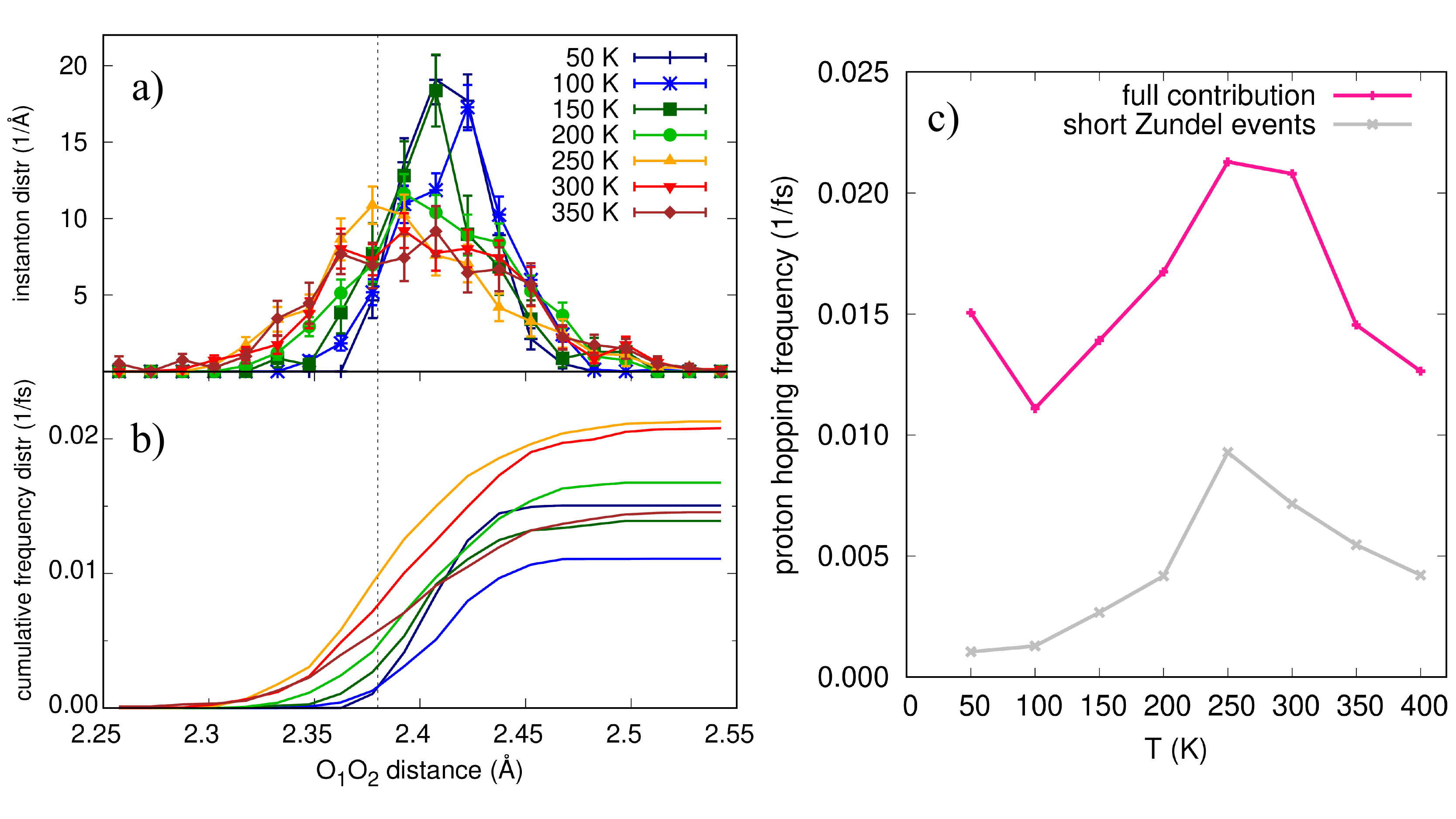}
    \caption{a) Instanton distribution resolved as a function of the $\textrm{d}_{\text{\scriptsize O}_1\text{\scriptsize O}_2}$ distance for different temperatures. b) Cumulative distribution of a) normalised by the occurrence frequency of the instanton (proton hopping) events during the PIMD simulations. c) Proton hopping frequency as a function of temperature, together with the contribution coming from the short Zundel configurations, with
    $\textrm{d}_{\text{\scriptsize O}_1\text{\scriptsize O}_2} < \textrm{d}_\textrm{symm} = 2.38$ \AA. The $\textrm{d}_\textrm{symm} $ value is reported as vertical dashed line in panels a) and b). Here, we report simulations performed also at 400 K, a temperature at which the cluster is still stable or meta-stable.
    }
    \label{figure:instantons}
\end{figure}

With the aim at resolving the contribution of the three different
regimes to the PT dynamics, we collect the instanton events and
compute their statistical distribution as a function of
$\text{d}_{\text{\scriptsize O}_1\text{\scriptsize O}_2}$. We plot the
instanton density distribution function in
Fig.~\ref{figure:instantons}(a) at various temperatures. To deepen our
analysis, we compute also the cumulative density distribution function
in Fig.~\ref{figure:instantons}(b), after normalising it based on the
algorithmic frequency of the instanton occurrences, as counted during
our QMC-PIMD simulations. Although this does not give direct access to
real-time
quantities, the ring-polymer MD with Langevin thermostat 
has been shown to yield
physically reliable
information
on frequencies and frequency variations\cite{hele2016}. Note that the coupling with the Langevin thermostat is kept constant across the full temperature range analysed here\cite{hele2016}.
The fully integrated frequency distribution gives the total proton hopping frequency, plotted in Fig.~\ref{figure:instantons}(c) as a function of temperature. This shows a clear maximum located in the $[250-300]$ K temperature range. Consequently, we expect the hydrated proton mobility to be optimal in a near-RT window, with a maximised Grotthus diffusion. To understand the source of this temperature ``sweet spot'', in the same panel (c) we plot the contribution to the total frequency of instanton events occurring in the short-Zundel region. This is yielded by the cumulative frequency distribution of panel (b) evaluated at the boundaries between short and elongated Zundel, i.e. at $\textrm{d}_{\text{\scriptsize O}_1\text{\scriptsize O}_2} = \textrm{d}_\textrm{symm}$. The short-Zundel contribution to the total frequency shows a peak of the same intensity as the total one in the same temperature range, clearly pointing to the key role played by thermally activated short-Zundel configurations to the PT dynamics. The short-Zundel arrangement enables instantaneous proton jumps between the two sides of the cation, since there is no barrier to cross. Thus, 
the ``sweet spot'' constitutes the best compromise between acquiring enough thermal 
energy to access short-distance configurations, boosted by NQEs, 
and controlling the amplitude of the chemical (covalent or H-) bonds fluctuations, that might trap the proton into an asymmetric well.
Indeed, at larger temperatures ($> 300$ K), 
the onset of distorted-Eigen and the corresponding fall of short-Zundel configurations localize the hydrated proton around its closest oxygen atom, thus reducing its shuttling probability.

Beside this PT mechanism, which is adiabatic in nature and driven by the synergy of ZPE and thermal effects, NQEs could also contribute to the proton diffusion by means of instantaneous tunneling, which can further accelerate the PT dynamics. By computing the root-mean-square (RMS) displacement correlation functions\cite{Chandler1994} over the instantons population, we verified that tunneling events could take place only in the distorted Eigen and in the intermediate temperature range (see Sec.~S6 of SI for a detailed analysis). This additional PT channel has however a marginal effect with respect to the main mechanism unveiled here. Indeed, Fig.~\ref{figure:instantons}(c) shows that the ``sweet spot'' is mainly due to PT events originating in short-Zundel configurations, where quantum tunneling is not relevant.

\section{Conclusions}
\label{conclusions}

Using unprecedentedly accurate QMC-PIMD simulations of the
H$_{13}$O$_6^+$ cation at finite temperature, we found a remarkably
low thermal expansion 
of the protonated water hexamer core. 
It stems from a cooperative action of both NQEs and thermal effects, which
leads to the emergent behaviour of short-Zundel species as PT booster, where the excess proton is perfectly shared between two neighbouring water molecules. 
The relevance of short-Zundel configurations is enhanced by NQEs, which instead penalize the distorted-Eigen states, having a larger ZPE. In the intermediate temperature range, comprising RT, the occurrence of short-Zundel events is maximised by thermal population, leading to a ``sweet spot'' in the PT dynamics. Around these temperatures, distorted-Eigen states can still contribute to PT with 
quantum tunneling processes, although occurring at much lower rates.
The cluster core spreads out again at larger temperatures, as soon as stronger thermal fluctuations favor the formation of more classical distorted-Eigen structures, where the proton gets strongly localised in one of the flanking molecules.

The short-Zundel quantum species is crucial for an efficient proton diffusion, 
as the shortness of its structure
enables a fast charge redistribution during the adiabatic PT process.
Very recent progress in ultrafast broadband two-dimensional (2D) IR spectroscopy\cite{Dahms2017,Fournier2018} allowed to probe the vibrational properties of protonated water at vibrational frequencies around the hydrated proton stretching mode, by measuring the lowest-lying excitations in the mid-infrared continuum\cite{Fournier2018}. These state-of-the-art experiments revealed a strongly inhomogeneous behaviour of the pump-probe spectra, implying large structural distributions in proton asymmetry and $\text{O}_1\text{O}_2$ distance. Therefore, the traditional ``Zundel limit''\cite{Zundel1968} needs to be revisited and extended, in order to cover the broad range of structures detected experimentally\cite{daly2017decomposition,carpenter2020}. In particular, the occurrence of qualitatively different short hydrogen-bond configurations, straightforwardly connected with the short-Zundel species described here, has been detected and highlighted in a recent fully solvated (HF$_2$)$^-$(H$_2$O)$_6$ experiment through femtosecond 2D IR spectroscopy in Ref.~\cite{dereka2021}. The present work crucially extends those findings by providing a temperature resolved analysis of the short hydrogen-bond events and by revealing their fundamental relation with the PT dynamics.

While proton transfer and proton transport occur in a variety of environments, from solutions to membrane proteins and fuel-cell membranes, the protonated water hexamer is the smallest cluster to incorporate most of the PT experimental features and solvation effects at the leading order. Our findings thus call for further efforts to explore the temperature behaviour of the proton dynamics and transport both in aqueous systems and in other environments, more specifically for biological systems around life conditions.

\section{Methods}
\label{methods}

\subsection{Zero-temperature electronic structure calculations}

 Before running finite-temperature calculations, we build a quantum Monte Carlo (QMC) variational wave function. The molecular dynamics (MD) will develop on the potential energy surface (PES) generated by the variational energy of this wave function. All zero- and finite-temperature calculations have been carried out with the TurboRVB code\cite{nakano2020}. We choose the variational ansatz such that the chemical accuracy in the binding energy of the Zundel ion and water dimer is reached. We highlight the fact that benchmarking the binding energy is much stricter than taking energy differences of geometries around the minimum, because the configurations involved in the binding energy are very different.
 
 The variational wave function $| \Psi_{\mathbf{q}} \rangle$ we used in our work is written as a Jastrow Antisymmetrised Geminal Power (JAGP) product\cite{Casula2004}
\begin{equation}
 \Psi_{\mathbf{q}}(\mathbf{x}_1,\dots,\mathbf{x}_{N_\mathrm{el}})  = J_{\mathbf{q}}(\mathbf{r}_1,\dots,\mathbf{r}_{N_\mathrm{el}}) \Psi_{AGP,\mathbf{q}}(\mathbf{x}_1,\dots,\mathbf{x}_{N_\mathrm{el}}).
\end{equation}
The set $\left\{ \mathbf{x}_i = (\mathbf{r}_i,\sigma_i) \right\}_{i=1,\dots,N_\mathrm{el}}$ represents spatial and spin coordinates of the $N_\mathrm{el}$ electrons. 

The JAGP wave function has been described extensively described elsewhere\cite{Casula2005,Sorella2007,michele_agp2,Dagrada2014}. Here, we report its main ingredients. The Bosonic Jastrow factor
is written as a product of one-body, two-body and three/four-body terms $J_{\mathbf{q}} = J_{1,\mathbf{q}}J_{2,\mathbf{q}}J_{3,\mathbf{q}}$. The one body term reads
\begin{eqnarray}
J_{1,\mathbf{q}}& = \exp \left ( - \sum\limits_i^{N_\mathrm{el}}\sum\limits_j^{N}(2Z_j)^{3/4}u\left((2Z_j)^{1/4}|\mathbf{r}_i - \mathbf{q}_j |\right)  \right )
\label{1body}
\end{eqnarray}
with $u(|\mathbf{r}-\mathbf{q}|) = \frac{1-e^{-b|\mathbf{r}-\mathbf{q}|}}{2b}$ and $b$ is a variational parameter, which satisfies the electron-ion Kato cusp conditions. 
$N$ the number of atoms and $Z_j$ the electric charge of the j-th atom.
In the protonated hexamer Hamiltonian, the hydrogen keeps the bare Coulomb potential, while for the oxygen atoms are replaced by
the Burkatzki-Filippi-Dolg (BFD) pseudopotential\cite{filippi_pseudo}, smooth at the electron-ion coalescence points. Thus, $J_{1,\mathbf{q}}$ is applied only to the hydrogen atom.
The two-body correlations are dealt with by the higher-order Jastrow factors. The two-body Jastrow factor is defined as 
\begin{equation}
J_{2,\mathbf{q}} = \exp \left ( \sum\limits_{i < j}^{N_\mathrm{el}}u(|\mathbf{r}_i-\mathbf{r}_j|) \right ),
\label{2body}
\end{equation}
where $u$ is a function of the same form as in Eq.~(\ref{1body}), but with a different variational parameter.  
The three-four body Jastrow factor is
\begin{equation}
J_{3,\mathbf{q}} = \exp \left( \sum\limits_{i<j}^{N_\mathrm{el}} \Phi_{J_{\mathbf{q}}}(\mathbf{r}_i,\mathbf{r}_{j}) \right),
\end{equation}
with 
\begin{equation}
\Phi_{J_{\mathbf{q}}}(\mathbf{r}_i,\mathbf{r}_j) = \sum\limits_{a,b}^{N}\sum\limits_{\mu,\nu}^{N^J_\mathrm{basis}}g_{\mu,\nu}^{a,b}\Psi_{a,\mu}^{J}(\mathbf{r}_i-\mathbf{q}_a)\Psi_{b,\nu}^{J}(\mathbf{r}_j-\mathbf{q}_b),
\label{jastrow_pairing}
\end{equation}
where $N^J_\mathrm{basis}$ is the number of the basis set functions $\Psi_{a,\mu}^{J}$. We used optimally contracted geminal embedded orbitals (GEOs)\cite{Sorella2015} as basis set, expanded over a primitive O(3s,2p,1d) H(2s,1p) Gaussian basis. The convergence study of the water dimer binding energy as a function of the Jastrow GEO expansion is reported in Sec.~S1.1 of the SI.

The Fermionic part of the wave function is expressed as an antisymmetrised product of the spin singlet geminals or pairing (AGP) functions $\Phi_{\mathbf{q}}(\mathbf{x}_i,\mathbf{x}_j)$:
\begin{equation}
\Psi_{AGP,\mathbf{q}}(\mathbf{x}_1,\dots,\mathbf{x}_{N_\mathrm{el}}) = \hat{A}\left[\Phi_{\mathbf{q}}(\mathbf{x}_1,\mathbf{x}_2),\dots,\Phi_{\mathbf{q}}(\mathbf{x}_{N_\mathrm{el}-1},\mathbf{x}_{N_\mathrm{el}})\right].
\end{equation}
The spatial part $\phi_{\mathbf{q}}(\mathbf{r}_i,\mathbf{r}_j)$ of the 
spin singlets $\Phi_{\mathbf{q}}(\mathbf{x}_i,\mathbf{x}_j)$ is expanded over $N^{AGP}_\mathrm{basis}$ optimally contracted GEOs,
such that
\begin{equation}
\phi_{\mathbf{q}}(\mathbf{r}_i,\mathbf{r}_j) = \sum\limits_{a,b}^{N}\sum\limits_{\mu,\nu}^{N^{AGP}_\mathrm{basis}}\lambda_{\mu,\nu}^{a,b}\bar{\Psi}_{a,\mu}^{AGP}(\mathbf{r}_i-\mathbf{q}_a)\bar{\Psi}_{b,\nu}^{AGP}(\mathbf{r}_j-\mathbf{q}_b).
\label{agp}
\end{equation}
The AGP GEOs are linear combination of primitive O(5s5p2d) H(4s2p) Gaussian basis functions. We highlight that the Fermionic part is fully optimised at the QMC level for each MD step, and not generated by on-the-fly DFT calculations.

The GEOs are very effective in reducing the total number of variational parameters, by keeping a high level of accuracy. This makes the wave function optimisation\cite{Sorella2001,Casula2004,Umrigar2007} much more efficient. 
Dealing with a compact wave function is very important, if one wants to use it as variational ansatz in a MD simulation, because in a MD framework the wave function needs to be optimised at every MD iteration. Indeed, the wave function optimisation is by far the most time consuming step of our QMC-driven MD approach.

Previous works on the Zundel ion\cite{Dagrada2014,Mouhat2017} found that the optimal balance between accuracy and computational cost for the determinantal part is reached by the O$[8]$H$[2]$ contracted GEO basis, in self-explaining notations. As the protonated water hexamer is a very similar system, in this work we used the same O$[8]$H$[2]$ GEO contraction for the AGP part. Moreover, we further simplified the variational wave function previously developed for the Zundel ion, by contracting also the Jastrow basis set, using the same GEO embedding scheme. We found that the O$[6]$H$[2]$ GEO basis set for the Jastrow factor is a very good compromise between accuracy and number of parameters, as we checked in the water dimer (see Sec.~S1.1 of the SI). The final accuracy is about 1 kcal/mol in the dissociation energy of the water dimer, and supposedly it is much higher around the stable geometries of water clusters. Thus, we used the O$[6]$H$[2]$ GEO basis set for the Jastrow factor, and the O$[8]$H$[2]$ GEO basis for the AGP part in all our subsequent MD simulations. For the protonated water hexamer, this results into a total number of 6418 variational parameters, comprising $g_{\mu,\nu}^{a,b}$ in Eq.~\ref{jastrow_pairing}, $\lambda_{\mu,\nu}^{a,b}$ in Eq.~\ref{agp}, the parameters of the homogeneous one-body (Eq.~\ref{1body}) and two-body (Eq.~\ref{2body}) Jastrow factors, and the linear coefficients of the Jastrow and determinantal basis sets.

\subsection{Finite-temperature calculations}

\subsubsection{Path Integral Langevin Dynamics}

At finite temperature, we carried out path integral Langevin dynamics simulations to include NQEs. To do so, we used the recently developed algorithm published in Ref.~\cite{Mouhat2017}, which
combines a path integral approach with very accurate Born-Oppenheimer (BO) forces computed by QMC. 
It is an efficient approach, alternative to the coupled electron ion Monte Carlo (CEIMC) developed by Ceperley and coworkers\cite{morales2010,liberatore2011,rillo2018}.
The intrinsic noise of the QMC force estimator is treated by the noise correction scheme developed in Refs.~\cite{Attaccalite2008,Mazzola2014,Mouhat2017}, which is based on the fulfilment of the fluctuation-dissipation theorem. This implies that the friction matrix $\gamma$ governing the dumped dynamics is related to the random force ${\boldsymbol \eta}$ via the $\alpha$ matrix:
\begin{eqnarray}
\alpha \left( {\mathbf{q}} \right) &=& 2 k_B T\gamma \left( {\mathbf{q}} \right), 
\label{fluctuation_dissipation}\\
\left\langle {{\eta _i}\left( t \right){\eta _j}\left( {t'} \right)} \right\rangle  &=& {\alpha _{i,j}}\left( {\mathbf{q}} \right)\delta \left( {t - t'} \right),
\label{stochastic_force}
\end{eqnarray}
with $\mathbf{q}$ the vector of nuclear coordinates. The $\mathbf{q}$-dependence comes from the QMC noise correction, implemented through the relations:
\begin{eqnarray}
\alpha \left( {\mathbf{q}} \right) & = & {\gamma_{BO}/(2 k_B T)} + {\Delta _0}{\alpha ^{{\text{QMC}}}}\left( {\mathbf{q}} \right) 
\label{alpha_matrix}
\\
\alpha _{i,j}^{^{{\text{QMC}}}}\left( {\mathbf{q}} \right) & = &
\left\langle {\left( {{\mathbf{f}_i}\left( {\mathbf{q}} \right) -
        \left\langle {{\mathbf{f}_i}\left( {\mathbf{q}} \right)}
        \right\rangle } \right)} \right\rangle 
\left\langle {\left( {{\mathbf{f}_j}\left( {\mathbf{q}} \right) 
     - \left\langle {{\mathbf{f}_j}\left( {\mathbf{q}} \right)} 
       \right\rangle } \right)} \right\rangle, 
\label{qmc_force_cov}
\end{eqnarray}
where $\alpha{^{{\text{QMC}}}}$ in Eq.~\ref{qmc_force_cov} is the covariance matrix of QMC ionic forces, measuring their stochastic fluctuations, and $\gamma_{BO}$ and $\Delta_0$ parameters taking values for an optimal Langevin dynamics. The random force ${\boldsymbol \eta}$ is then used to thermalise the system to a target temperature, according to the Langevin thermostat of Eq.~\ref{fluctuation_dissipation}.

The quantum particles are described by necklaces extended in the imaginary time interval $[0,\hbar \beta]$, with $\beta=1/(k_B T)$, following the quantum-to-classical isomorphism. This imaginary time interval is divided into $M$ slices, the so-called ``beads'', leading to an effective classical system of $NM$ particles. The path integral Langevin dynamics we developed in Ref.~\cite{Mouhat2017} is very efficient, because the quantum harmonic forces and the Langevin thermostat - thermalising the quantum degrees of freedom - are evolved together by means of an exact propagator, without Trotter breakup. The evolution of quantum particles in a thermal bath \`a la Langevin represents a \emph{quantum} Ornstein-Uhlenbeck dynamics. Therefore, we dubbed our integration scheme as ``path integral Ornstein Uhlenbeck dynamics (PIOUD)''. The algorithm is detailed in Ref.~\cite{Mouhat2017}.

In order to evolve the system and to sample the thermal quantum partition function, nuclear forces must be estimated at each iteration by computing the gradients $\mathbf{f}=-\boldsymbol{\nabla}_{\mathbf{q}}V(\mathbf{q})$. The potential energy surface $V(\mathbf{q})$ is evaluated by VMC, namely: 
\begin{equation}
V(\mathbf{q}) = \frac{\langle \Psi_{\mathbf{q}}|H(\mathbf{q})|\Psi_{\mathbf{q}}\rangle}{\langle \Psi_{\mathbf{q}}|\Psi_{\mathbf{q}}\rangle},
\label{V_wave_function}
\end{equation}
where $|\Psi_{\mathbf{q}}\rangle$ is the QMC wave function, which minimizes the expectation value of $H(\mathbf{q})$ for each bead configuration $\mathbf{q}^{(k)}$, according to the Born-Oppenheimer approximation.

The electronic variational wave function depends on the coordinates $\mathbf{q}^{(k)}$ of the $k$-th bead in two ways. Directly, through the explicit dependence on the ion positions provided by the localised basis set, and in an indirect way, through the wave function parameters optimised at each $\mathbf{q}^{(k)}$, in compliance with the Born-Oppenheimer approximation. While the former dependence is of leading order, the latter can be neglected in a first approximation. A clever way to do so is to average the optimal parameters across different beads, by gaining a significant fraction of statistics in the QMC energy minimisation. More precisely, the beads are gathered in groups of $N_\textrm{groups}$ members each, to share the same set of wave function parameters. This is called ``bead grouping approximation'', and it has been introduced in Ref.~\cite{Mouhat2017}.

Once the number of groups $N_\textrm{groups}$ is set for the beads, the electronic wave function is optimised at each new ionic position generated by the dynamics. This is done by energy minimisation via the most advanced optimisation techniques\cite{Sorella2007,Umrigar2007}. Between two consecutive steps of ion dynamics, one needs to perform $N_\textrm{opt}$ steps of energy minimisation, in an iterative fashion. $N_\textrm{opt}$ must be large enough to converge the wave function for each new ionic configuration. Thus, this parameter is tuned such that the BO approximation is fulfilled, and the dynamics follows the correct PES along the PIOUD trajectories.

During the dynamics, the GTO exponents in both the Jastrow and the AGP part of the wave function are kept frozen to make the simulation stable. Due to the continuity of the nuclear trajectories, the number of energy minimisation steps is significantly smaller than the one required for a wave function optimisation from scratch. 

There is a set of sensitive parameters one needs to tune to have stable and unbiased simulations. They are:
\begin{itemize}
    \item[(i)] Convergence for quantum effects set by $M$. In our calculations, we used M$=32$ for T$=300-400$~K and M$=64$ for T$=200$~K. The bead grouping approximation is made with $N_\textrm{group}=1$. Therefore the whole ring shares the same wave function parameters, except for the nuclear coordinates;
\item[(ii)] Time step $\delta t$ for the integration of the equations of motion. We used $\delta t = 1$ fs for a controlled time integration error, yielding a difference between the virial and primitive estimators of the quantum kinetic energy below a $25$ mHa threshold along all the trajectories;
\item[(iii)] A stable target temperature is reached despite the QMC noise by setting the parameters $\gamma_{BO}$ and $\Delta_0$, defining the $\alpha$ matrix in Eq.~\ref{alpha_matrix}. We used $\gamma_{BO}=0$ and $\Delta_0 = 0.5 ~ \delta t$, optimal values for other similar systems, such as the Zundel ion\cite{Mouhat2017}. Indeed, the simulation is efficient when the damping in the BO sector is minimised. The thermalisation of the system is guaranteed thanks to the optimal damping condition\cite{Ceriotti2010} 
applied to
the internal modes of the ring-polymer, 
with the damping parameter of
the center-of-mass translational modes
set to $0.231$ fs$^{-1}$;
\item[(iv)] To enforce the fulfilment of the BO approximation, we allowed for $N_\textrm{opt}=5$ iterative wave function optimisation steps at each MD iteration, such that the electronic energy minimum is reached within the statistical accuracy for every ionic configuration, and the PES is sampled without stochastic bias. 
\end{itemize}

\subsubsection{Classical Langevin Dynamics}

For classical MD calculations, we used an improved variant of the original algorithm developed by Attaccalite and Sorella\cite{Attaccalite2008}. This variant has been detailed in Refs.~\cite{Mazzola2014,Mouhat2017}. It includes a better integration scheme, involving a Langevin noise touching both coordinates and momenta, which are therefore correlated.

As in the PIOUD calculations, relevant parameters are the time step $\delta t = 1$ fs, the QMC noise correction parameters $\gamma_{BO} =0.2$ fs$^{-1}$ and $\Delta_0 = \delta t$, and the number of QMC optimisation steps per nuclear iteration $N_\textrm{opt}=5$.

\section*{Acknowledgments}
{\small
FM, RV, AMS and MC acknowledge Dominik Marx for useful discussions.
MP, RV, AMS and MC thank the CNRS for the 80PRIME doctoral grant allocation.  
FM and MC acknowledge computational resources provided by the PRACE
projects number 2016133322 and 2017153936. MC thanks GENCI
for providing computational resources at TGCC and IDRIS under the grant number 0906493, and the Grands Challenges DARI for allowing calculations on the Joliot-Curie Rome HPC cluster under the project number gch0420. TM and MC are grateful to the European Centre of Excellence in Exascale Computing TREX-Targeting Real Chemical Accuracy at the Exascale, which partially supported this work.
This project has received funding from the European Unions Horizon 2020 Research and Innovation program under Grant Agreement No. 952165.
}

\clearpage

\clearpage

\setcounter{figure}{0}
\setcounter{page}{1}
\setcounter{section}{0}
\setcounter{table}{0}
\renewcommand{\thepage}{S\arabic{page}} 
\renewcommand{\thesection}{S\arabic{section}}  
\renewcommand{\thetable}{S\arabic{table}}  
\renewcommand{\thefigure}{S\arabic{figure}}

\section*{Supplementary information}

\section{Zero-temperature electronic structure calculations}

\subsection{Preparation and optimisation of quantum Monte Carlo (QMC) wave function}

 As described in the Methods Section, before running finite-temperature calculations, we optimize a QMC variational wave function  $| \Psi_{\mathbf{q}} \rangle$ at zero temperature, written as a  Jastrow Antisymmetrised Geminal Power (JAGP) ansatz\cite{Casula2004}. During the finite-temperature molecular dynamics, energy and forces are computed at the variational Monte Carlo (VMC) level, based on the variational optimisation of the QMC wave function. 
 Both Jastrow and AGP expansions are developed over a primitive O(3s2p1d) H(2s1p) 
and O(5s5p2d) H(4s2p) Gaussian basis functions, respectively. The primitive basis sets are then contracted using the 
geminal embedded orbitals (GEOs) scheme\cite{Sorella2015}.

\begin{figure}[!htb]
\centering
\includegraphics[width=0.8\textwidth]{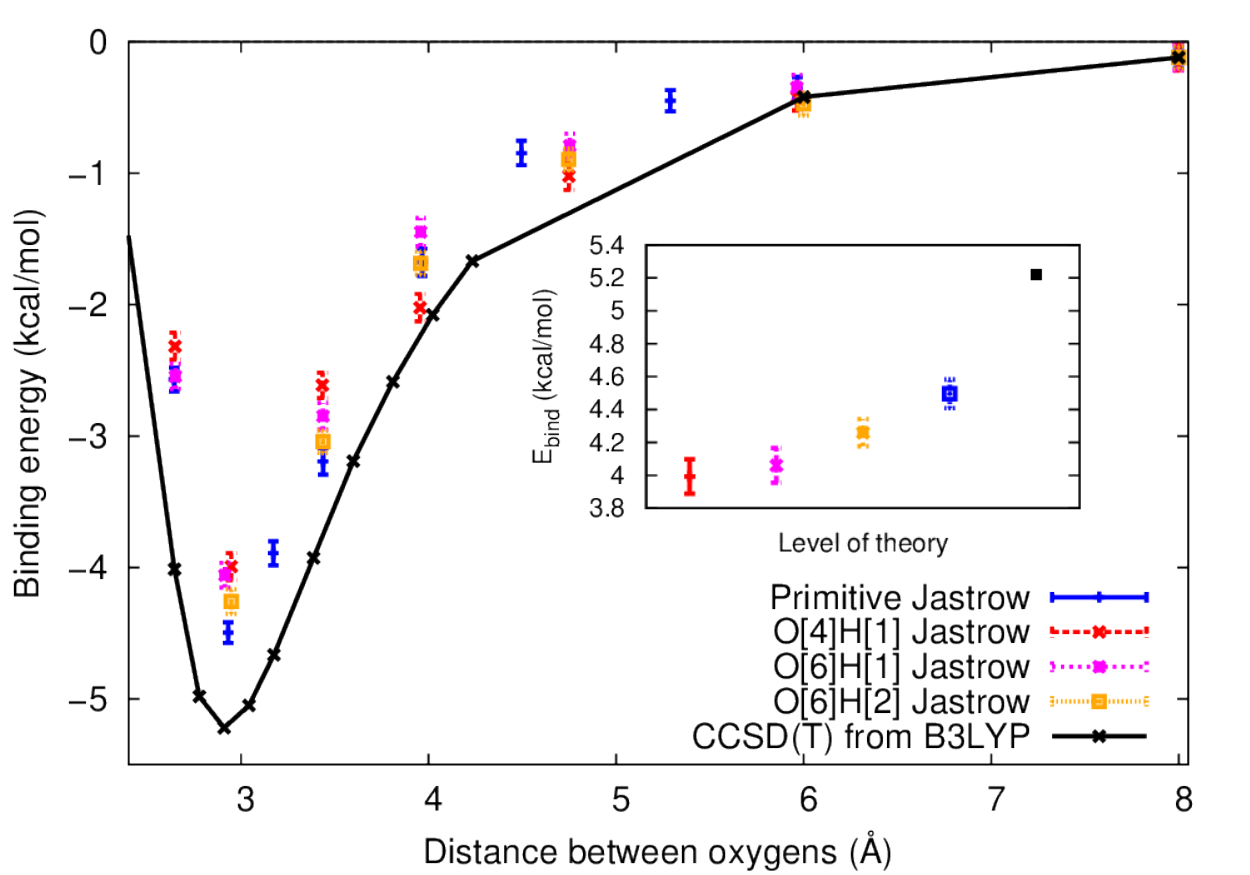}
\caption{\label{figure:jas} Water dimer dissociation energy curve as a function of the oxygen-oxygen distance obtained by VMC, using different contracted basis sets in the Jastrow factor of a Jastrow-Slater wave function. Each trial wave function is built using the same basis set for the determinantal part, which is optimised together with the various Jastrow factors tested here. The black curve indicates the reference CCSD(T) result.}
\end{figure}

Previous works on the Zundel ion\cite{Dagrada2014,Mouhat2017} found that the optimal balance between accuracy and computational cost for the determinantal part is reached by the O$[8]$H$[2]$ contracted GEO basis, in self-explaining notations. As the protonated water hexamer is a very similar system, in this work we used the same O$[8]$H$[2]$ GEO contraction for the AGP part. Moreover, we further simplified the variational wave function previously developed for the Zundel ion, by contracting also the Jastrow basis set, using the same GEO embedding scheme. We tried different contraction sets, and tested them on the water dimer dissociation energy curve, as reported in Fig~\ref{figure:jas}. The water dimer is a stringent benchmark for the quality of our wave function, as it has a chemical complexity similar to the Zundel ion, with the main difference of being charge neutral. Charge neutrality allows us to directly probe the Jastrow capability of controlling charge fluctuations in the system, a fundamental property when coupled with the AGP determinantal part\cite{eric_size_cons}.

As shown in Fig.~\ref{figure:jas}, we find a systematic improvement as the number of GEOs orbitals increases, with the O$[6]$H$[2]$ set yielding energies very close to the Jastrow primitive basis set reference at all oxygen-oxygen distances. As reported in Tab.~\ref{table:Jparameters}, this is obtained with a number $p$ of variational parameters significantly smaller than the one of the primitive basis set expansion. Thus, we used the O$[6]$H$[2]$ GEO basis set for the Jastrow factor, and the O$[8]$H$[2]$ GEO basis for the AGP part in all our subsequent molecular dynamics (MD) simulations. For the protonated water hexamer, this results into a total number of 6418 variational parameters, comprising $g_{\mu,\nu}^{a,b}$, $\lambda_{\mu,\nu}^{a,b}$, the parameters of the homogeneous one-body and two-body Jastrow factors, and the linear coefficients of the Jastrow and determinantal basis sets (see Methods Section for a detailed description of the wave function parameters).

\begin{table}[!htp]
 \centering
 \hspace*{-10mm}
\begin{tabular}{|c|c|c|}
\hline
      Basis set & $p$ &  E$_{\text{bind}}$ (kcal/mol)     \\
\hline
\hline
Primitive Jastrow and primitive determinant & $6303$ & $ 4.46(8)$  \\
Primitive Jastrow and O$[8]$H$[2]$ GEO determinant & $2089$   & $ 4.40(8)$  \\
O$[6]$H$[2]$ GEO Jastrow and O$[8]$H$[2]$ GEO determinant & $1283$ & $4.26(8)$  \\
\hline 
\end{tabular}
\caption {\label{table:Jparameters} Water dimer binding energies for QMC variational wave functions obtained with different types of basis set contractions. The corresponding number $p$ of variational parameters is also reported.}
\end{table}

A more extended description of the variational wave function can be found in Ref.~\cite{Dagrada2014}.

\subsection{Comparison between QMC and other electronic structure methods}

To probe the microscopic rearrangement which triggers proton transfer (PT) processes in
the protonated water hexamer at finite temperature, it is necessary to
determine the potential energy surface (PES), in particular around 
the equilibrium geometry of the cluster.
At variance with the Zundel cation, the protonated
hexamer minimum energy configuration is asymmetric, implying that the
hydrated proton is not equally shared between the two central water
molecules. 

\begin{figure}[!htb]
\centering
\includegraphics[width=0.8\textwidth]{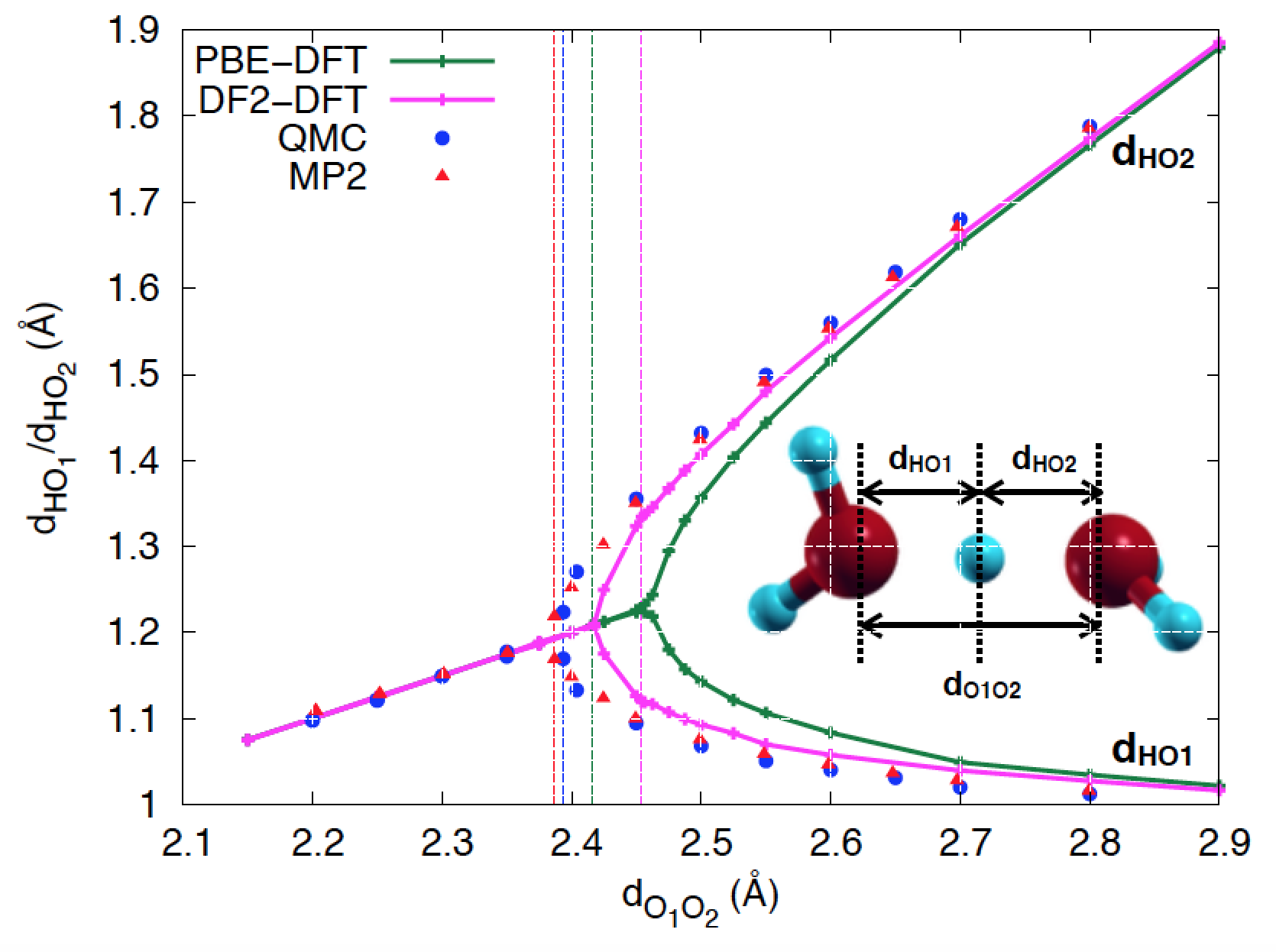}
\caption { Equilibrium position of the excess proton H - expressed as distance d$_{{\text{\scriptsize H}}{\text{\scriptsize O}}_1}$ 
(d$_{{\text{\scriptsize H}}{\text{\scriptsize O}}_2}$) 
from the flanking oxygen atom O$_1$ (O$_2$) - as a function of 
the separation d$_{{\text{\scriptsize O}}_1{\text{\scriptsize O}}_2}$ 
between the two central oxygen atoms, reported for different computational methods. Vertical dashed lines indicate the equilibrium d$_{\text{O}_1\text{O}_2}$ for each method.  
\label{figure:staticgeohex}
}
\end{figure}

In Fig.~\ref{figure:staticgeohex}, we report the equilibrium position of the hydrated proton H$^+$ - expressed as distance d$_{{\text{\scriptsize H$^+$}}{\text{\scriptsize O}}_1}$ 
(d$_{{\text{\scriptsize H$^+$}}{\text{\scriptsize O}}_2}$) 
from the flanking oxygen atom O$_1$ (O$_2$) - as a function of 
the distance d$_{{\text{\scriptsize O}}_1{\text{\scriptsize O}}_2}$ 
between the two central oxygen atoms. Fig.~\ref{figure:staticgeohex} shows the equilibrium geometries at constrained d$_{{\text{\scriptsize O}}_1{\text{\scriptsize O}}_2}$ obtained by various methods: density functional theory (DFT) with the PBE functional (dark green), DFT modified to include dispersive van der Waals (vdW) interactions in the DF2 implementation\cite{Lee2010} (magenta), variational Monte Carlo (blue circles) and M\o ller-Plesset (MP2) (red triangles). The d$_{{\text{\scriptsize O}}_1{\text{\scriptsize O}}_2}$ distance corresponding to the global minimum is indicated by a vertical dashed line for each method.

The PBE functional predicts a short-Zundel-like symmetric global minimum, which is erroneous. More generally, this functional gives a poor description of the proton location when one stretches the d$_{{\text{\scriptsize O}}_1{\text{\scriptsize O}}_2}$ distance. After inclusion of dispersion effects in the DF2 functional, the geometric properties of the system are significantly improved, displaying a better agreement with both QMC and MP2. Nevertheless, the predicted 
equilibrium d$_{{\text{\scriptsize O}}_1{\text{\scriptsize O}}_2}$ is too large, which leads to an overly asymmetric cluster. Therefore, we expect the DF2 static barriers to be inaccurate, which is problematic in the perspective of studying PT at finite temperature by this method. Instead, MP2 and QMC are in excellent agreement, especially around d$_{\text{O}_1\text{O}_2} = 2.4$ \AA.

In Tab.~\ref{table_geometry_hex_core_dist} we report the equilibrium geometries yielded by PBE, DF2, MP2 and VMC, by showing the relevant distances involving the Zundel core of the protonated water hexamer. The VMC minimum geometry is in a very good agreement with the MP2 one, with an accurate description of the excess proton localisation, as previously seen in Fig.~\ref{figure:staticgeohex}. The largest discrepancy between VMC and MP2 is for the $\overline{OH}$ intramolecular distances, which are slightly shorter in VMC. However, 
the overall evolution of equilibrium position of the hydrated proton H$^{+}$ as a function of the oxygen-oxygen distance predicted by QMC follows well the one obtained by MP2, as one can see from Fig.~\ref{figure:staticgeohex}. 

\begin{table}[!htp]
\centering
\hspace*{-8mm}
\begin{tabular}{|c|c|c|c|c|c|c|c|}
\hline
\multicolumn{1}{|c|}{Theory}  & \multicolumn{1}{c|}{$\overline{O_1O_2}$}  & \multicolumn{1}{c|}{$\overline{O_1H^+}$} & \multicolumn{1}{c|}{$\overline{H^+O_2}$}  & \multicolumn{1}{c|}{$\overline{O_1H_1}$} & \multicolumn{1}{c|}{$\overline{O_1H_2}$} &  \multicolumn{1}{c|}{$\overline{O_2H_3}$}  &  \multicolumn{1}{c|}{$\overline{O_2H_4}$}  \\
\hline
DFT-PBE &  2.4156  & 1.2078  &  1.2078  & 0.9935 & 0.9935 & 0.9935 & 0.9935 \\
DFT-DF2  & 2.4541 & 1.1196 & 1.3346 & 0.9913 & 0.9911 & 0.9797 & 0.9797  \\
MP2 & 2.3867 & 1.1690  & 1.2188 & 0.9877 & 0.9878  & 0.9847 & 0.9848 \\
VMC & 2.3930(5) & 1.1555(5) & 1.2375(5) & 0.9800(8) & 0.9798(8) & 0.9752(8) & 0.9748(8)  \\
\hline
\end{tabular}
\caption {Geometric properties (distances in \AA) of the core of the protonated water hexamer minimum. Comparison
between different computational methods.
 \label{table_geometry_hex_core_dist}
 }
\end{table}
\vspace*{3mm}

Due to the central
distorted H-bond, the hydrated proton motion from one flanking water
molecule to another is conditioned by the necessary energy it should
acquire to go across the PT static barrier.
This quantity is defined as the energy difference between the
equilibrium structure (asymmetric) and the symmetrised one, where the
proton is located at the mid-point of the considered
$\text{d}_{\text{O}_1\text{O}_2}$ distance. 
PT static barriers have been estimated at $\text{d}_{\text{O}_1\text{O}_2} = 2.45$ \AA \, 
and $2.5$ \AA. The barriers, converted into effective temperatures, are reported in
Tab.~\ref{table:staticbis} for different levels of theory. 

\begin{table}[!htp]
\centering
\begin{tabular}{|c|c|c|}
\hline
      $\text{d}_{\text{O}_1\text{O}_2}$ (\AA) & 2.45 \AA     & 2.50 \AA             \\
\hline
DFT-DF2  & 39 & 483 \\
CCSD(T)  & 141  & 431 \\
MP2 & 85 & 327          \\
VMC  & $195 \pm 25$ &$562 \pm 27$  \\
\hline
\end{tabular}
\caption {\label{table:staticbis} Static symmetrisation barriers (in
  Kelvin) of the H$_{13}$O$_{6}^+$ cation at different
  $\text{d}_{\text{O}_1\text{O}_2}$ distances for various computational
  methods.}  
\end{table}

DFT-DF2 seems unable to predict accurate PT barriers due to the misplacement of the global energy minimum. This further motivates the use of correlated methods to describe the 
electronic PES, such as coupled cluster CCSD(T) and MP2 theories.
Their values are closer to the barriers predicted by VMC. They differ between each other by 0.1-0.2 kcal/mol, a difference within chemical accuracy. However, the QMC method exhibits a milder scaling with the system size. Therefore, the QMC approach is certainly the best candidate to
perform fully \emph{ab initio} MD simulations of small protonated water
clusters at an affordable computational cost.

\section{Finite-temperature calculations: QMC-driven classical and quantum Langevin dynamics}
\label{simulations}

We simulated the protonated water hexamer by treating the electrons at the VMC level (JAGP wave function) and the nuclei considered as quantum particles as well, within a path integral (PI) formalism. To understand the impact of quantum effects, simulations with classical nuclei have also been performed. The Langevin dynamics (LD) algorithms used have been described in the Methods Section.

In Tab.~\ref{simulations_table}, we report the list of VMC+PILD simulations done. For their importance, particularly long simulations are performed for the quantum case at temperatures of 50, 100, 200, and 300 K. 

\begin{table}[h]
  \caption{Summary 
   of the simulations carried out in this work. In both classical and quantum calculations, a time step $\delta t$ of 1 ft is used for all temperatures. 
}
\label{simulations_table}
\begin{tabular}{| c | c | c | c |}
  \hline
  & \multicolumn{2}{c|}{\makebox[100pt][r]{quantum simulations}} &  \makebox[100pt][r]{classical simulations} \\
  \hline
  \makebox[50pt][c]{$T$(K)} & \makebox[120pt][c]{$N_{\textrm{beads}}$} & \makebox[120pt][c]{$N_{\textrm{iterations}}$} & \makebox[120pt][c]{$N_{\textrm{iterations}}$} \\
  \hline 
  50       & 128    & 35282  &  - \\
  100     & 128    &  52184  & 21454 \\
  150     & 64      &  11218 & - \\
  200     & 64       &  32553 & 20478 \\
  250     & 32      &  23912 & 24154 \\
  300      &  32      &  31929 & 22656 \\
  350      &  32      &  18489 & 26481 \\
  400      &  32      &  23026 & 27517 \\
 \hline
\end{tabular}
\end{table}

\section{Structural properties of protonated water clusters}

\subsection{Role of solvation: Zundel ion versus protonated water hexamer}

\begin{figure}[!htb]
\centering
\begin{tabular}{cc}
    \includegraphics[width=0.495\textwidth]{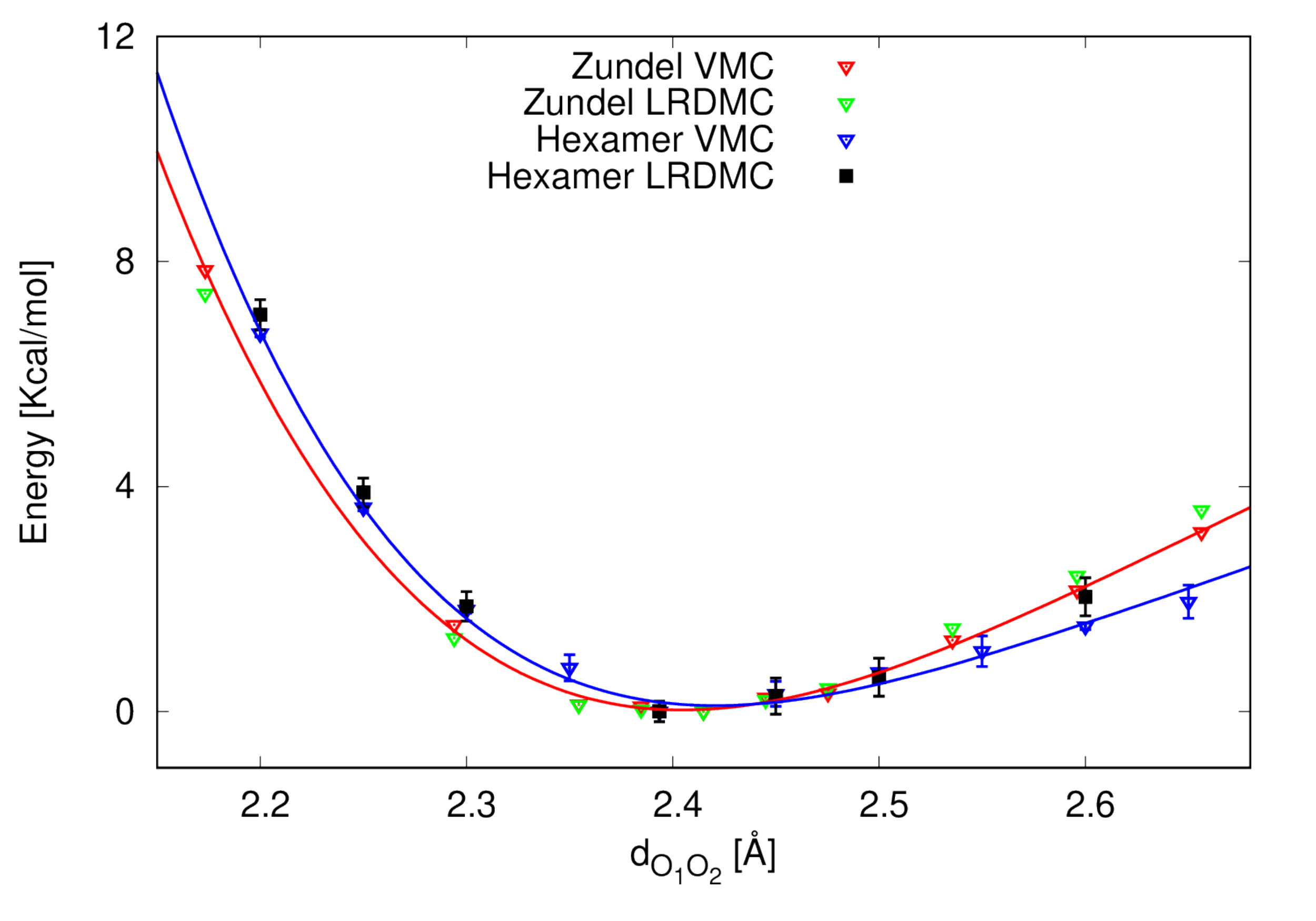} &
     \includegraphics[width=0.505\textwidth]{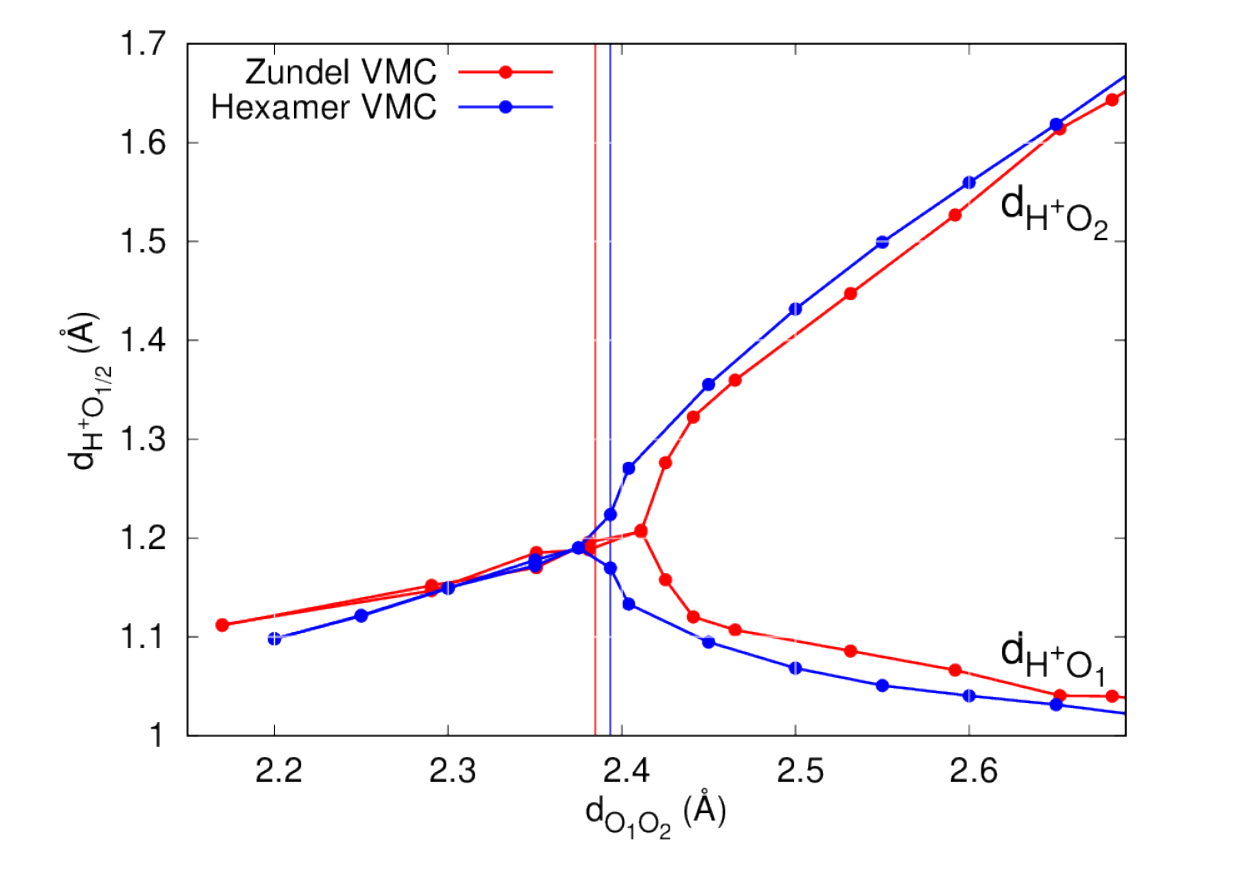}
\end{tabular}
\caption{\label{figure:concludesolv} Comparison of the protonated water dimer and hexamer 
$V_{\text{\scriptsize O}_1\text{\scriptsize O}_2}$ potential (left)
and equilibrium geometry (right) as a function of the (central) oxygen-oxygen distance d$_{\text{\scriptsize O}_1\text{\scriptsize O}_2}$. Vertical dashed lines indicate the corresponding equilibrium d$_{\text{\scriptsize O}_1\text{\scriptsize O}_2}$. Notice that VMC and lattice regularised diffusion Monte Carlo (LRDMC)\cite{Casula2005} energies are in nice statistical agreement for d$_{\text{\scriptsize O}_1\text{\scriptsize O}_2}$ $\in [2.3, 2.6]$ \AA, the phase-space range explored by our MD simulations.}
\end{figure}

As mentioned in the main text of the paper, the protonated water hexamer is the smallest water cluster including the full first-solvation shell, due to the presence of a Zundel core surrounded by 4 solvating H$_2$O molecules. To quantify the impact of the solvation shell on the Zundel core, we compare in Fig.~\ref{figure:concludesolv} the 
$\text{O}_1$-$\text{O}_2$ potential, $V_{\text{\scriptsize O}_1\text{\scriptsize O}_2}$
(left), and the corresponding classical equilibrium geometry (right) of the two clusters at various d$_{\text{\scriptsize O}_1\text{\scriptsize O}_2}$ (distance between the 2 central oxygen atoms). At short d$_{\text{\scriptsize O}_1\text{\scriptsize O}_2}$, the slope of the protonated hexamer $V_{\text{\scriptsize O}_1\text{\scriptsize O}_2}$ is slightly 
larger
than the Zundel one, due to a greater electrostatic repulsion because of steric hindrance.
At large d$_{\text{\scriptsize O}_1\text{\scriptsize O}_2}$, the protonated hexamer PES 
is
softer than the Zundel one, because the solvating H$_2$O molecules
enhance the polarisability of the core atoms. As explained in the
paper, the balance between short- and long-range repulsion, once
supplemented with the zero-point energie (ZPE), is key to quantify the
relative abundance of short-Zundel and distorted-Eigen configurations,
and thus, it allows for a quantitative understanding of the PT
mechanism.

We also find the H$_{13}$O$_6^+$ equilibrium d$_{\text{\scriptsize O}_1\text{\scriptsize O}_2}$, represented by a vertical dashed line in Fig.~\ref{figure:concludesolv}, to be $\sim 0.05$ \AA \, larger than the H$_{5}$O$_2^+$ one. More importantly, at variance with the Zundel cation which is centrosymmetric, the protonated water hexamer equilibrium geometry is \emph{asymmetric} with classical ions. This fundamental symmetry modification of the PES
is induced by 
solvation effects, which 
tend to stabilize the hexamer into its 
elongated-Zundel
configuration.
This can rationalise some THz/FTIR absorption spectroscopy 
fingerprints of the solvated proton\cite{Decka2015}, which have been related to a fast inter-conversion between the (distorted-)Eigen and (short-)Zundel forms.

Furthermore, 
it is noteworthy that at fixed d$_{\text{\scriptsize O}_1\text{\scriptsize O}_2}$, the predicted barriers are larger in H$_{13}$O$_{6}^+$ (Tab.~\ref{table:staticbis}) than in the Zundel cation\cite{Dagrada2014}. This is clearly due to the additional price that must be paid for the rearrangement of the molecules in the solvation shell during the PT process.

\subsection{H$_{13}$O$_6^+$ bidimensional distribution functions}

\begin{figure}[!h]
\centering 
\begin{tabular}{lr} 
\includegraphics[width=0.5\columnwidth]{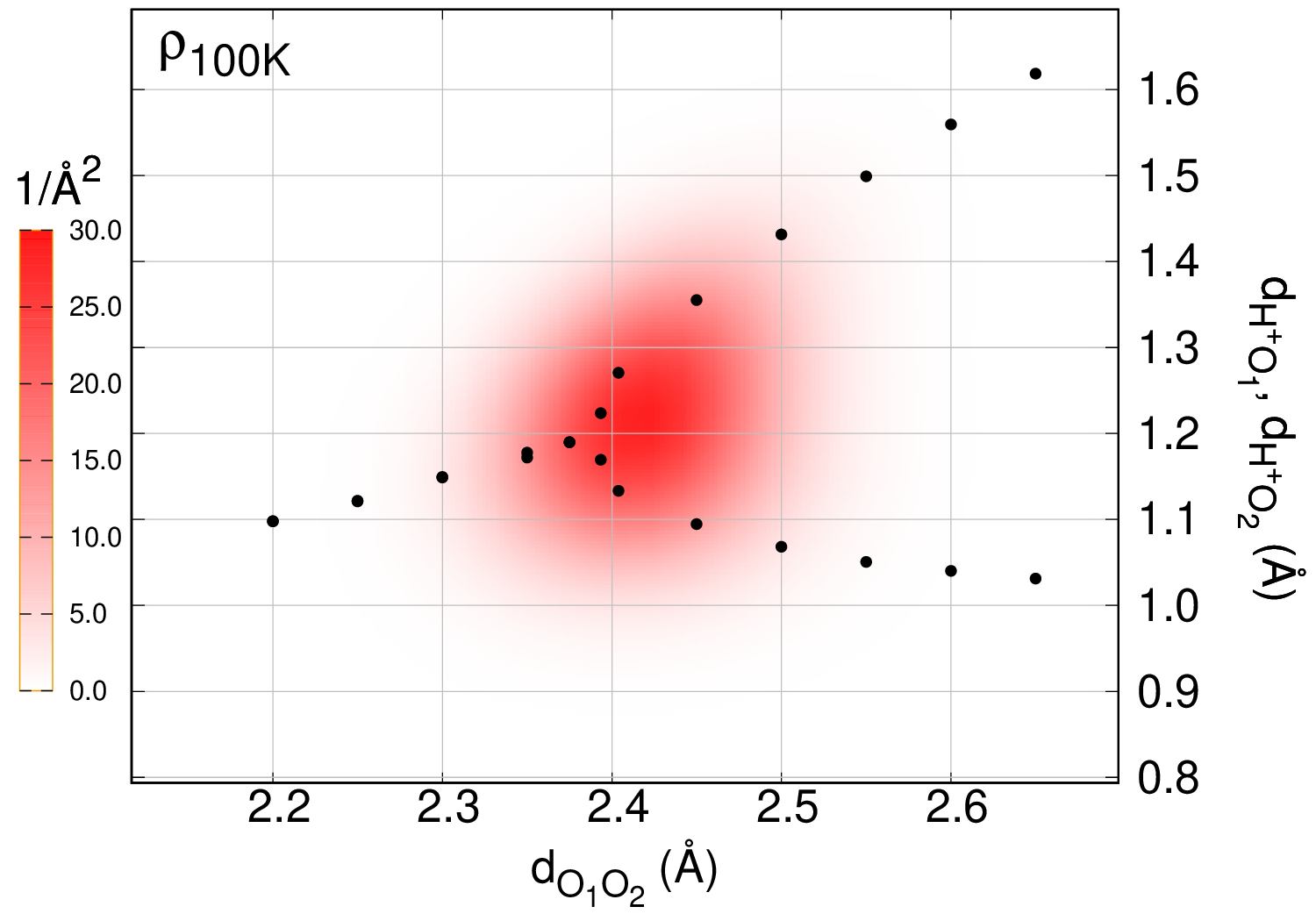} &
\includegraphics[width=0.5\columnwidth]{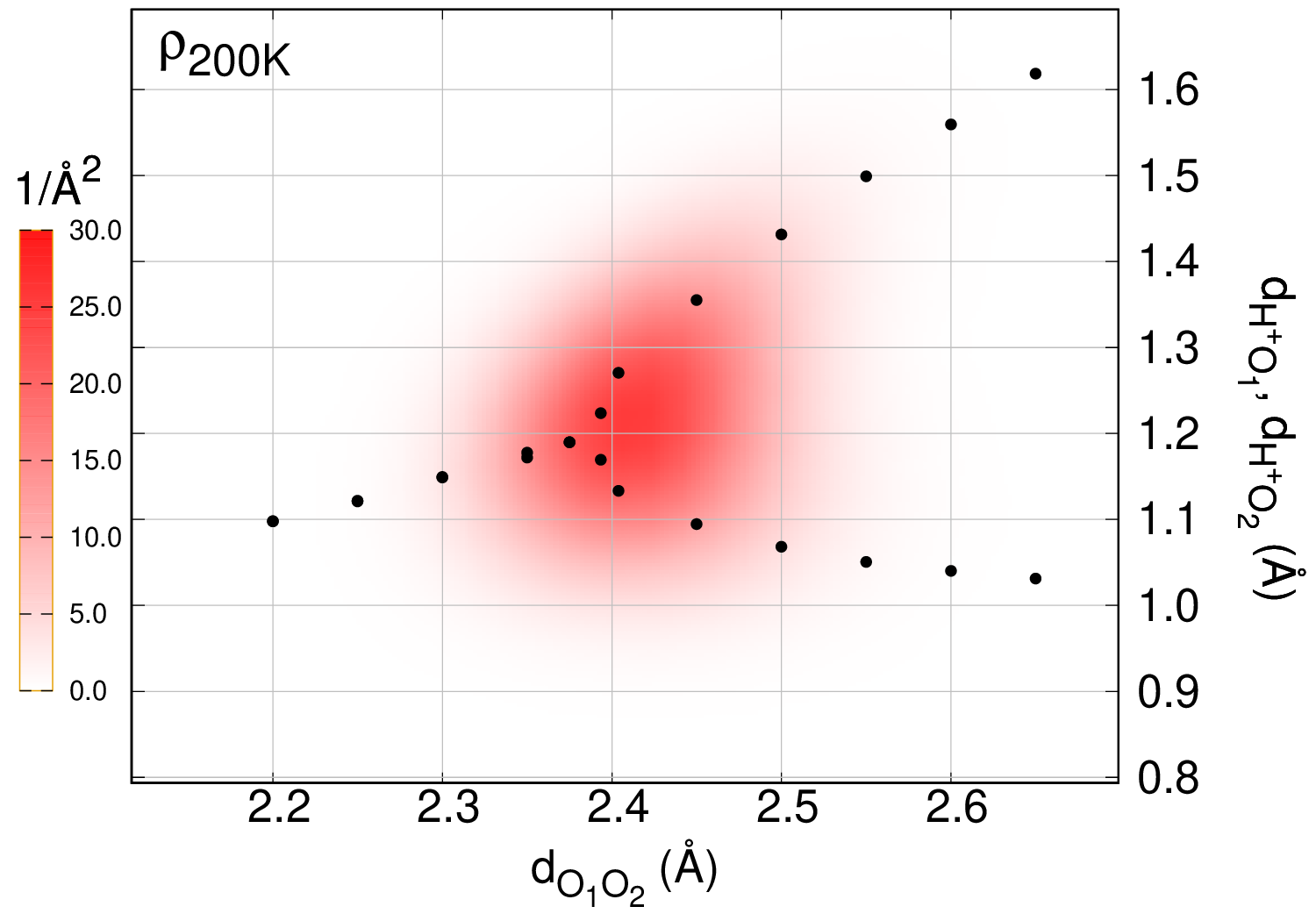} \\
\includegraphics[width=0.5\columnwidth]{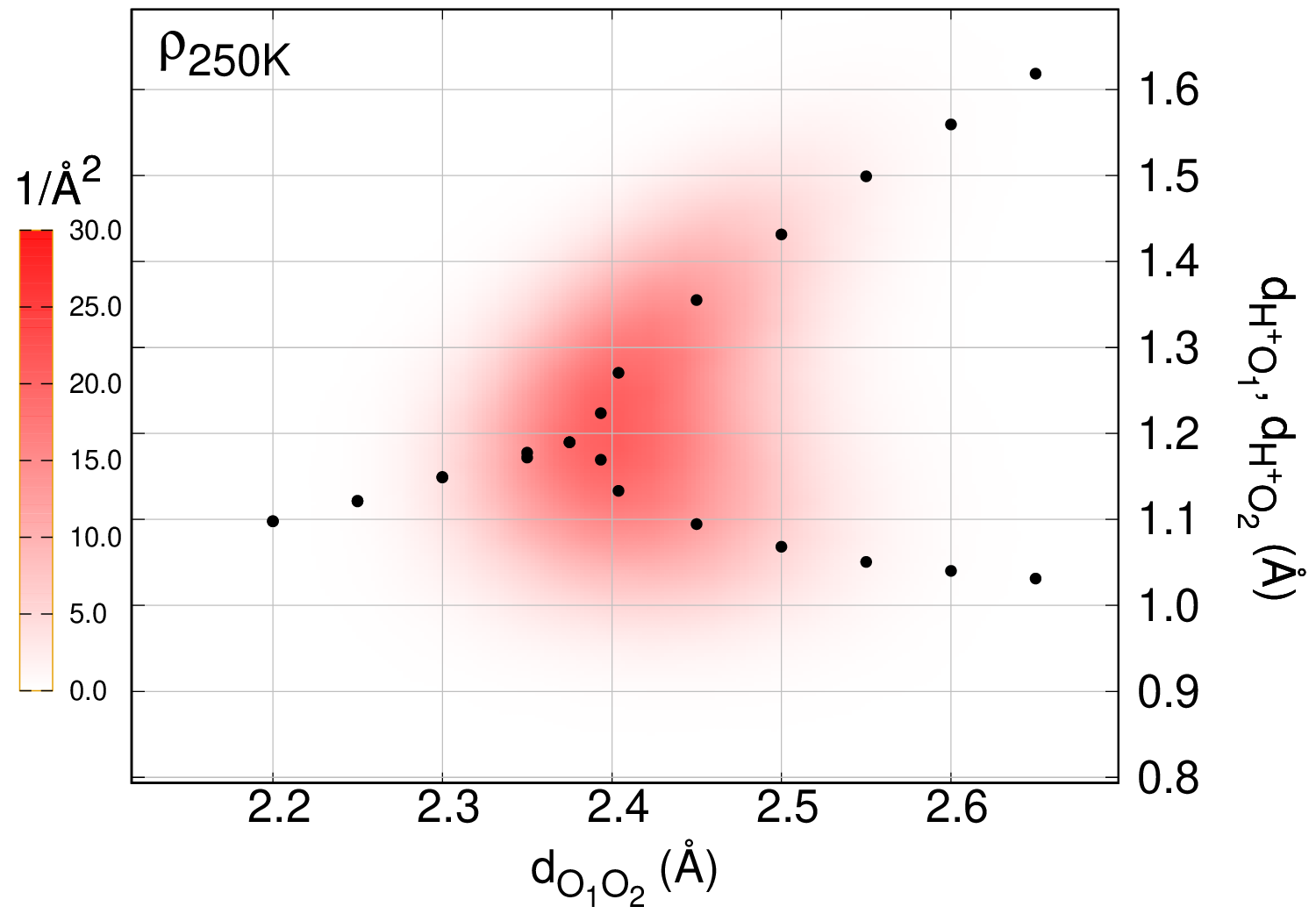} &
\includegraphics[width=0.5\columnwidth]{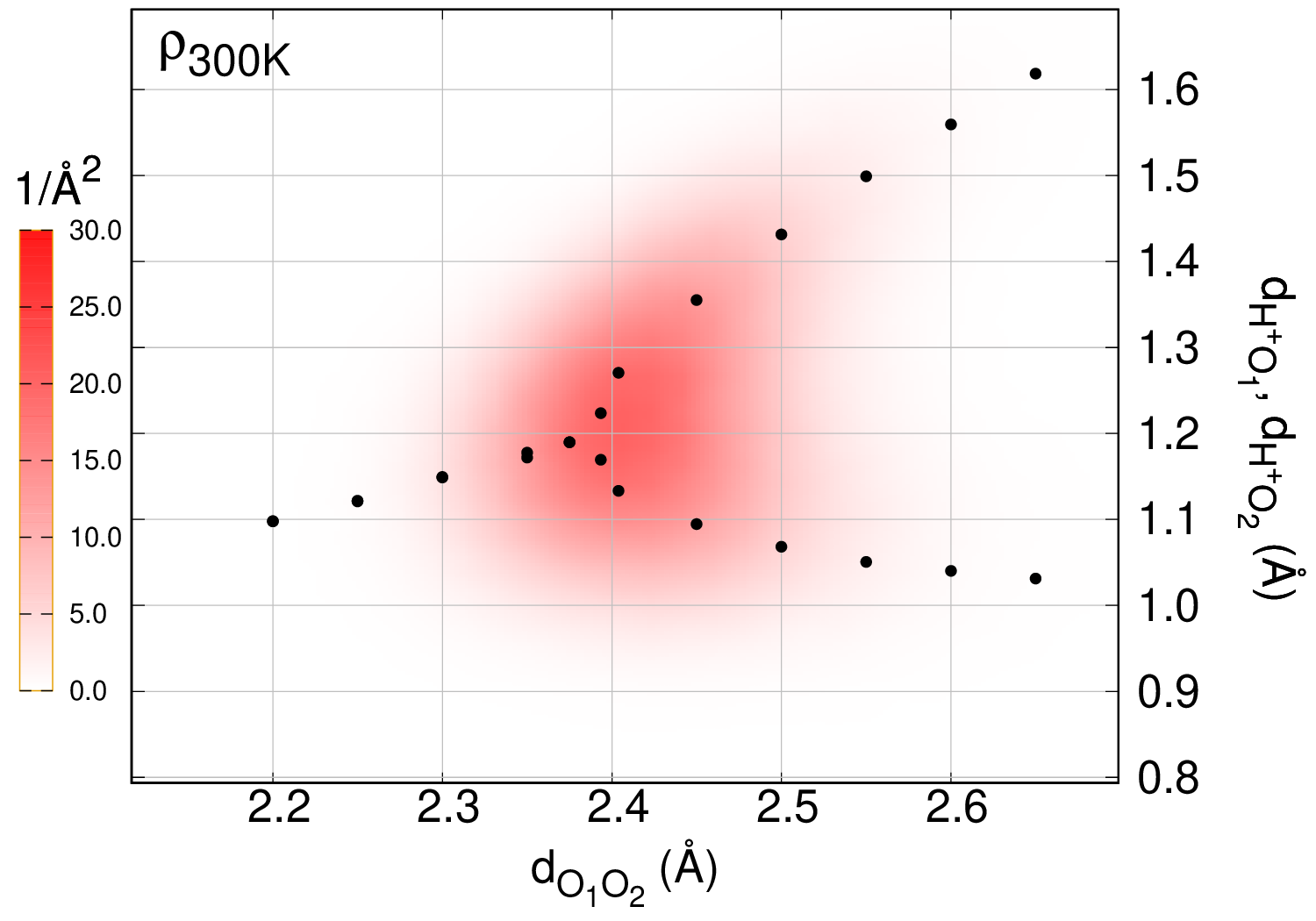} \\
\includegraphics[width=0.5\columnwidth]{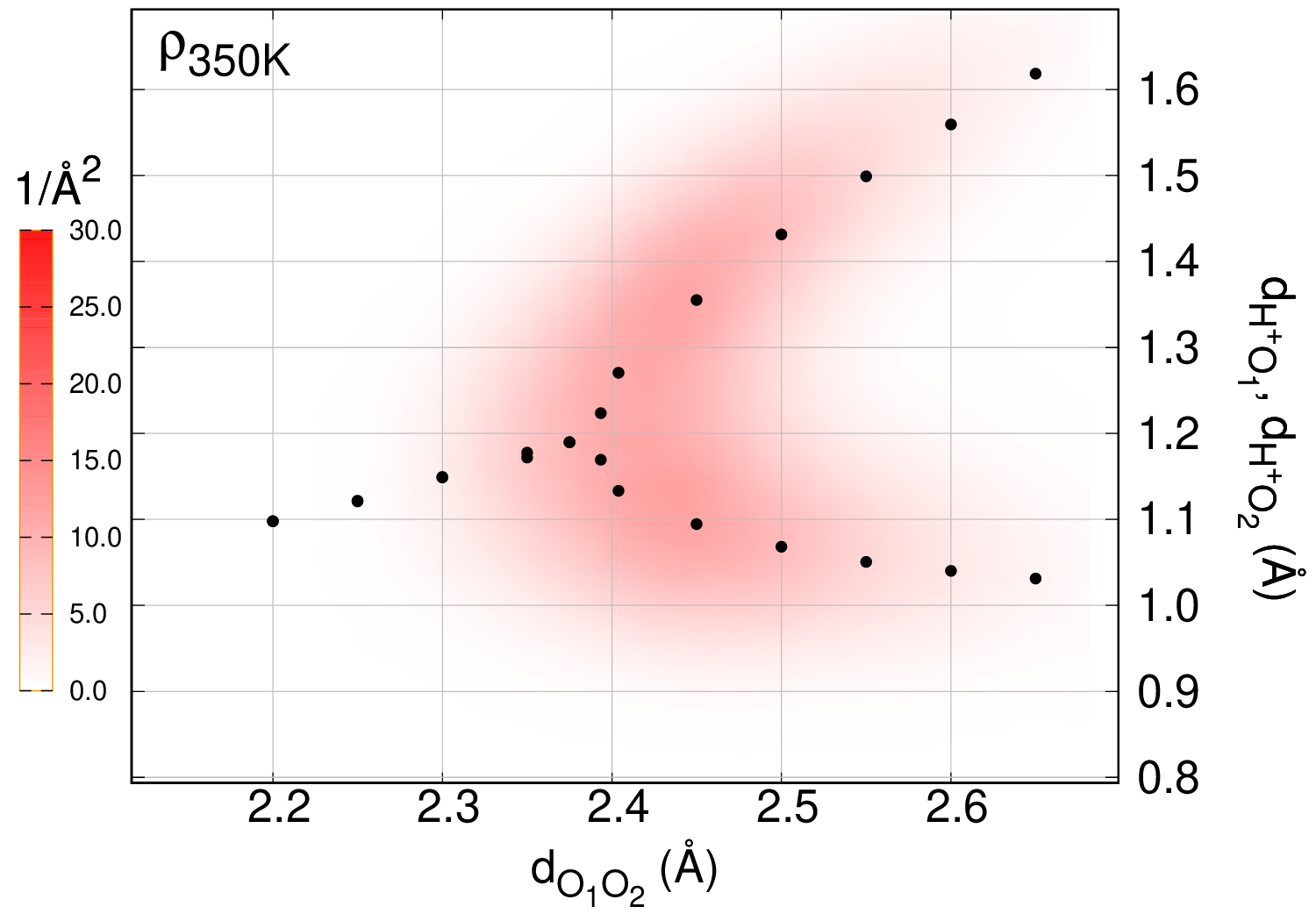} &
\\  
\end{tabular}
\caption{$\rho_{2D}$ computed from VMC-PIMD simulations at different temperatures.
}
\label{g2d_quantum}
\end{figure}

\begin{figure}[!h]
\centering 
\begin{tabular}{lr} 
\includegraphics[width=0.5\columnwidth]{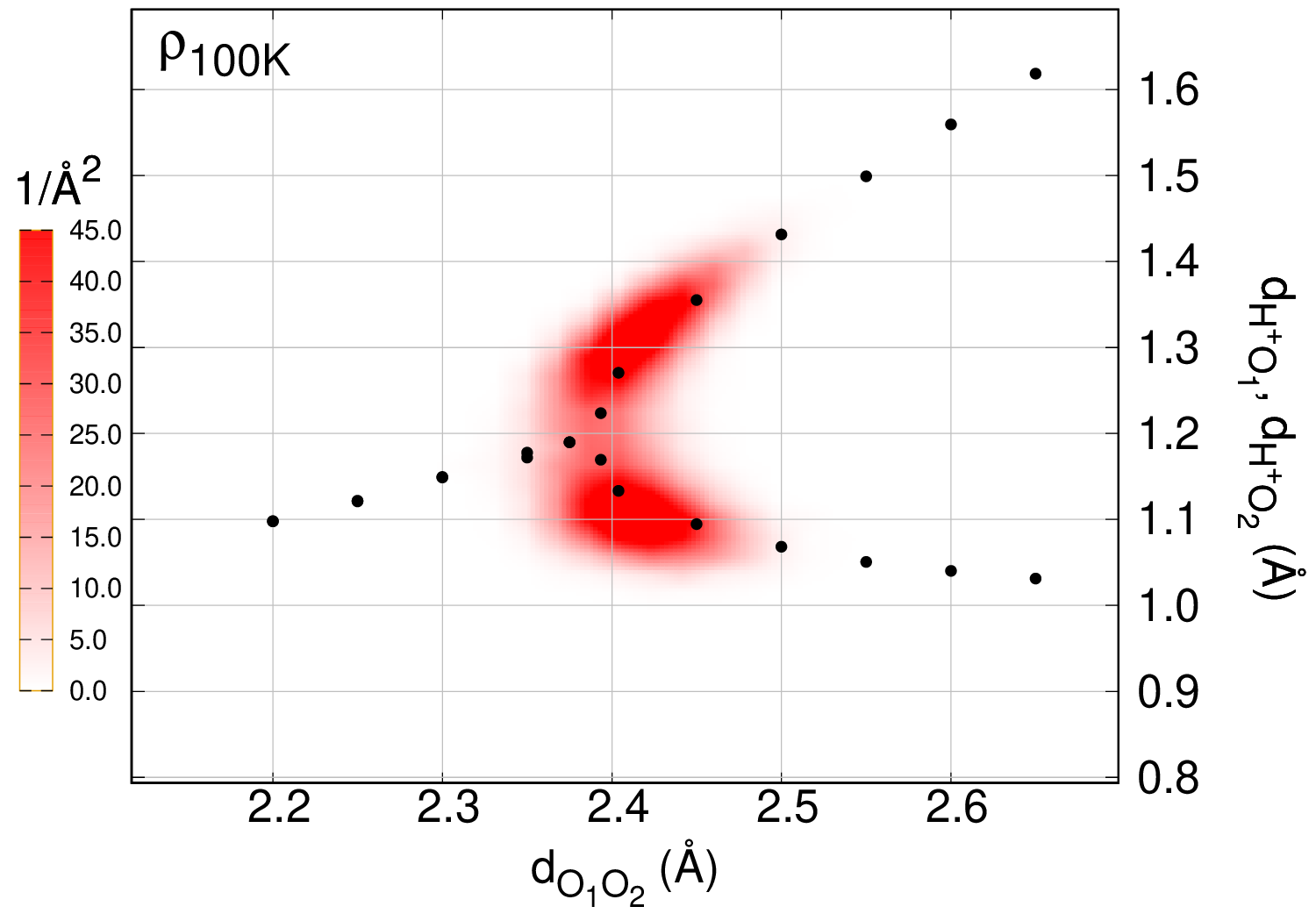} &
\includegraphics[width=0.5\columnwidth]{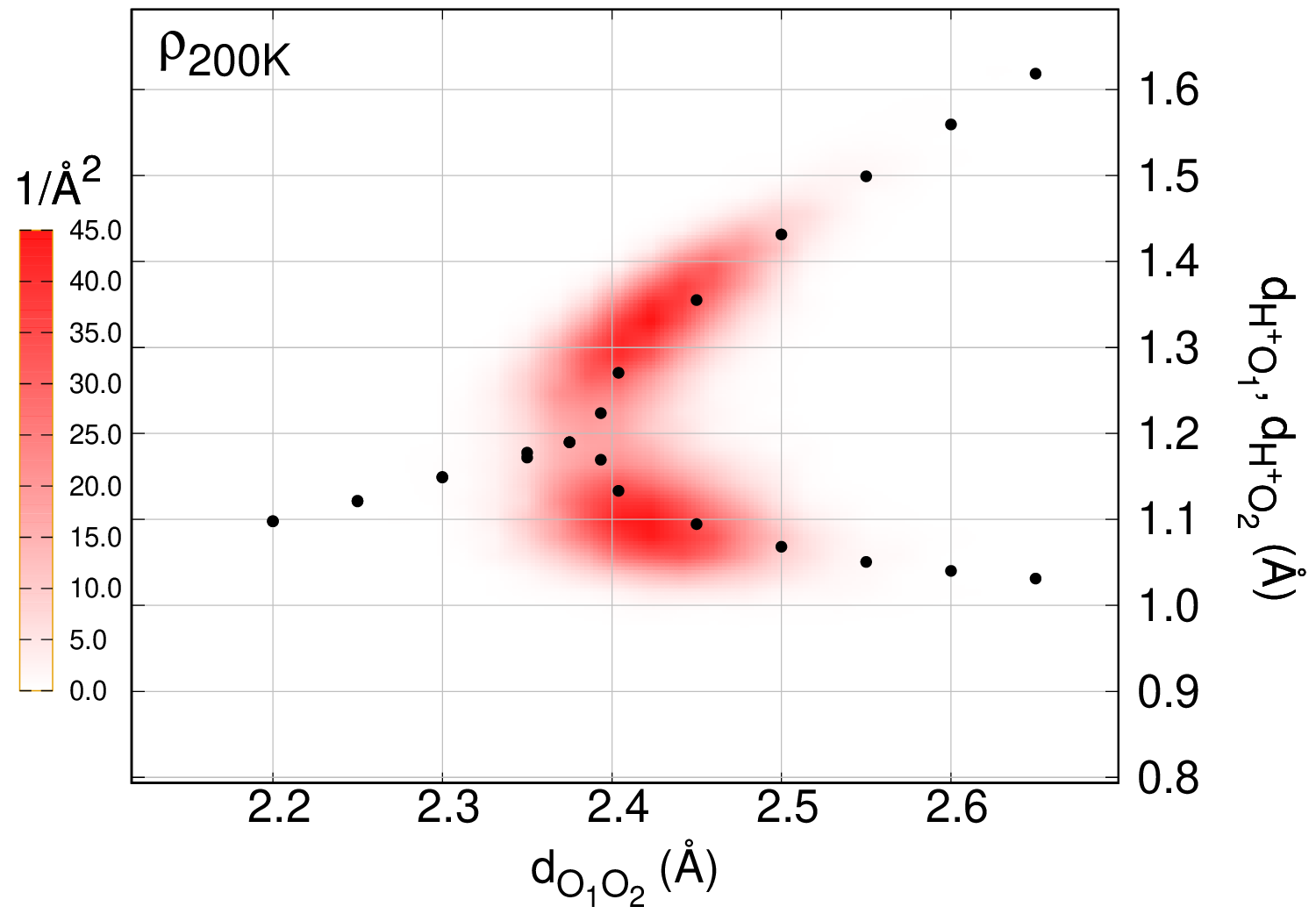} \\
\includegraphics[width=0.5\columnwidth]{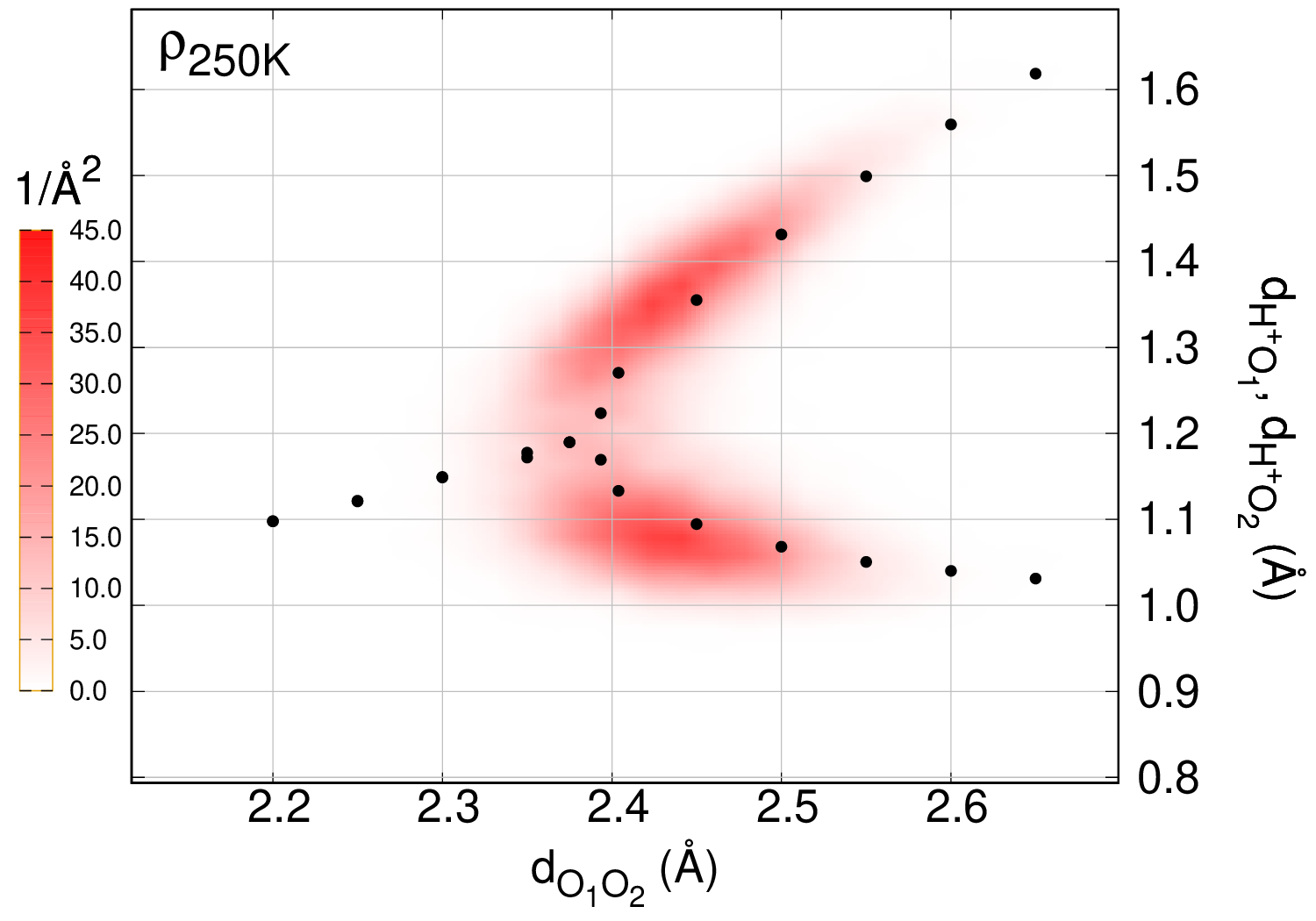} &
\includegraphics[width=0.5\columnwidth]{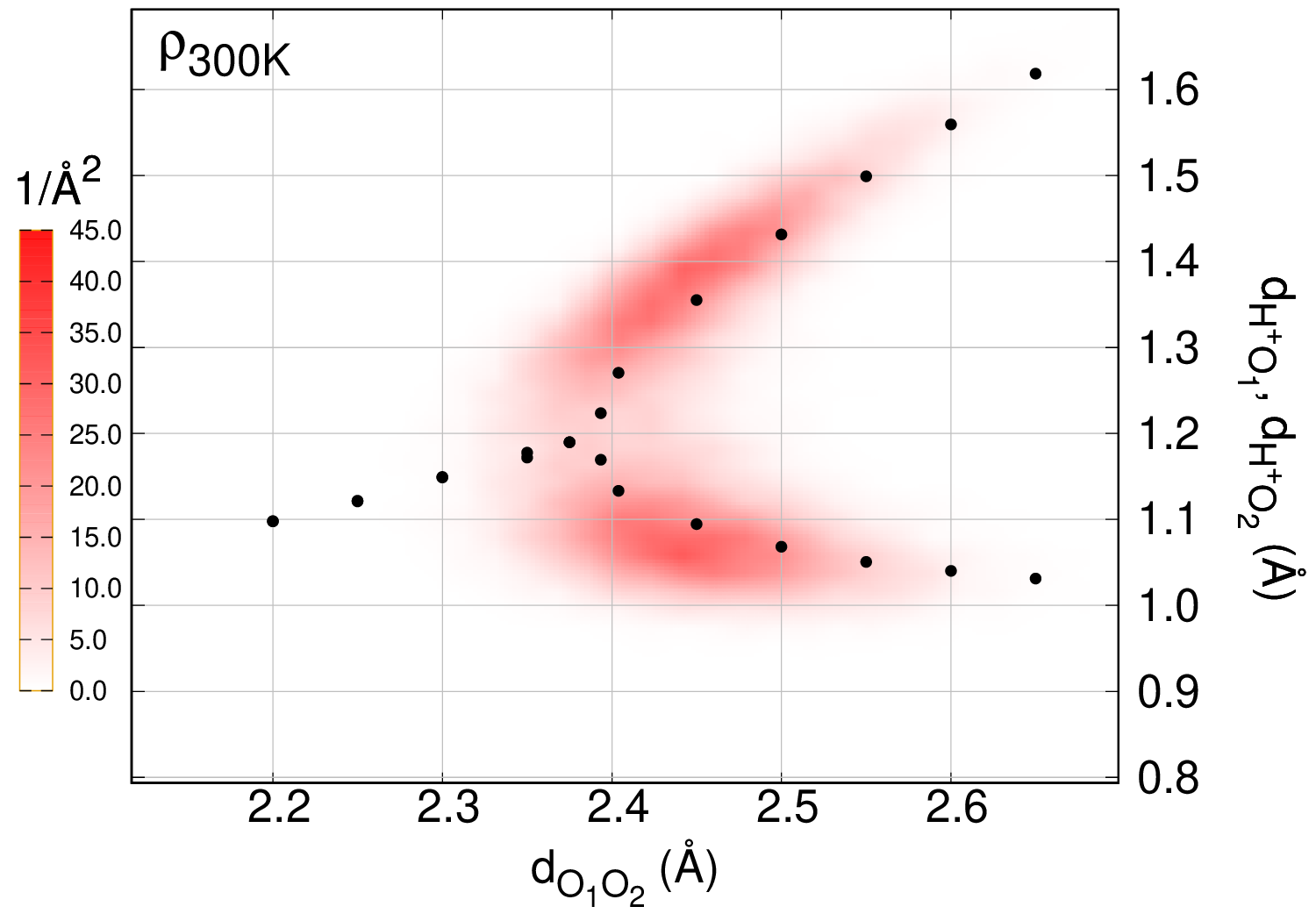} \\
\includegraphics[width=0.5\columnwidth]{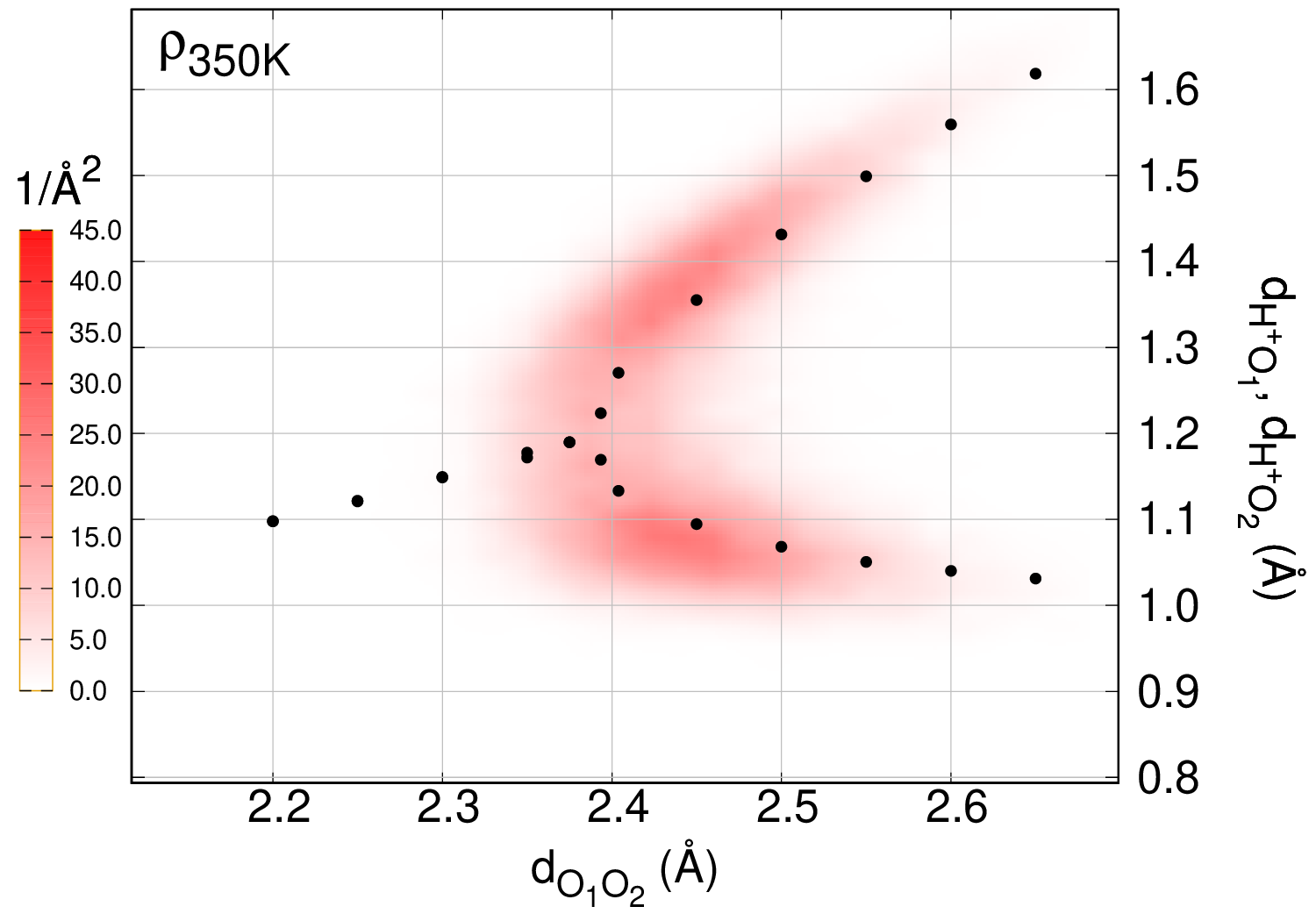} &
\\  
\end{tabular}
\caption{$\rho_{2D}$ computed from VMC-MD simulations, with classical nuclei, at different temperatures.
}
\label{g2d_classical}
\end{figure}

We report some finite-temperature properties of the H$_{13}$O$_6^+$ system, studied by looking
at the bidimensional distributions functions $\rho_{2D}$, which correlate
the distance between the central proton and the neighbouring oxygen
atoms ($d_\textrm{H$^+$O$_1$}$, $d_\textrm{H$^+$O$_2$}$) with
$d_\textrm{O$_1$O$_2$}$. They are shown in Figs.~\ref{g2d_quantum} and
\ref{g2d_classical}, for quantum and classical simulations, respectively.

\newpage

\newpage

The difference between quantum and classical distributions is striking. The impact of nuclear quantum effects (NQEs) on the temperature-dependence of H$_{13}$O$_6^+$ structural properties and PT mechanism are analysed and detailed in the main text, based on the proton density distributions shown in Figs.~\ref{g2d_quantum} and \ref{g2d_classical}.

\section{Towards an accurate modeling of the potential energy surface (PES)}
\label{towards_PES}

We exploit the calculation of VMC forces not only to perform QMC-driven classical and quantum LD, but also to extract the 
best PES fitting functional form
for the excess proton and for the water-water
interaction in the Zundel core. The final goal is to derive the two-dimensional (2D) model potential $V_\textrm{2D}=V_\textrm{2D}(\text{d}_{\text{\scriptsize O}_1\text{\scriptsize O}_2},\delta)$, where d$_{\text{\scriptsize O}_1\text{\scriptsize O}_2}$ is the distance between the two central oxygen atoms and $\delta$ is the proton sharing coordinate, referenced to the midpoint of the $\text{O}_1\text{H}^+\text{O}_2$ complex: $\delta \equiv \tilde{d}_{\text{\scriptsize O}_{1/2}\text{\scriptsize H}^+} -\text{d}_{\text{\scriptsize O}_1\text{\scriptsize O}_2}/2$, with $\tilde{d}_{\text{\scriptsize O}_{1/2}\text{\scriptsize H}^+}$ the $\text{O}_{1/2}$-$\text{H}^+$ distance projected onto the O$_1$O$_2$ direction. The projection of the full interatomic potential on the restricted 2D manifold is done by integrating the other degrees of freedom over the thermal partition function, sampled during the MD dynamics, i.e. $V_\textrm{2D}(\text{d}_{\text{\scriptsize O}_1\text{\scriptsize O}_2},\delta) \equiv \langle V(q_1, q_2, \mathbf{q}_{N-2}) \delta(q_1-\text{d}_{\text{\scriptsize O}_1\text{\scriptsize O}_2}) \delta(q_2-\delta) \rangle $, where $\langle \ldots \rangle$ is the average over the partition function of the classical/quantum statistical ensemble at fixed temperature, and $V$ is the 3$N$-dimensional potential depending on the generalised nuclear coordinates of the full system. Analogously, one can define the one-dimensional (1D) potential acting between O$_1$ and O$_2$ as $V_\textrm{1D}=V_\textrm{1D}(\text{d}_{\text{\scriptsize O}_1\text{\scriptsize O}_2}) \equiv \langle V(q_1, q_2, \mathbf{q}_{N-2}) \delta(q_1-\text{d}_{\text{\scriptsize O}_1\text{\scriptsize O}_2}) \rangle $, according to previous notations. Derivatives of the previous potentials with respect to $\text{d}_{\text{\scriptsize O}_1\text{\scriptsize O}_2}$ and/or $\delta$ can be defined in the same way. For instance, $\partial V_\textrm{2D}/ \partial \delta \equiv \langle \partial V(q_1, q_2, \mathbf{q}_{N-2})/\partial q_2 ~ \delta(q_1-\text{d}_{\text{\scriptsize O}_1\text{\scriptsize O}_2}) \delta(q_2-\delta) \rangle $, and $\partial V_\textrm{1D}/ \partial \text{d}_{\text{\scriptsize O}_1\text{\scriptsize O}_2} \equiv \langle \partial V(q_1, q_2, \mathbf{q}_{N-2})/\partial q_1 ~ \delta(q_1-\text{d}_{\text{\scriptsize O}_1\text{\scriptsize O}_2}) \rangle $.

Given these definitions, we can proceed with the calculations of the corresponding quantities with the aim at modeling the potentials $V_\textrm{1D}$ and $V_\textrm{2D}$. To do so, we will integrate the other degrees of freedom using the classical Boltzmann distribution in $\langle \ldots \rangle$, as generated by the QMC-driven classical Langevin dynamics at 100, 250 and 350 K. Employing the classical partition function has the advantage that the potentials \emph{sampled} in this way will tend to the original PES of the system as $\beta \rightarrow \infty$, while the quantum partition function will lead to averaged potentials biased by quantum fluctuations even in the zero temperature limit.
To compute these quantities from an MD sampling, the $\delta$-functions in their definitions above are replaced by bins, whose size is given by the spacing between neighbouring points. 

\begin{figure}[!h]
\centering 
\begin{tabular}{lr} 
\includegraphics[width=0.4\columnwidth]{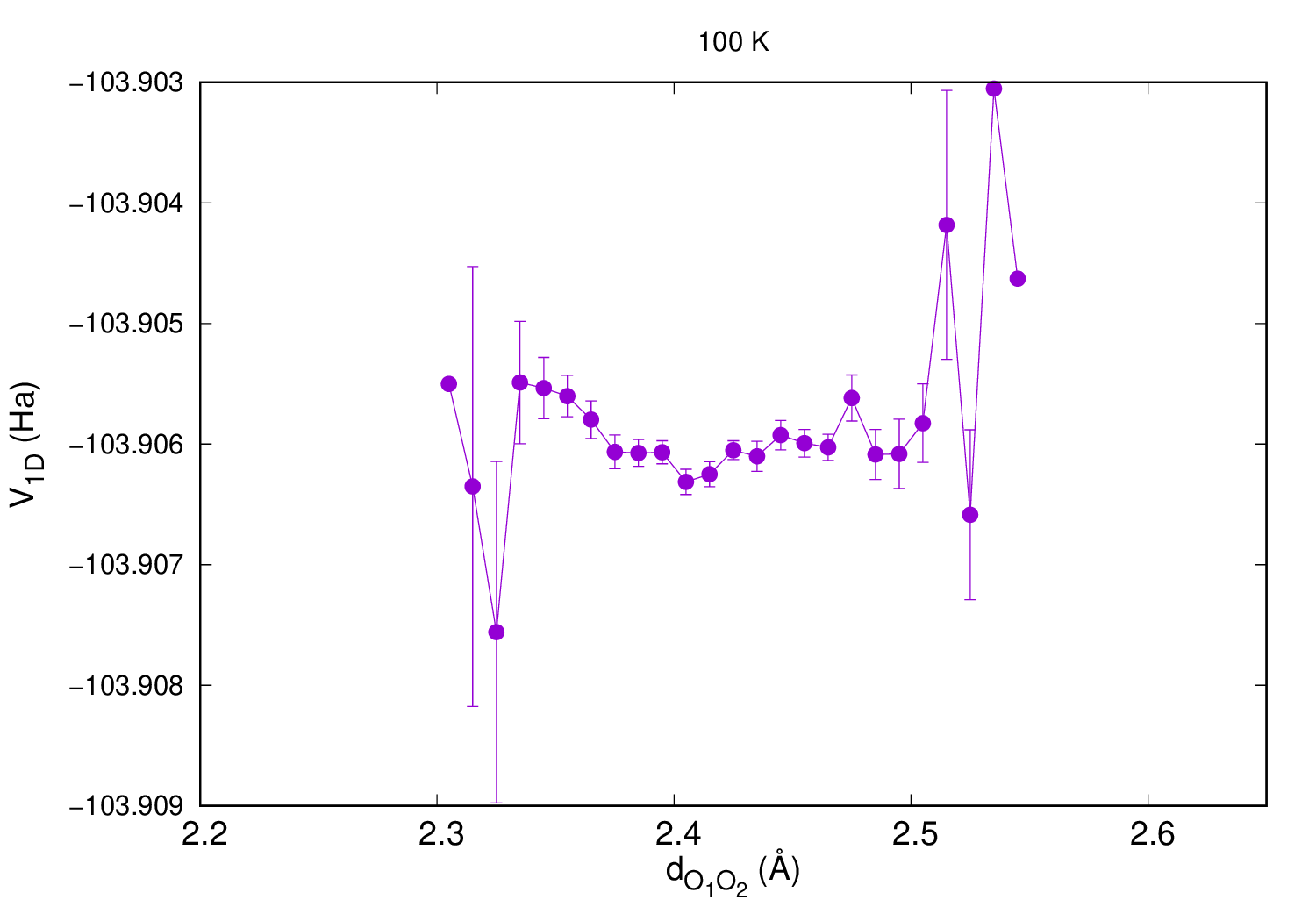}
  &
\includegraphics[width=0.4\columnwidth]{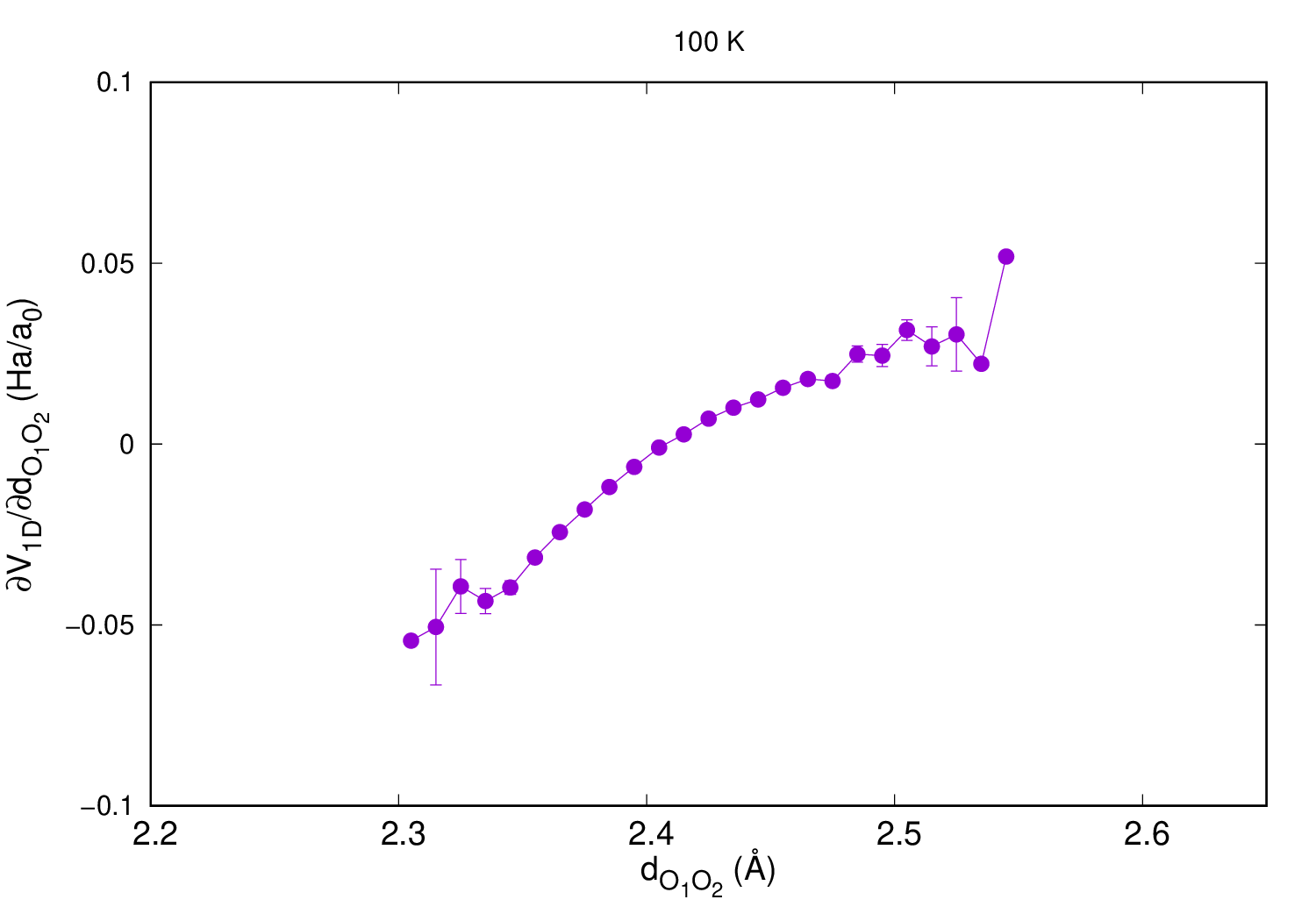}
  \\
\includegraphics[width=0.4\columnwidth]{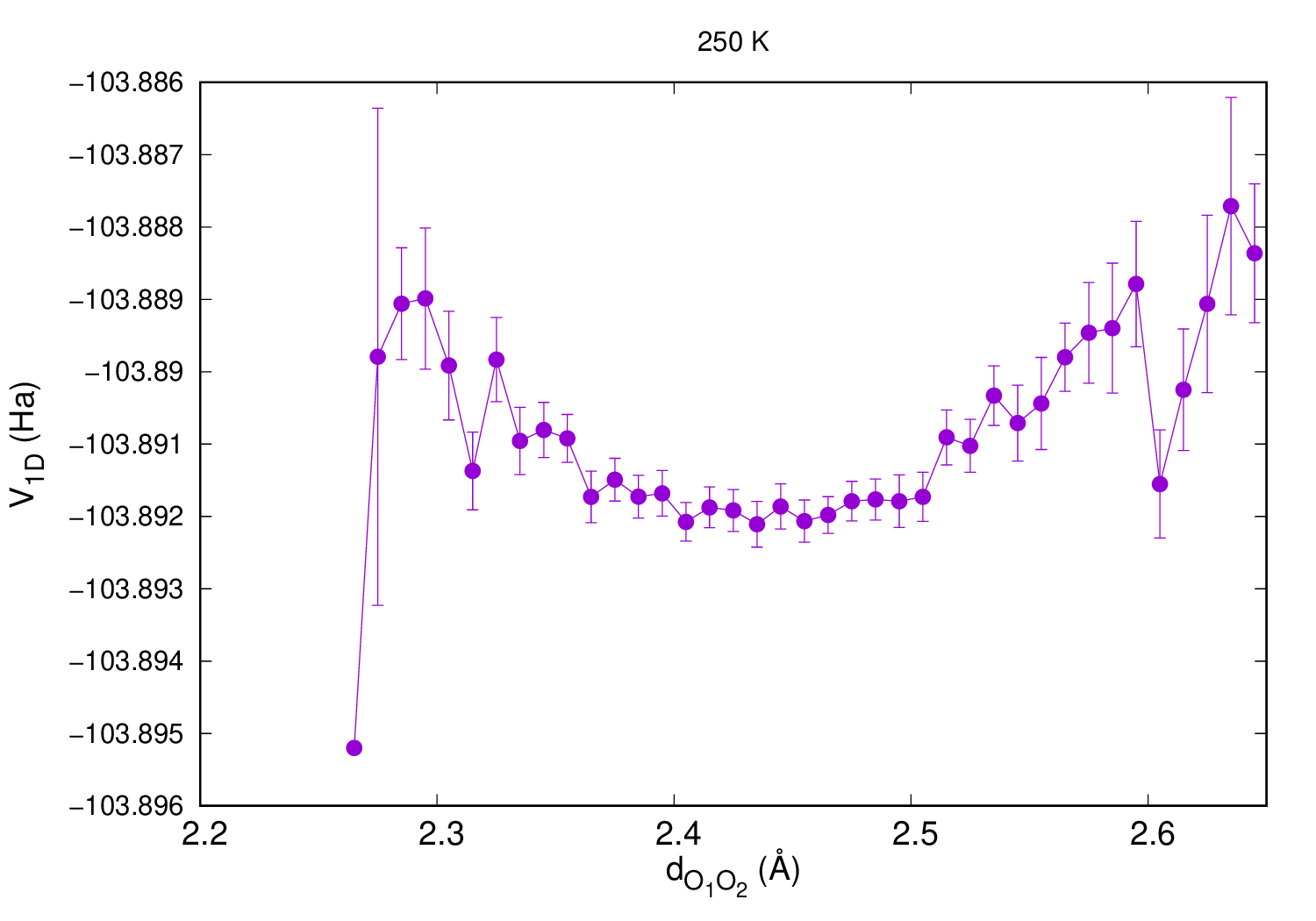}
  &
\includegraphics[width=0.4\columnwidth]{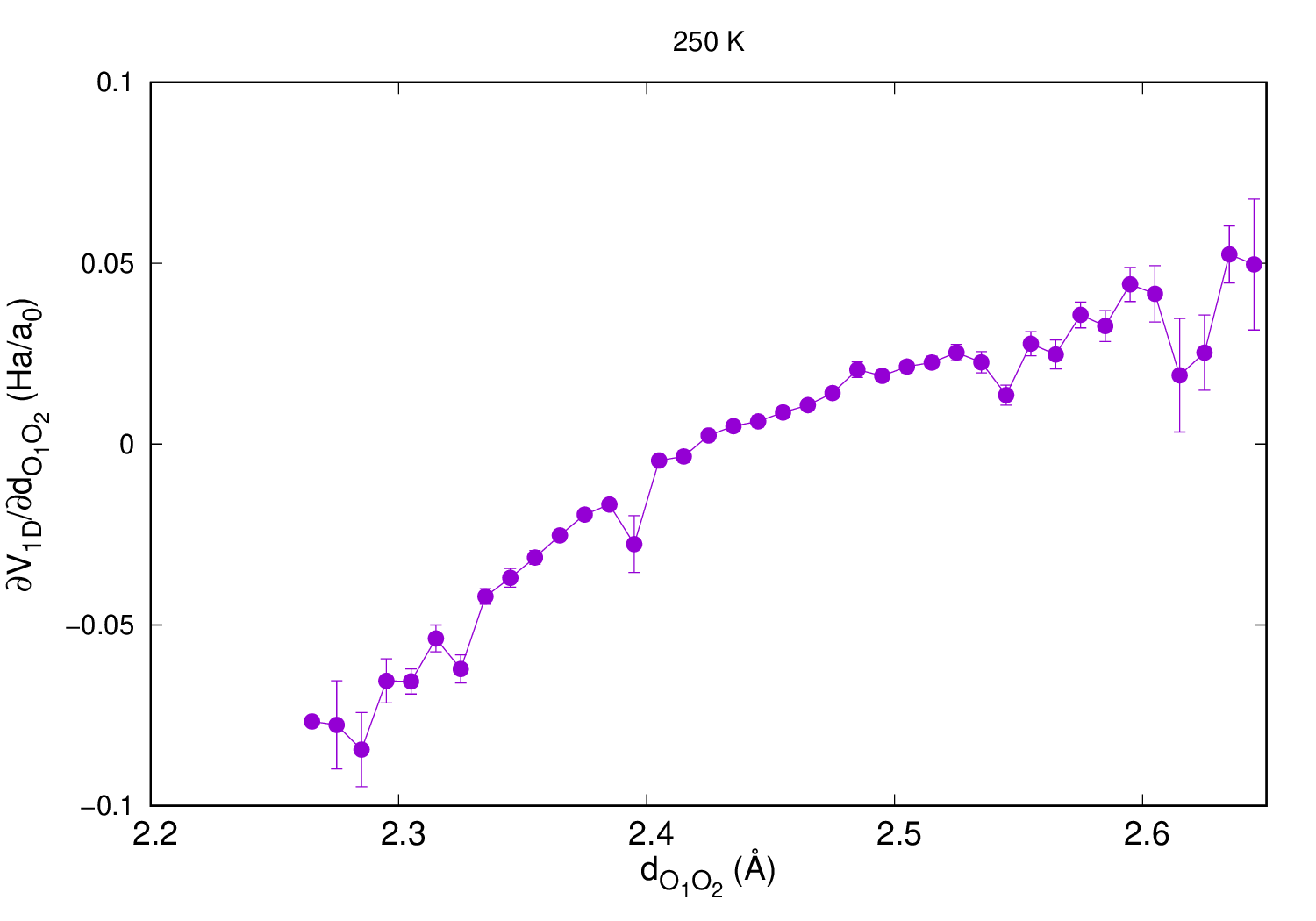}
  \\
\includegraphics[width=0.4\columnwidth]{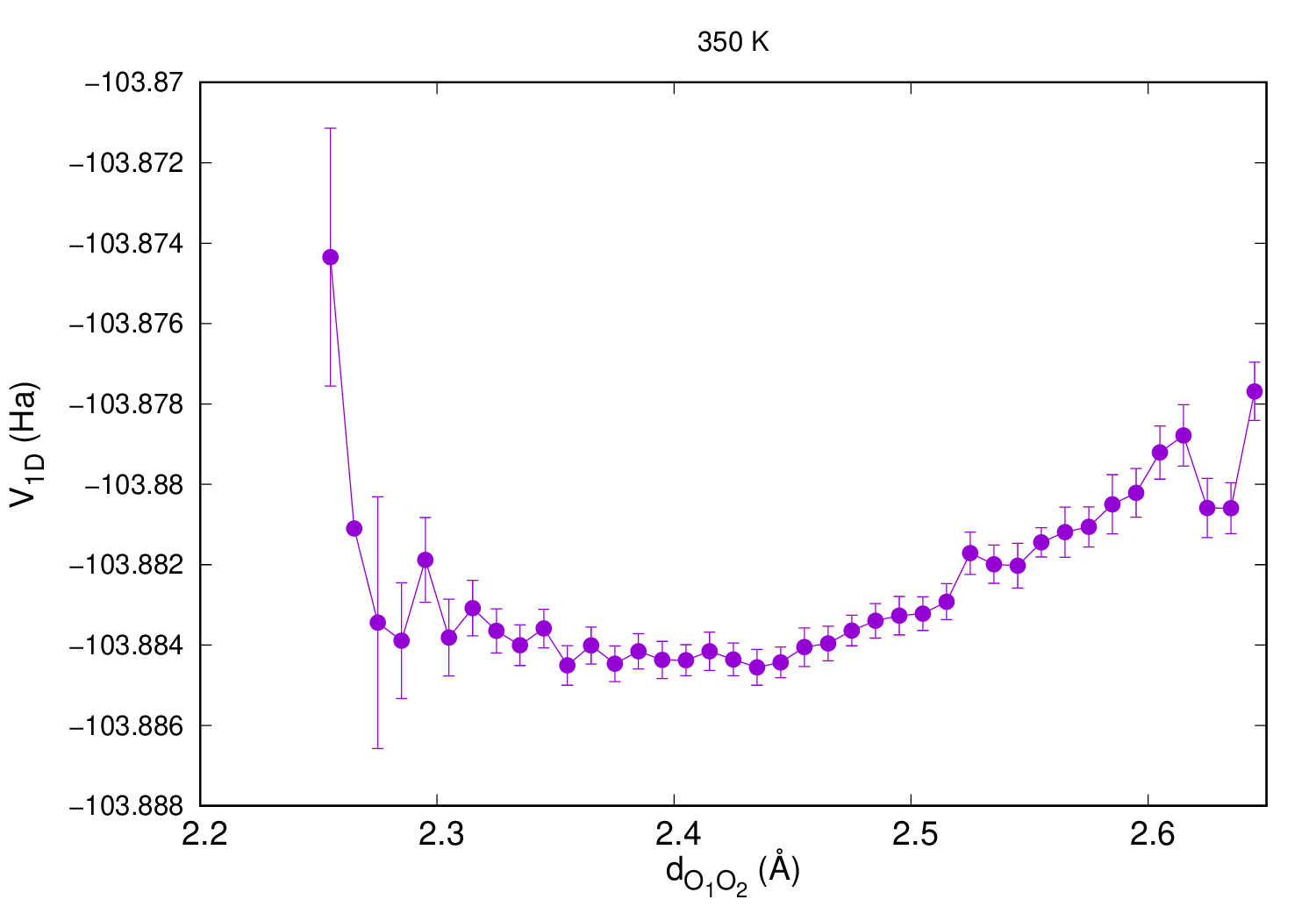}
  &
\includegraphics[width=0.4\columnwidth]{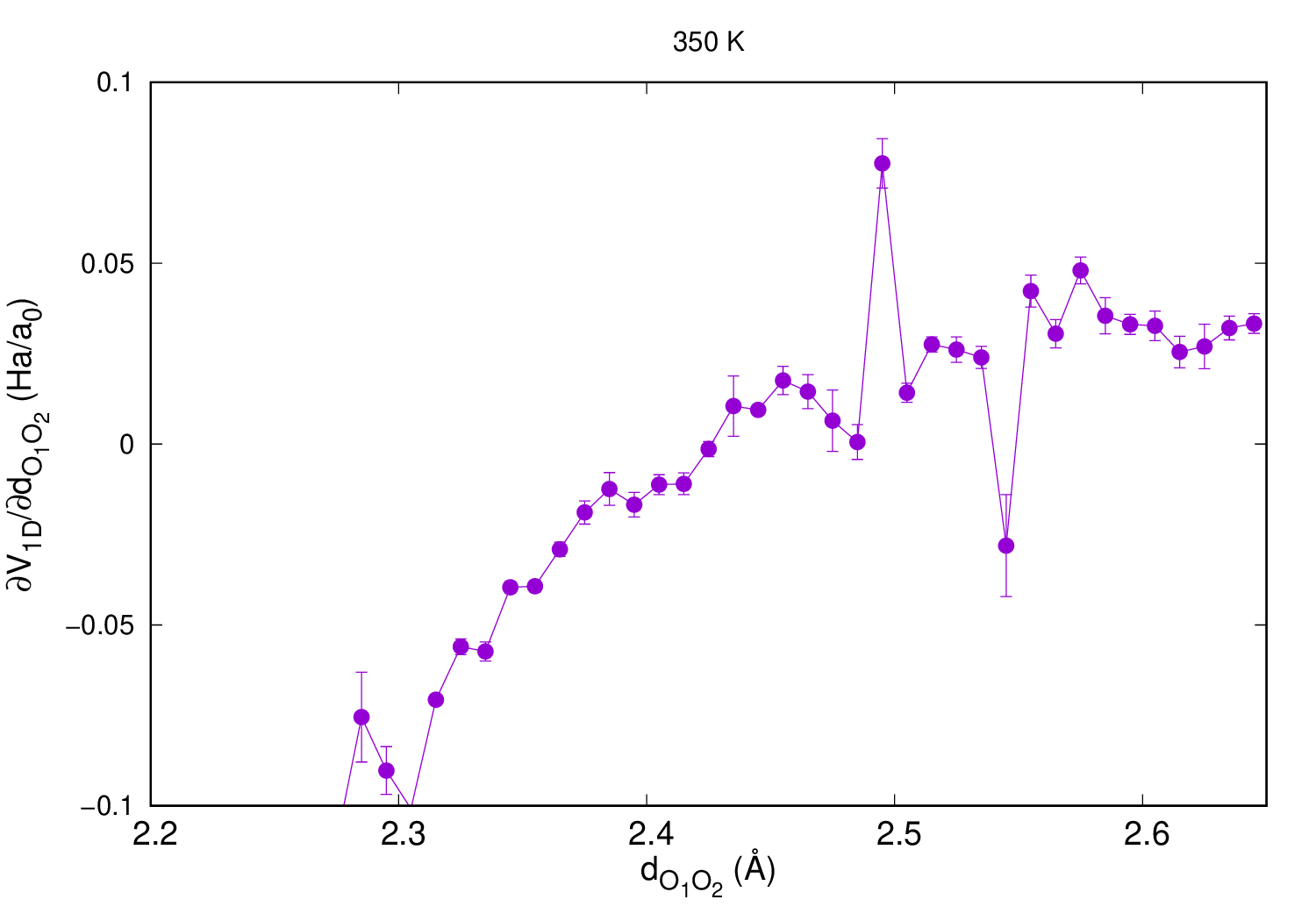}  
\\
\end{tabular}
\caption{Left-hand side: total energy variation of the cluster as a
  function of the  $d_\textrm{O$_1$O$_2$}$ distance ($V_\textrm{1D}$). Right-hand side:
  sum of the energy gradients with respect to $\mathbf{q}_\textrm{O$_1$}$ and $\mathbf{q}_\textrm{O$_2$}$ variations projected along the O$_1$O$_2$
  direction, for classical simulations at different temperatures. This
  corresponds to $\partial V_\textrm{1D}/ \partial \text{d}_{\text{\scriptsize O}_1\text{\scriptsize O}_2}$, resulting in the force that drives the O$_1$-O$_2$ stretching mode.
}
\label{forceOO_classical}
\end{figure}

In Fig.~\ref{forceOO_classical}, we study the $V_\textrm{1D}(\text{d}_{\text{\scriptsize O}_1\text{\scriptsize O}_2})$ potential depending on the water-water distance $\text{d}_{\text{\scriptsize O}_1\text{\scriptsize O}_2}$ (left column), and its derivative $\partial V_\textrm{1D}/ \partial \text{d}_{\text{\scriptsize O}_1\text{\scriptsize O}_2}$ (right column). 
As one can see, the energy profile, at
the left-hand side, is much more noisy than the behavior of its gradient, from
where we can extract a precise value of the equilibrium
$d_\textrm{O$_1$O$_2$}$ distance, and the evolution of the potential around the minimum. This shows the advantage of computing QMC forces in order to determine the 
PES, and suggests that a robust way of deriving the $V_\textrm{1D}$ potential is by fitting and integrating its derivatives, rather than by directly fitting the energies. 

\newpage

\begin{figure}[!h]
\centering 
\begin{tabular}{lr} 
\includegraphics[width=0.45\columnwidth]{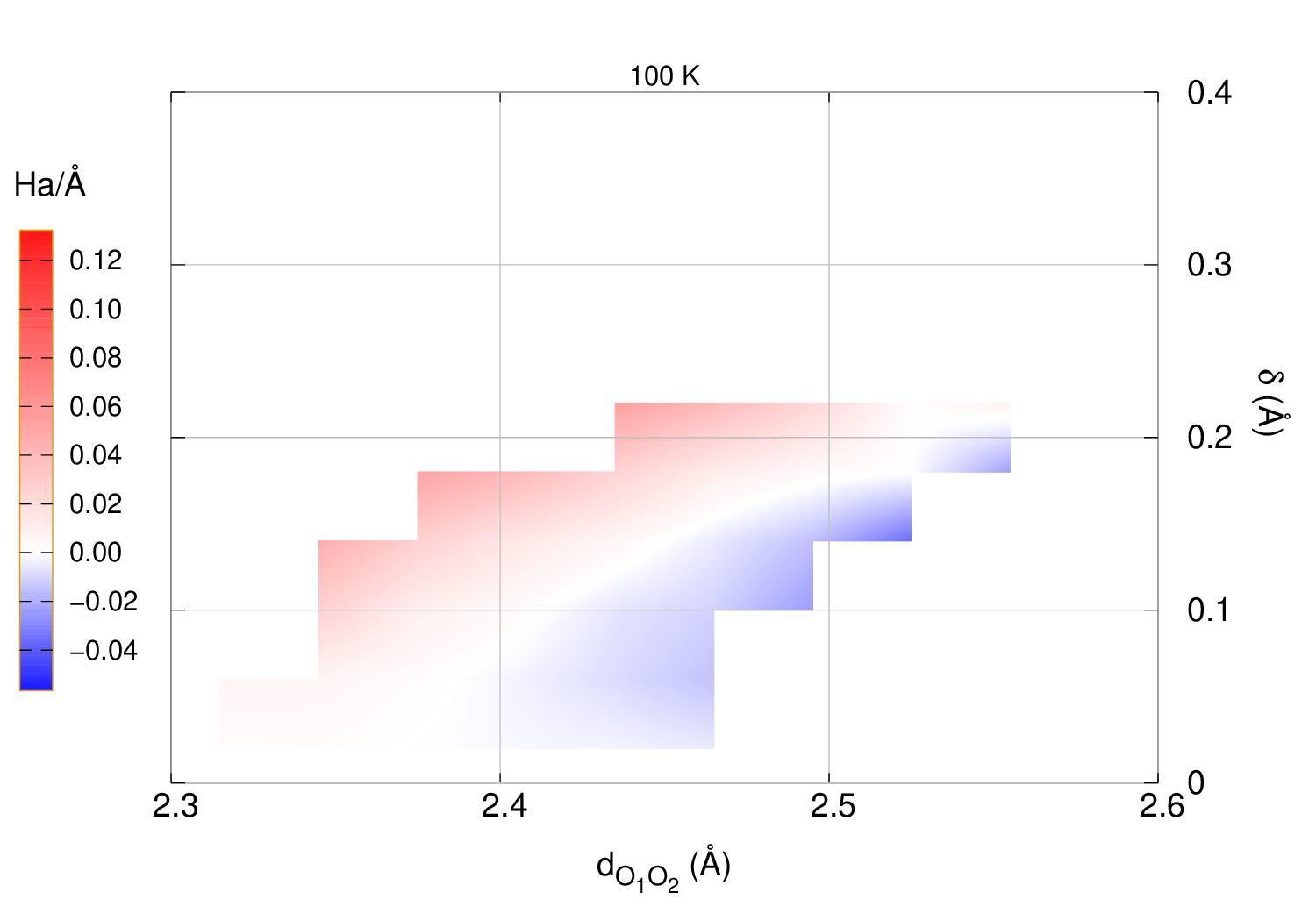}
  &
\includegraphics[width=0.45\columnwidth]{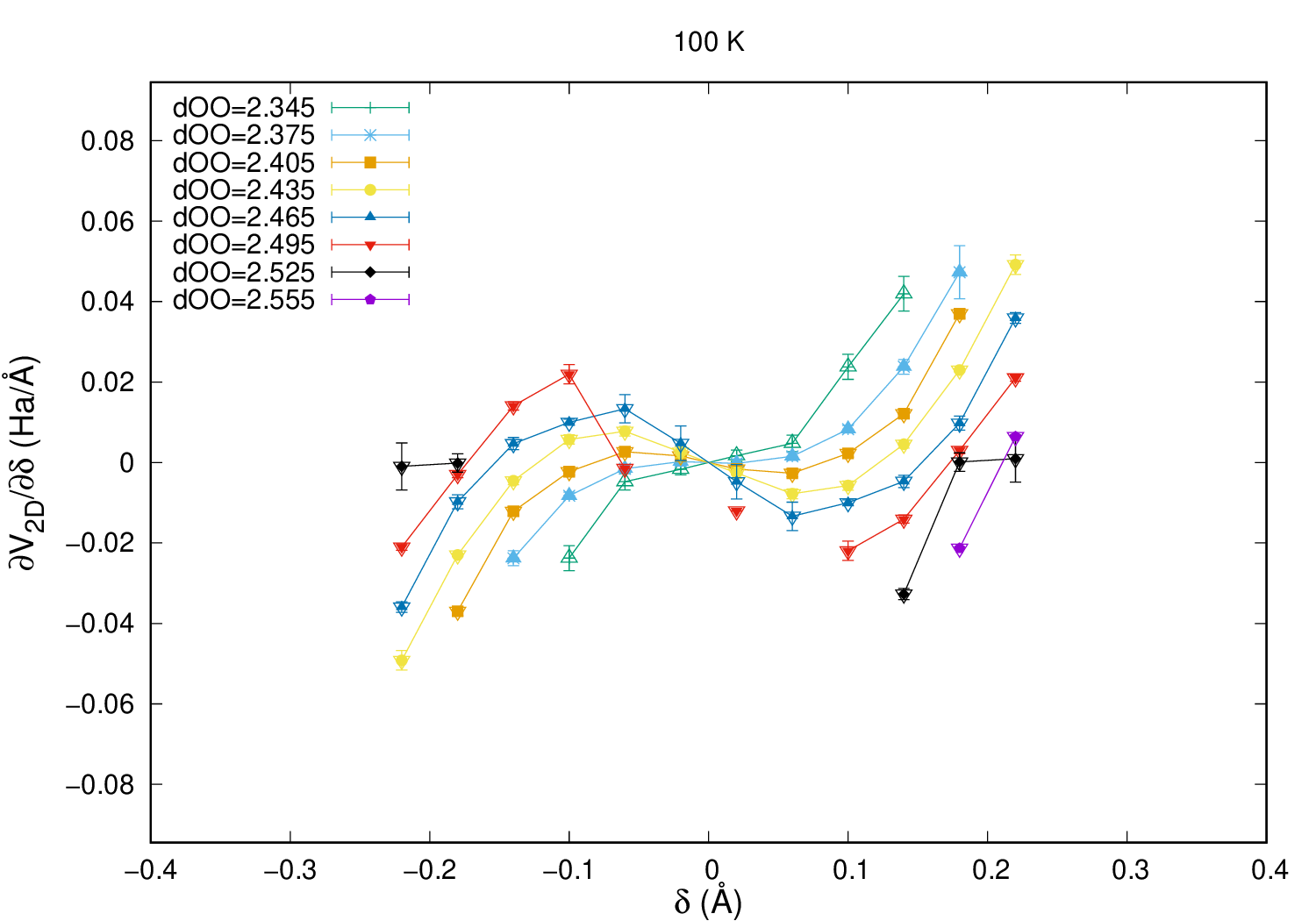}
  \\
\includegraphics[width=0.45\columnwidth]{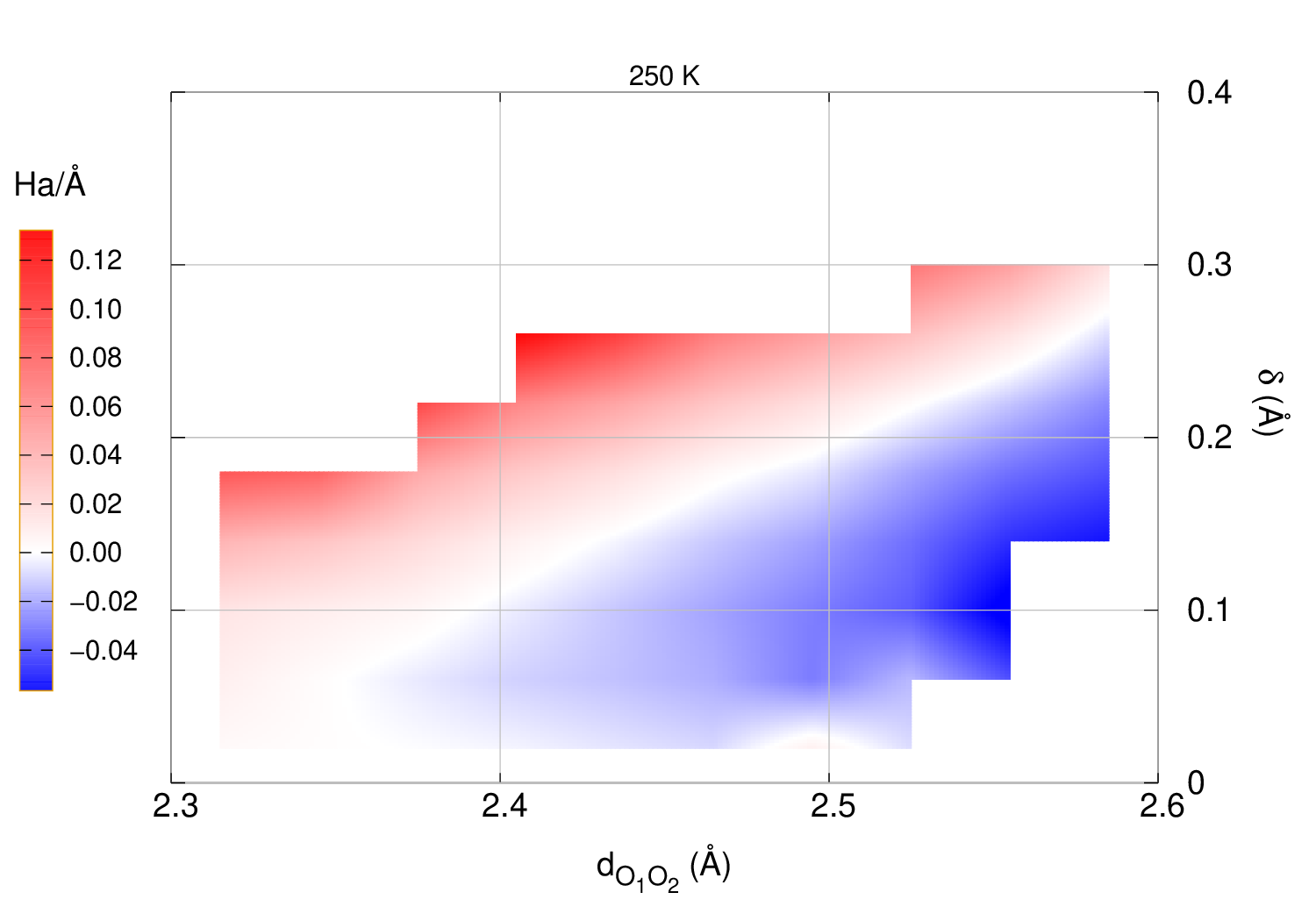}
  &
\includegraphics[width=0.45\columnwidth]{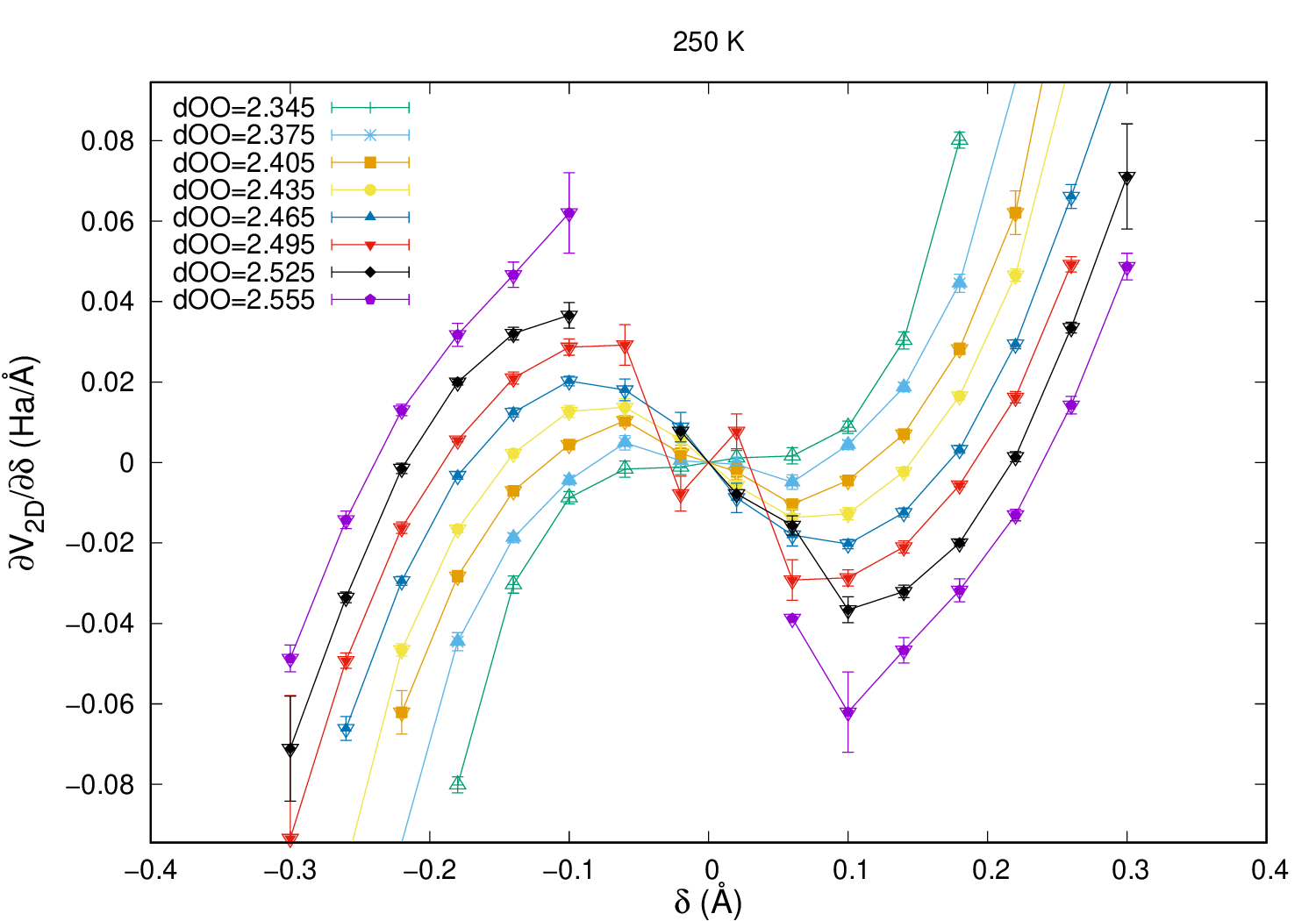}
  \\
\includegraphics[width=0.45\columnwidth]{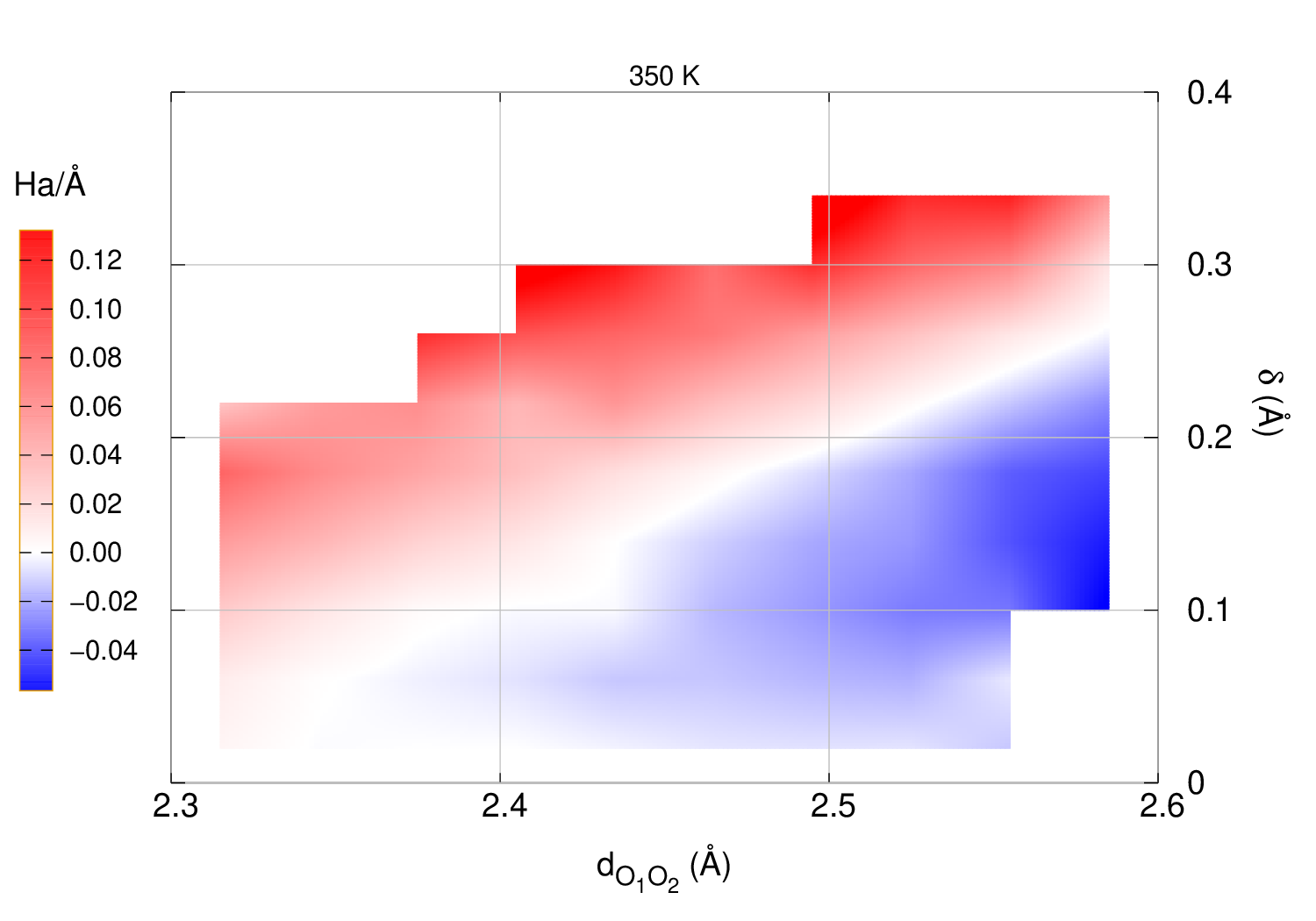}
  &
\includegraphics[width=0.45\columnwidth]{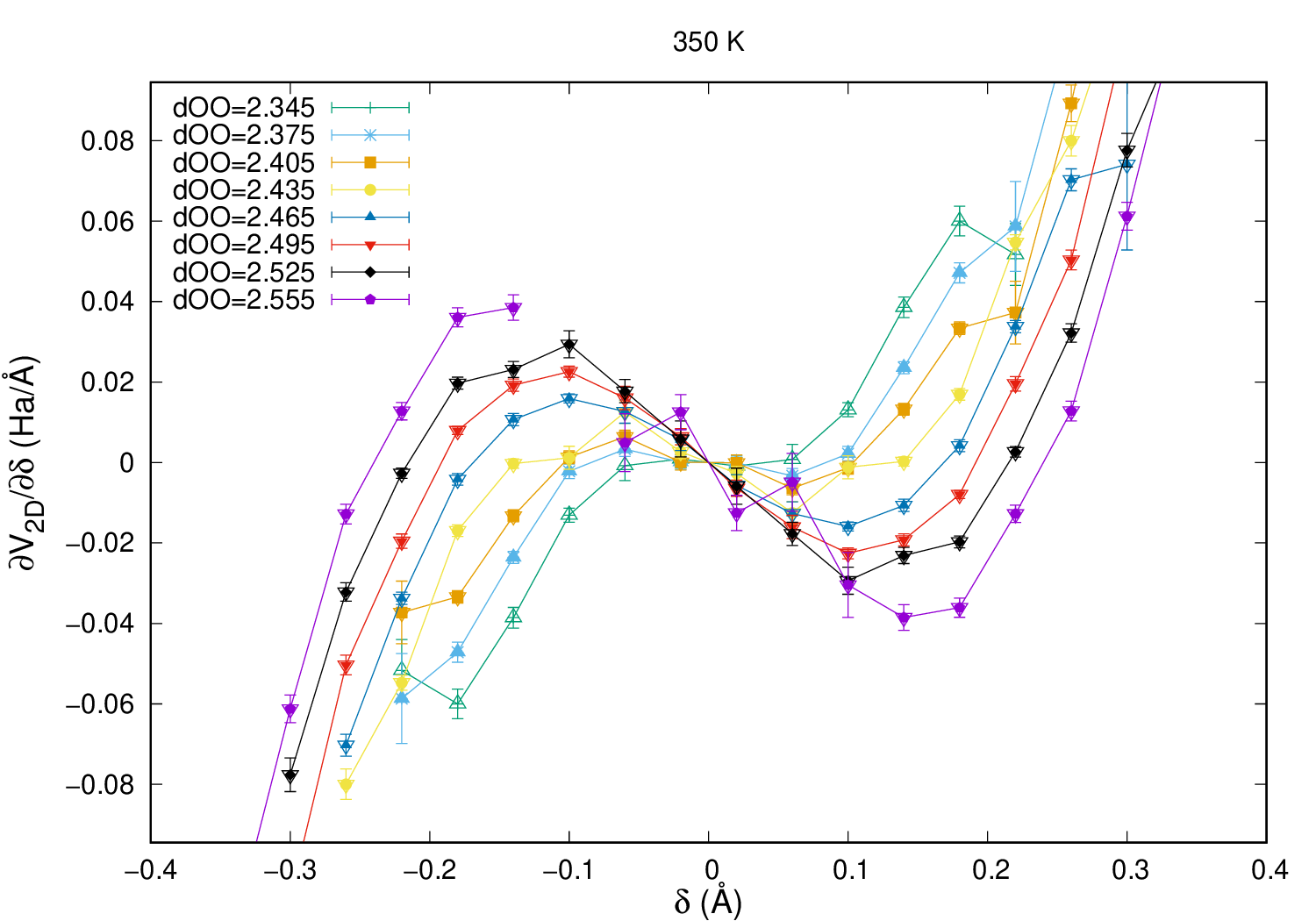}  
\\
\end{tabular}
\caption{Left column: contour plot of $\partial V_\textrm{2D}/ \partial \delta$ as a function of both 
$\text{d}_{\text{\scriptsize O}_1\text{\scriptsize O}_2}$ and $\delta$. Right column: superposition of $\partial V_\textrm{2D}/ \partial \delta$, plotted as a function of
$\delta$ at various (fixed) $\text{d}_{\text{\scriptsize O}_1\text{\scriptsize O}_2}$ values.
The force acting on H$^+$ projected along the O$_1$O$_2$ direction is given by $-\partial V_\textrm{2D}/ \partial \delta$. Notice that the size of the $(\text{d}_{\text{\scriptsize O}_1\text{\scriptsize O}_2},\delta)$ space accessible by MD to sample these quantities increases as a function of the temperature.
}
\label{2DforceH_classical}
\end{figure}

In Fig.~\ref{2DforceH_classical}, we study the $V_\textrm{2D}(\text{d}_{\text{\scriptsize O}_1\text{\scriptsize O}_2},\delta)$ potential depending on the proton coordinate $\delta$, at various (fixed) $d_\textrm{O$_1$O$_2$}$ distances. In the left column, we show $\partial V_\textrm{2D}/ \partial \delta$ in a contour plot as a function of both $\text{d}_{\text{\scriptsize O}_1\text{\scriptsize O}_2}$ and $\delta$. Positive (negative) values of $\partial V_\textrm{2D}/ \partial \delta$ are coloured in red (blue). The white region indicates the extrema of the 2D-PES. The classical proton is clearly asymmetric for $\text{d}_{\text{\scriptsize O}_1\text{\scriptsize O}_2} \gtrsim 2.37$ \AA, with a minimum departing from the $\delta=0$ axis. In the right column, the same information is provided
by superposing $\partial V_\textrm{2D}/ \partial \delta$ plotted as a function of $\delta$ and taken at fixed $\text{d}_{\text{\scriptsize O}_1\text{\scriptsize O}_2}$ distances.

\newpage

\section{Projected two-dimensional PES}
\label{2D-PES}

Using the data obtained in Sec.~\ref{towards_PES}, let us determine an analytic form for the 
$V_\textrm{2D}(d_\textrm{O$_1$O$_2$},\delta)$ potential, which 
depends on both $d_\textrm{O$_1$O$_2$}$ and $\delta$ coordinates. This
will take into account the variation of the proton-oxygen potential along the proton shuttling mode as
the distance between the two inner water molecules varies.

\begin{figure}[!htb]
\centering 
  \includegraphics[width=0.5\columnwidth]{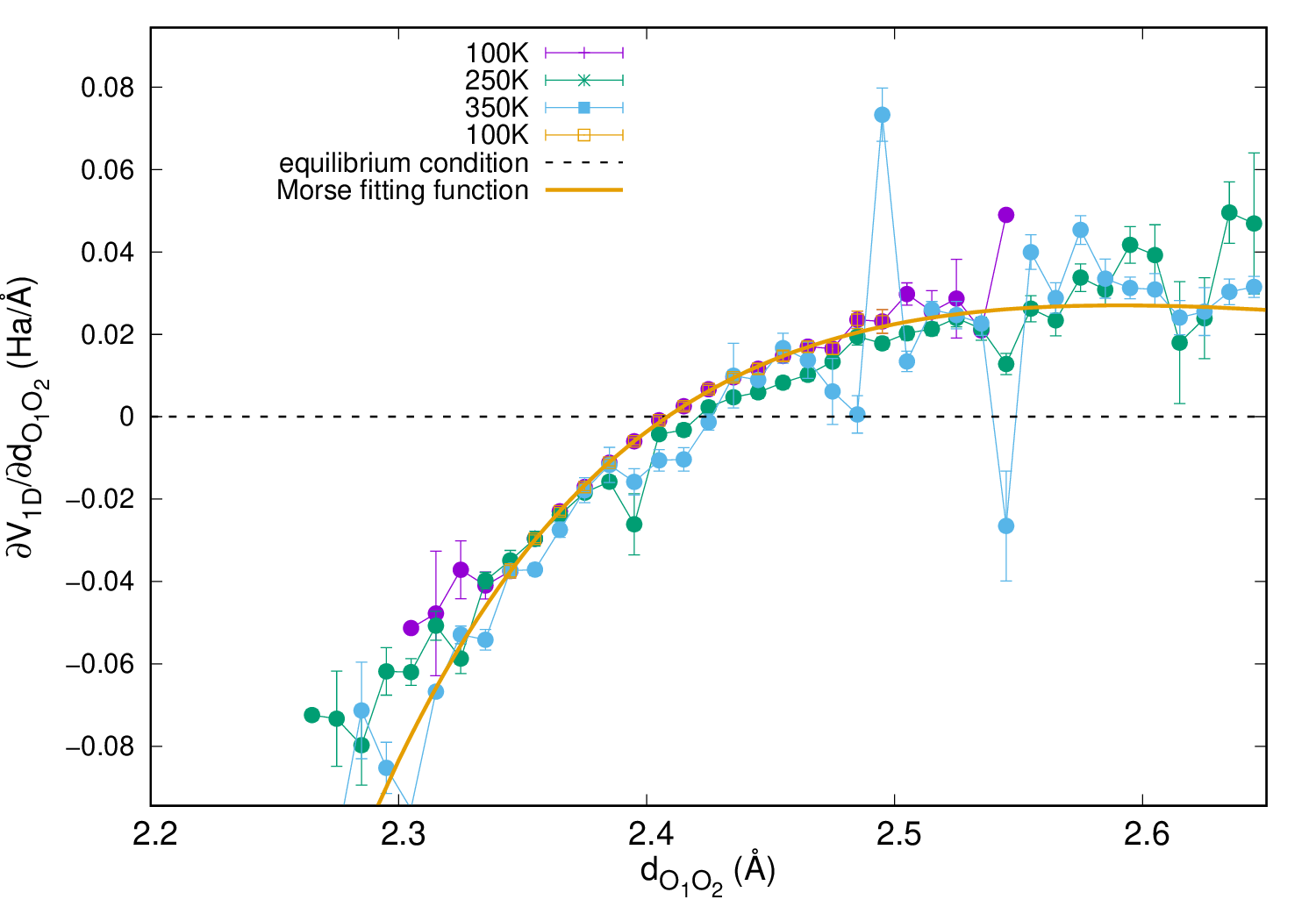}
  \caption{Fit of the QMC estimates of $\partial V_\textrm{1D}/ \partial \text{d}_{\text{\scriptsize O}_1\text{\scriptsize O}_2}$ at 100 K. As fitting function we have used the derivative of the $V_\textrm{1D}$ Morse potential, as defined in Eq.~\ref{Morse_pot}.}
\label{Morse_fit}
\end{figure}

We first derive the $V_\textrm{1D}$ potential between the two water molecules, which
depends only on the $d_\textrm{O$_1$O$_2$}$ stretching coordinate, by fitting the
derivatives shown in
Fig.~\ref{forceOO_classical}, for the simulation at 100 K, which yields
less noisy datapoints than the one at higher temperatures. As fitting function, we choose the Morse potential, such that:
\begin{equation}
V_\textrm{1D}(x)=b_\textrm{Morse} \left(\exp(-2 c_\textrm{Morse}
  (x-d_\textrm{Morse}))-2 \exp(-c_\textrm{Morse} (x-d_\textrm{Morse}))\right) 
- b_\textrm{Morse},
\label{Morse_pot}
\end{equation}
where we have chosen to set the zero of energy at the potential
minimum. The results of the fit are plotted in Fig.~\ref{Morse_fit},
together with the potential derivatives evaluated by classical MD
driven by QMC forces at 100 K, 250 K and 350 K.

From this analysis, the estimated equilibrium distance between two water
molecules in the Zundel core of the protonated water hexamer is
2.408 \AA\ at 100 K, in good agreement with the analysis based on the radial
distribution function reported in Fig.~2 of the main text.

While the data derived from QMC-MD simulations are less noisy at 100 K, they however explore a smaller phase space, due to a probability density distribution more localised in the $(d_\textrm{O$_1$O$_2$},\delta)$ space at lower temperatures. This turns out to be a problem, if one aims at estimating the behavior of the $V_\textrm{2D}(d_\textrm{O$_1$O$_2$},\delta)$ potential not only around its equilibrium geometry but also over its tails. A way to overcome this issue within the projection framework described in Sec.~\ref{towards_PES}, is to sample the projected potential from QMC-MD simulations carried out at higher temperatures. As clearly shown in Fig.~\ref{2DforceH_classical}, at 250 K and 350 K the $V_\textrm{2D}$ behavior can be evaluated on a much larger window in both $\delta$ and $d_\textrm{O$_1$O$_2$}$ directions. Moreover, we can increase the statistics of higher temperatures datapoints by averaging the 250 K and 350 K estimates. 

\begin{figure}[!htb]
\centering 
  \includegraphics[width=0.5\columnwidth]{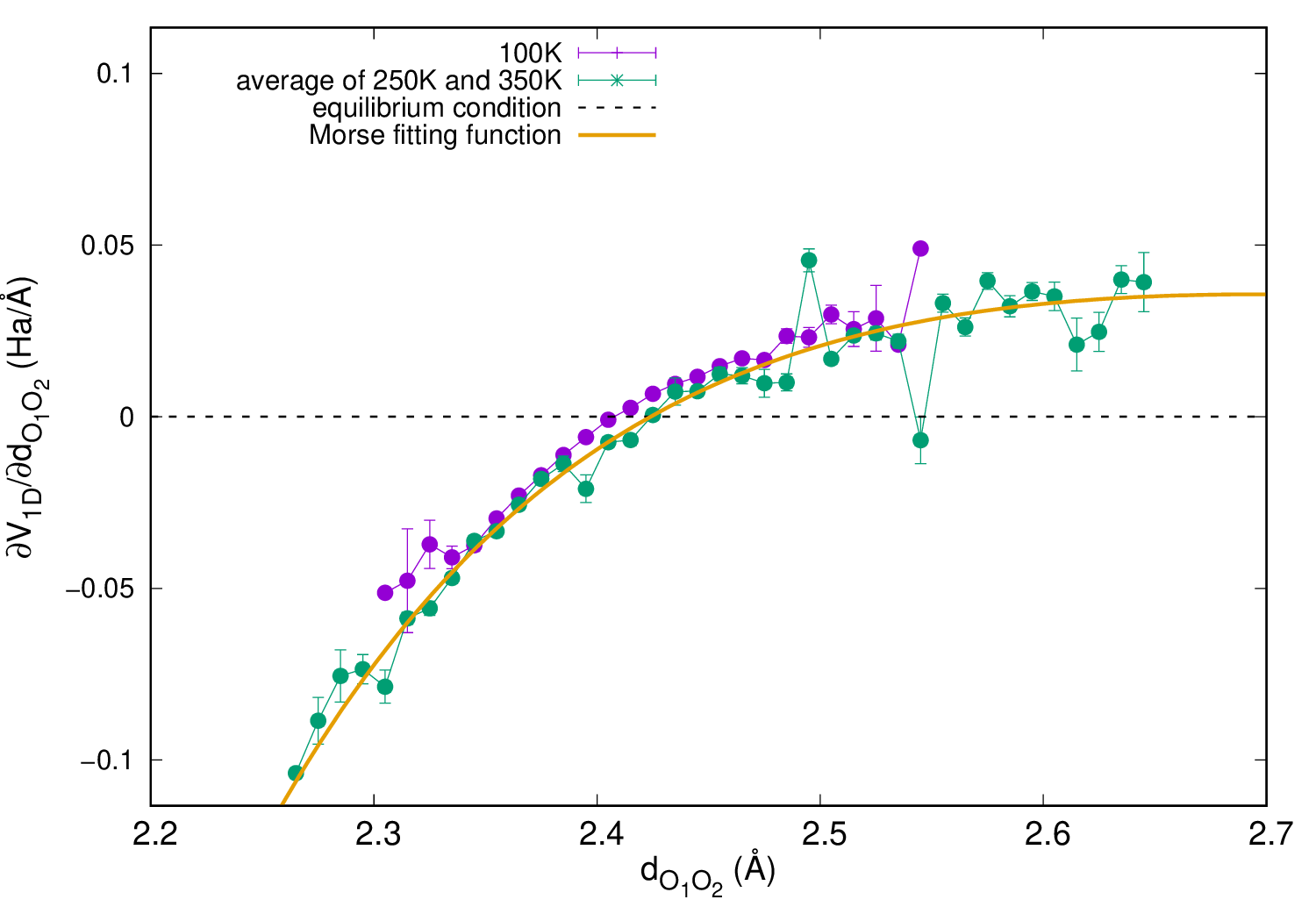}
  \caption{Fit of the QMC forces using $\partial V_\textrm{1D}/\partial x$ as
    fitting function, where the Morse potential $V_ \textrm{1D}(x)$ is
    defined in Eq.~\ref{Morse_pot} and the fitted dataset is taken by averaging 
    the outcome of 250 K and 350 K simulations.}
\label{Morse_fit_av}
\end{figure}

We then fit the $V_\textrm{1D}$  potential in
Eq.~\ref{Morse_pot} by using a dataset 
averaged 
over 250 K and 350 K. 
The corresponding Morse potential fit 
is reported in Fig.~\ref{Morse_fit_av}. From this analysis, the equilibrium distance between two water
molecules in the Zundel core of the protonated water hexamer is
2.425\AA\ at $\approx$ 300 K, again in a satisfactory agreement with the analysis
based on the radial distribution function reported in
Fig.~2 of the main text. Fitting 
over these datapoints
leads to an increase of the equilibrium distance by
0.16\AA\ as the temperature is raised from 100 K to $\approx$ 300 K. The
average based on the radial distribution function of the full VMC-MD simulations yields a cluster expansion
of $\approx$ 0.25\AA.

\begin{figure}[!htb]
\centering 
  \includegraphics[width=0.5\columnwidth]{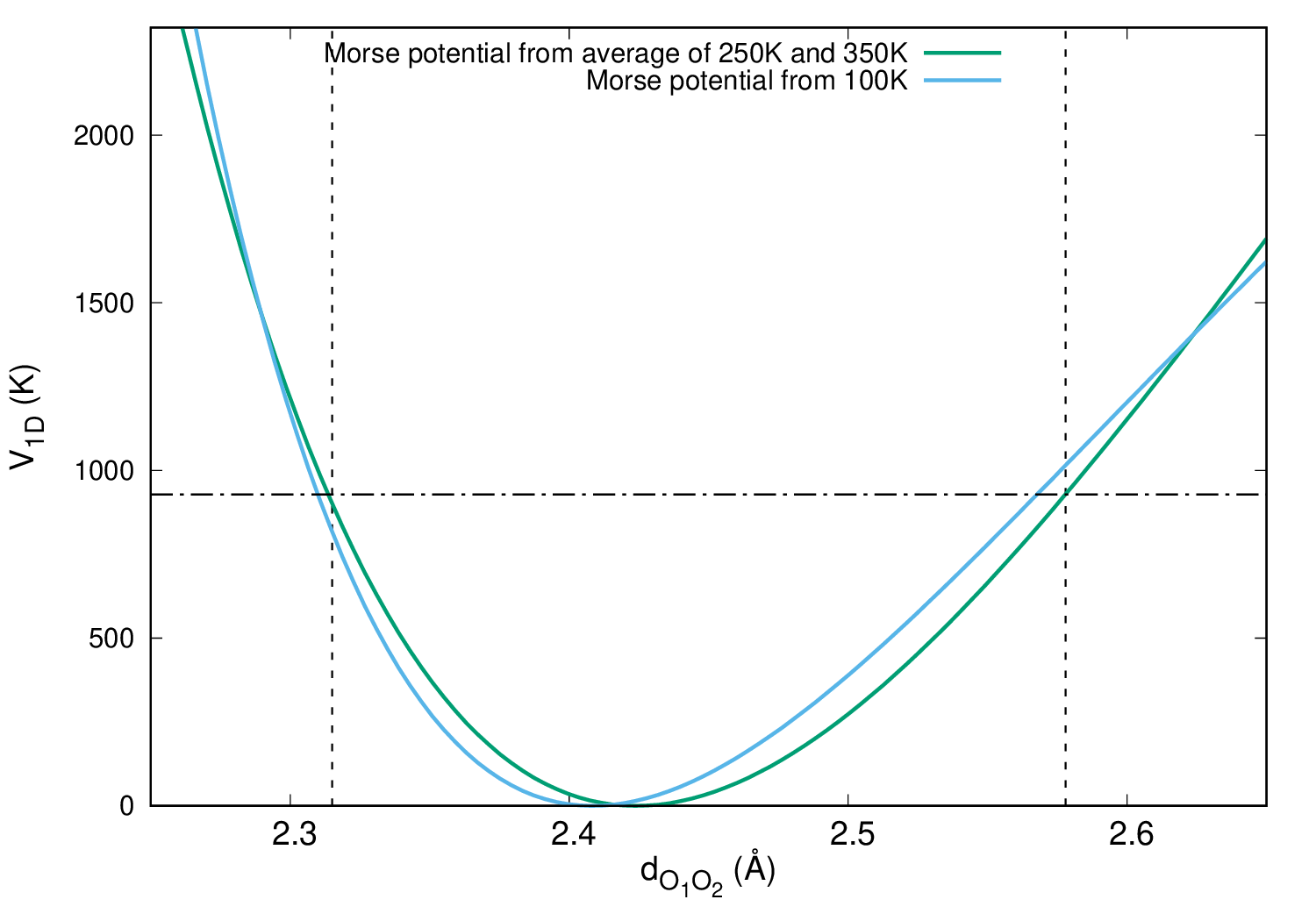}
  \caption{
    $V_\textrm{1D}$ determined from classical MD at 100 K and from
    the averaged dataset of classical MD at 250 K and 350 K. The energy is expressed in Kelvin. The horizontal and vertical dashed lines are guides for the eye. 
  }
\label{Morse_compT}
\end{figure}

The Morse potential determined from points computed at
100 K and the one from points averaged over 250 K and 350 K
are plotted in Fig.~\ref{Morse_compT}, where the two fitting functions
are superimposed. The ZPE analysis described in the main text of the paper is carried out using the dataset averaged over 250 K and 350 K. As one can see in Fig.~\ref{Morse_compT}, in the energy range below 1000 K, the two curves are just shifted from one another, having nearly the same curvature around the minimum. Therefore, the conclusions on the ZPE effect reached by using a model potential projected at larger temperatures would not be different from the ones one could reach using the potential derived at 100 K.

\begin{figure}[!h]
\centering 
\begin{tabular}{lr} 
\includegraphics[width=0.32\columnwidth]{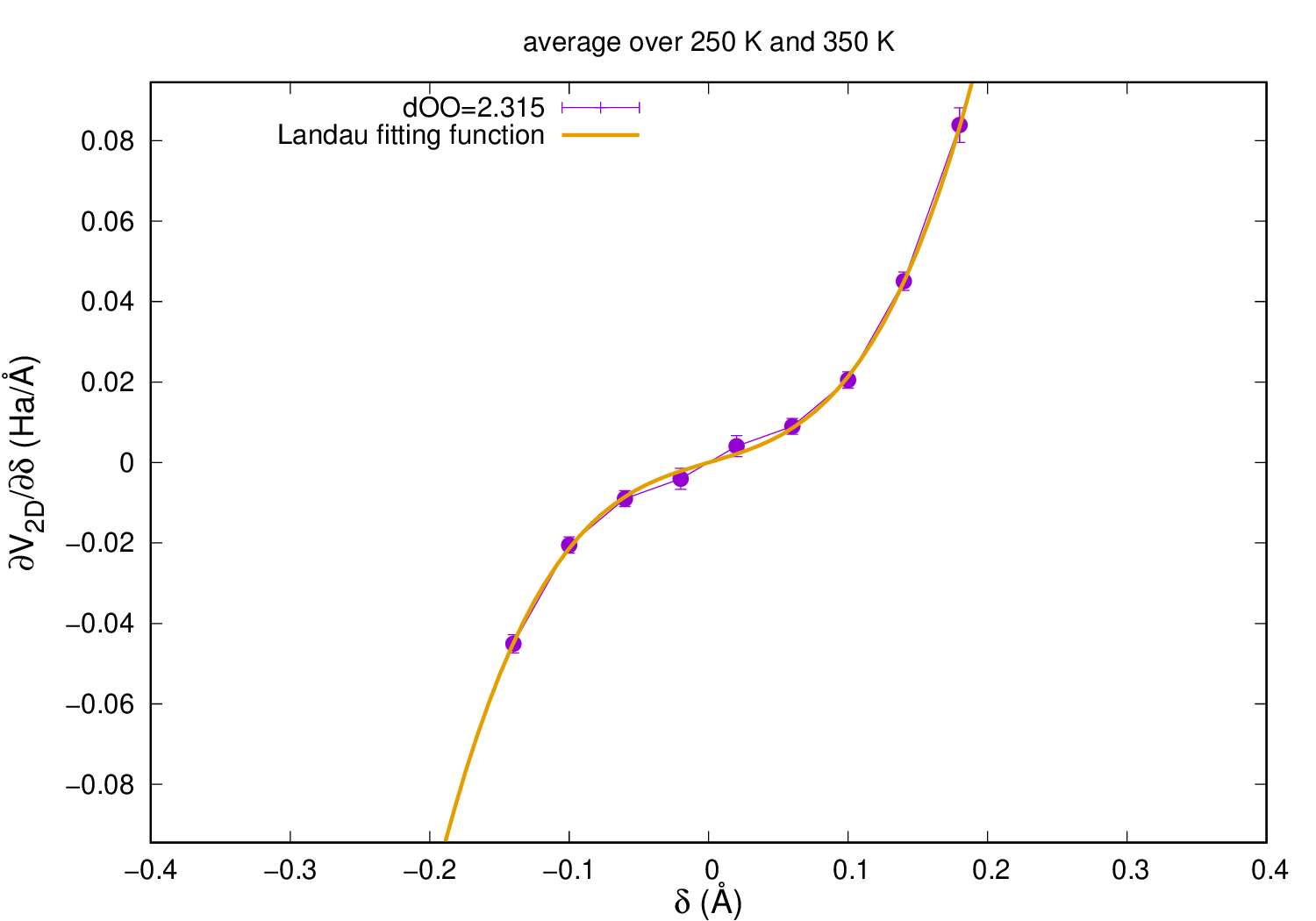}
  &
\includegraphics[width=0.32\columnwidth]{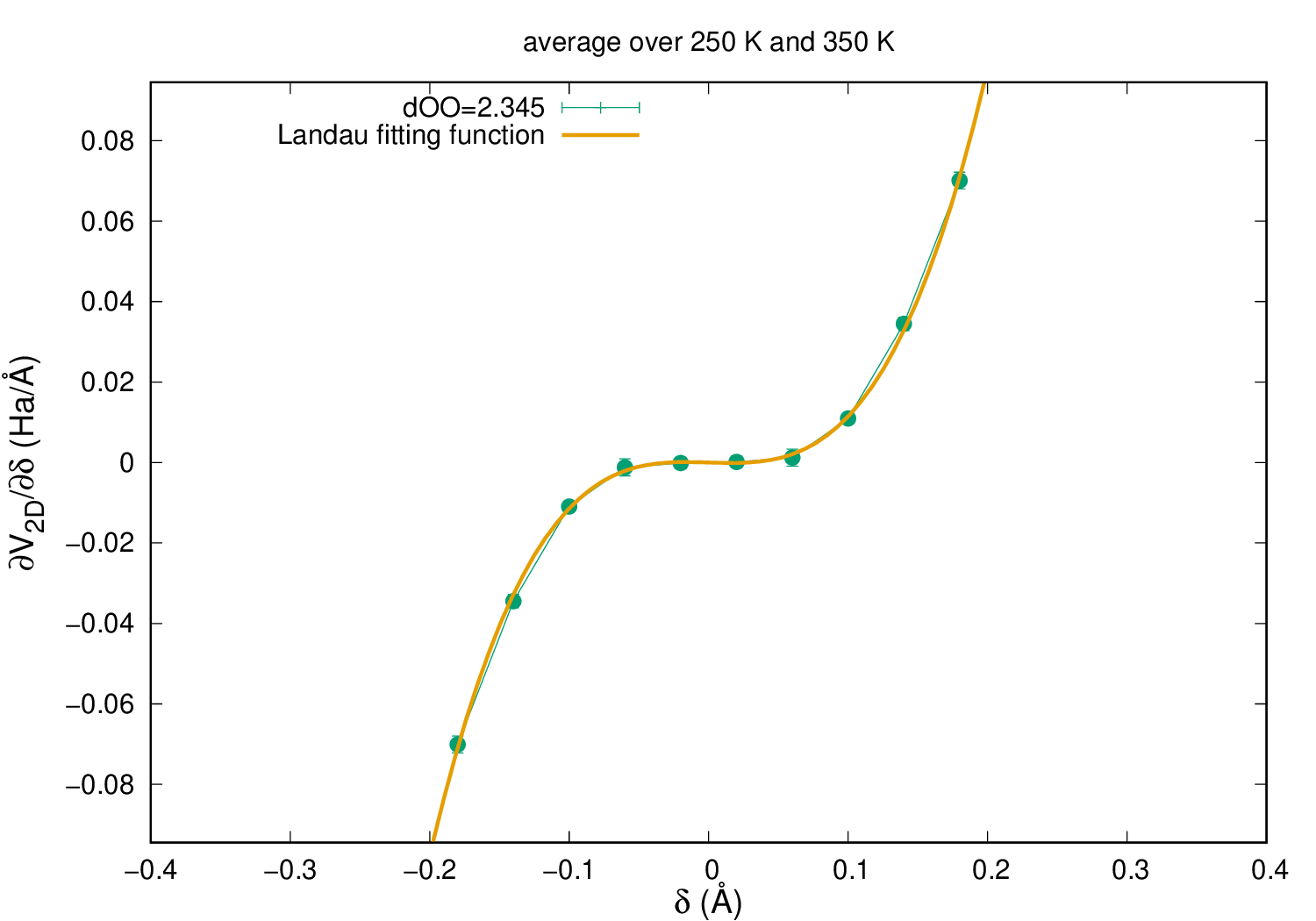}
  \\
\includegraphics[width=0.32\columnwidth]{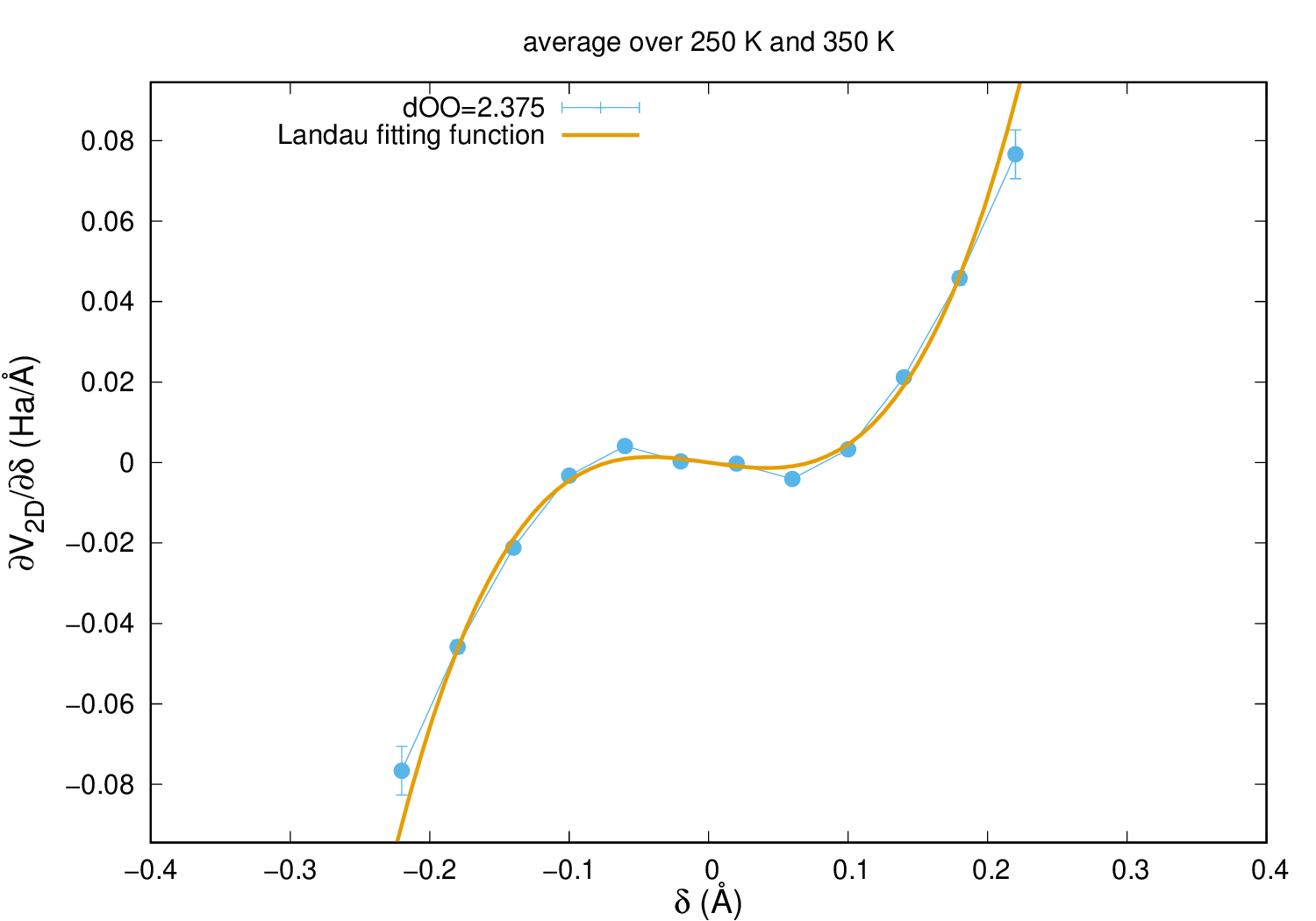}
  &
\includegraphics[width=0.32\columnwidth]{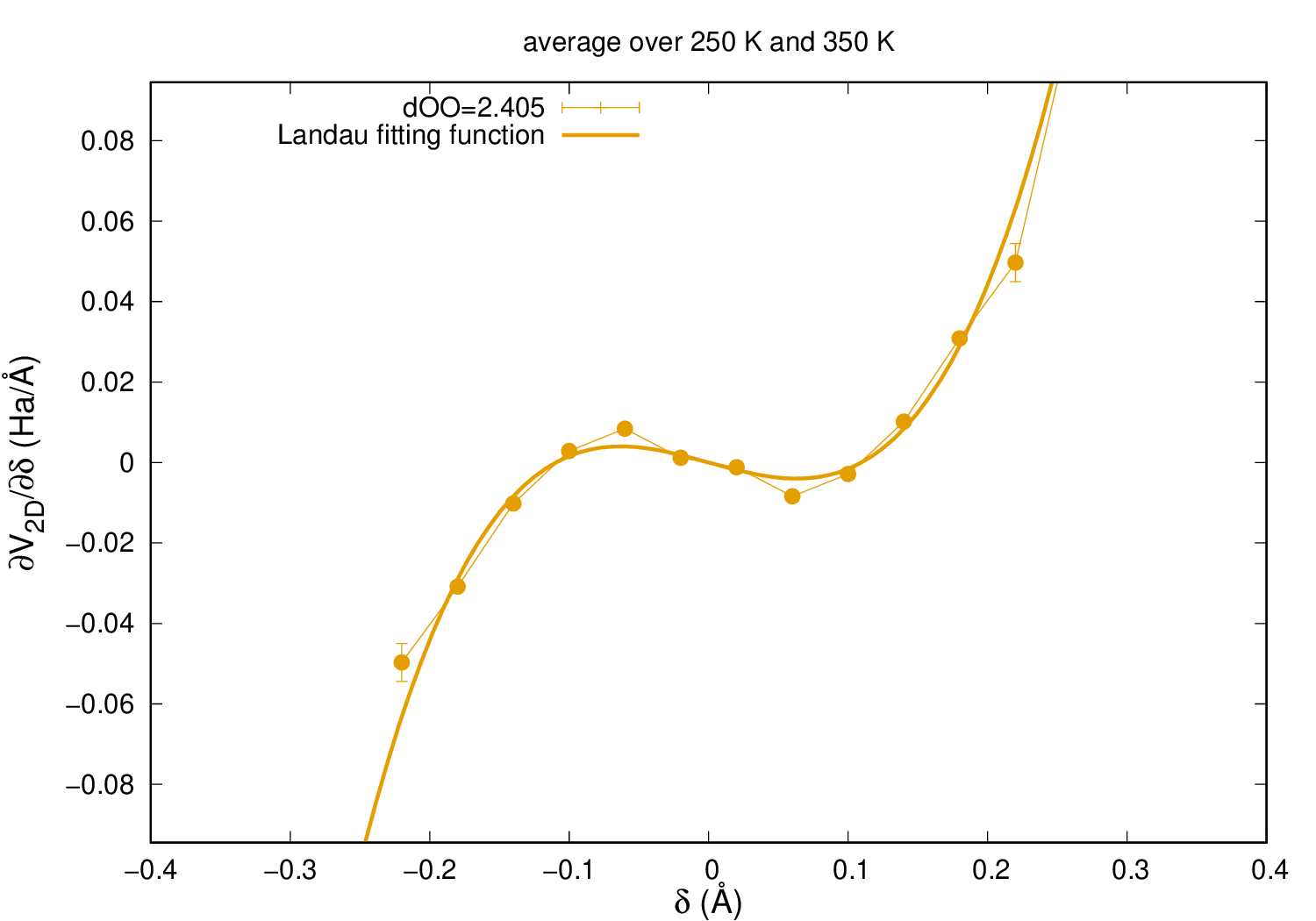}
  \\
\includegraphics[width=0.32\columnwidth]{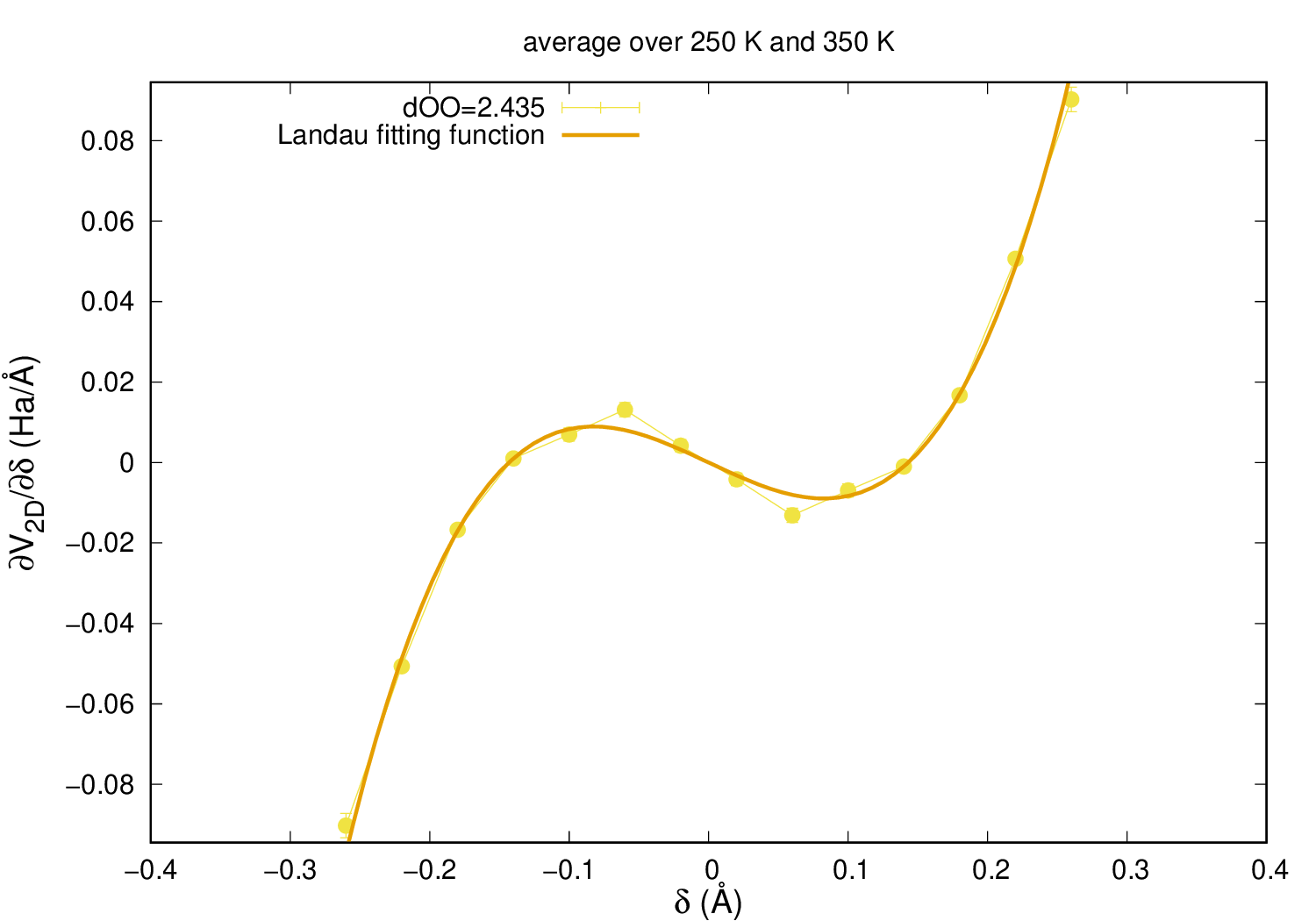}
  &
\includegraphics[width=0.32\columnwidth]{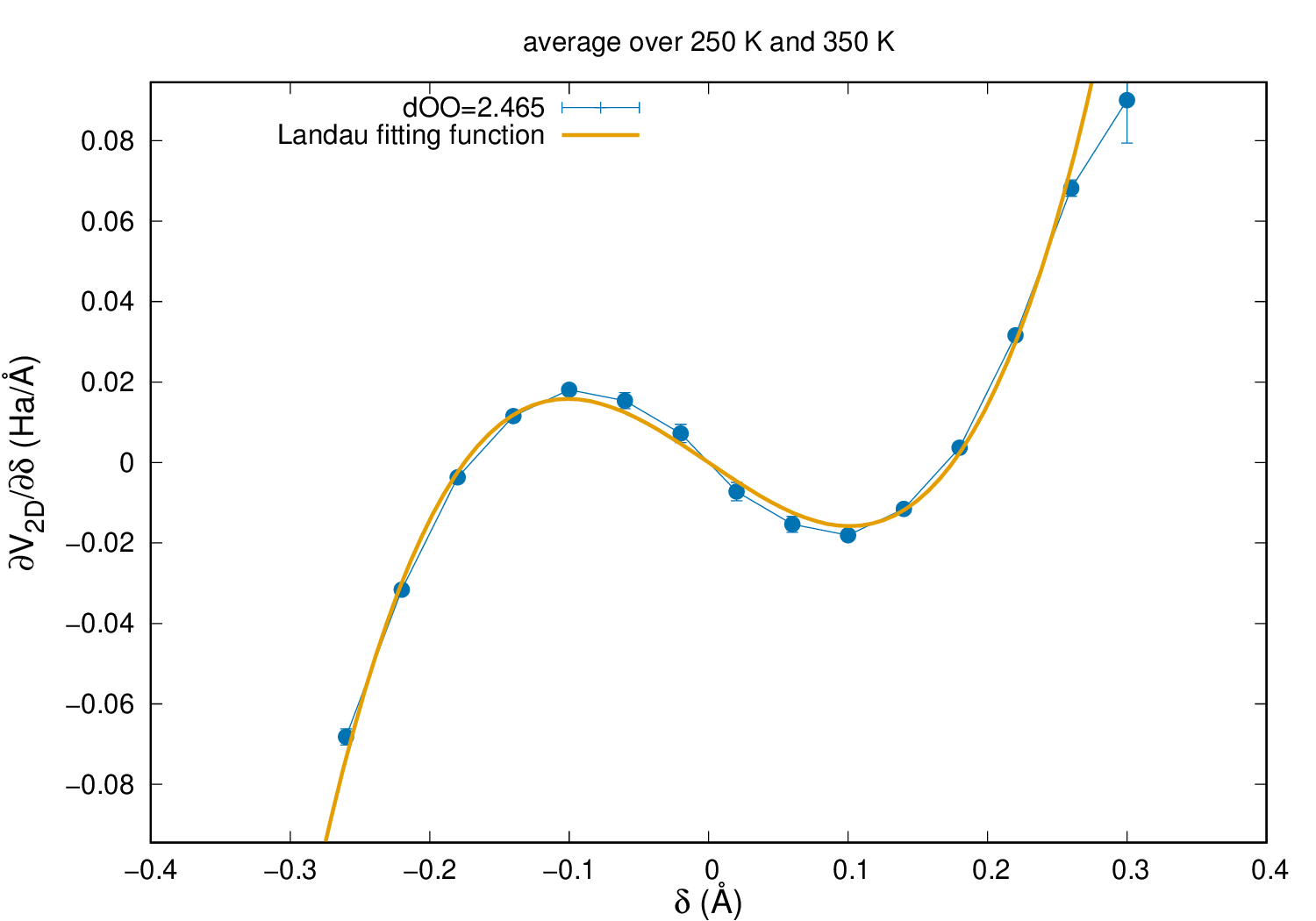}  
\\
\includegraphics[width=0.32\columnwidth]{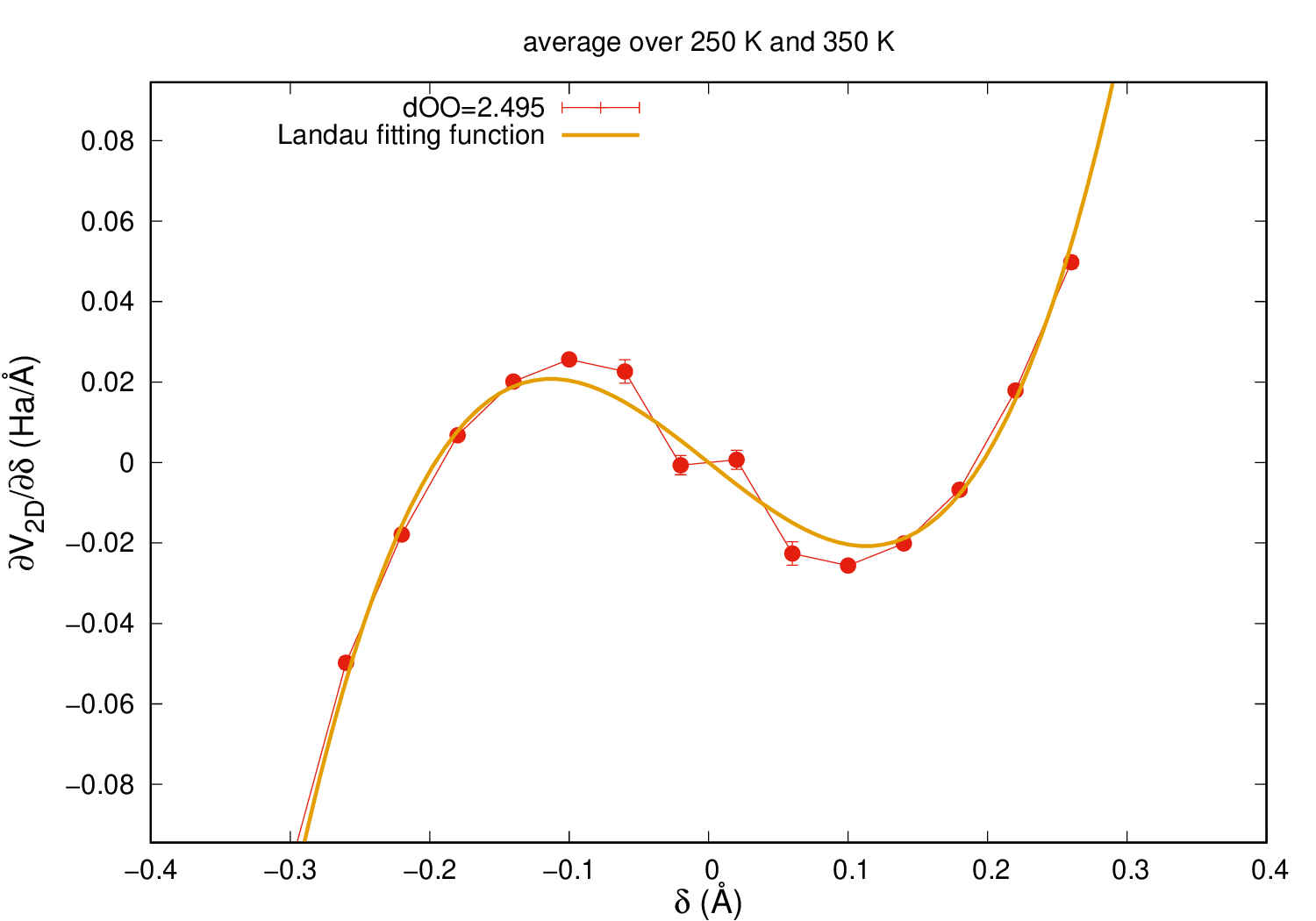}
  &
\includegraphics[width=0.32\columnwidth]{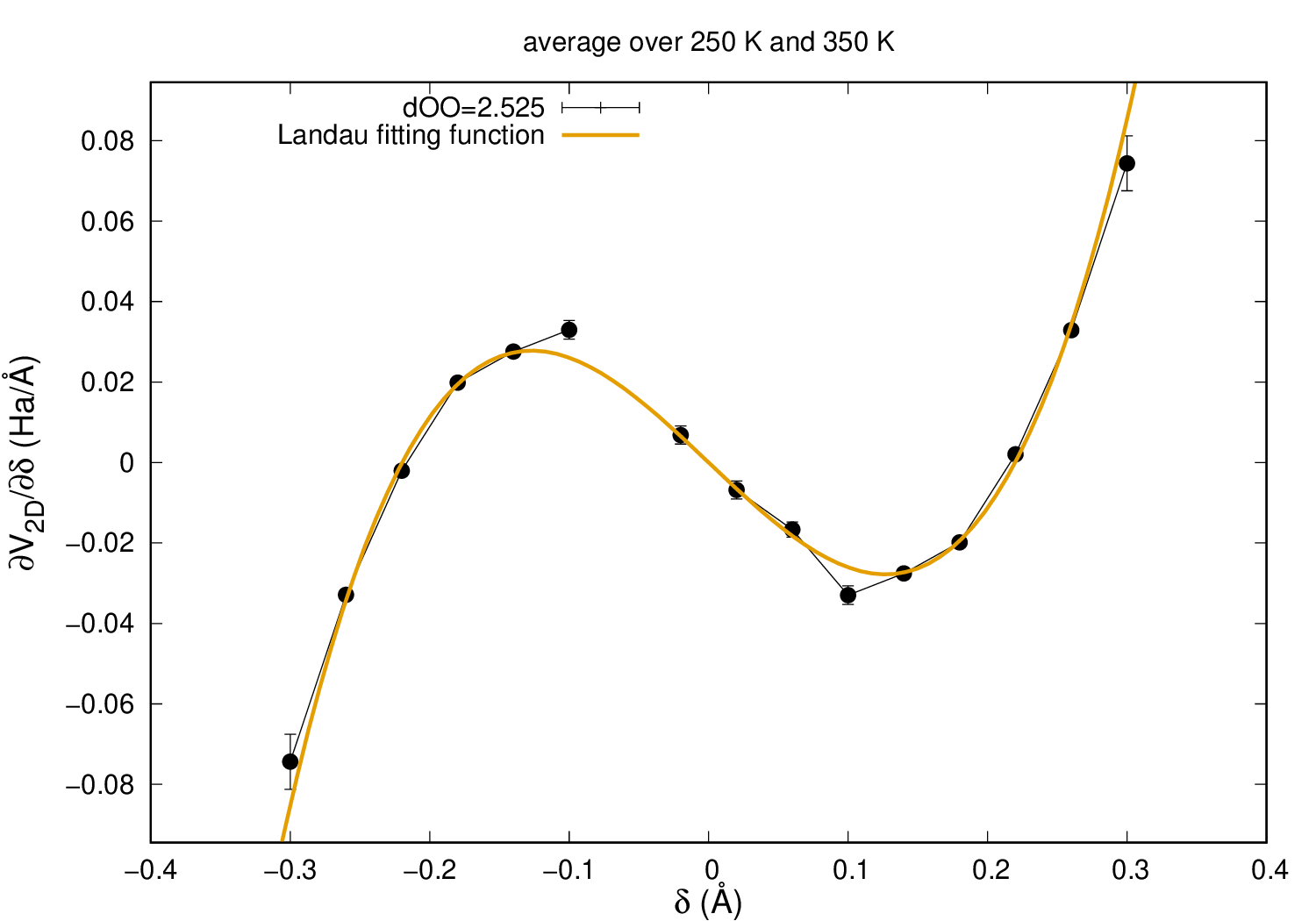}  
  \\
\includegraphics[width=0.32\columnwidth]{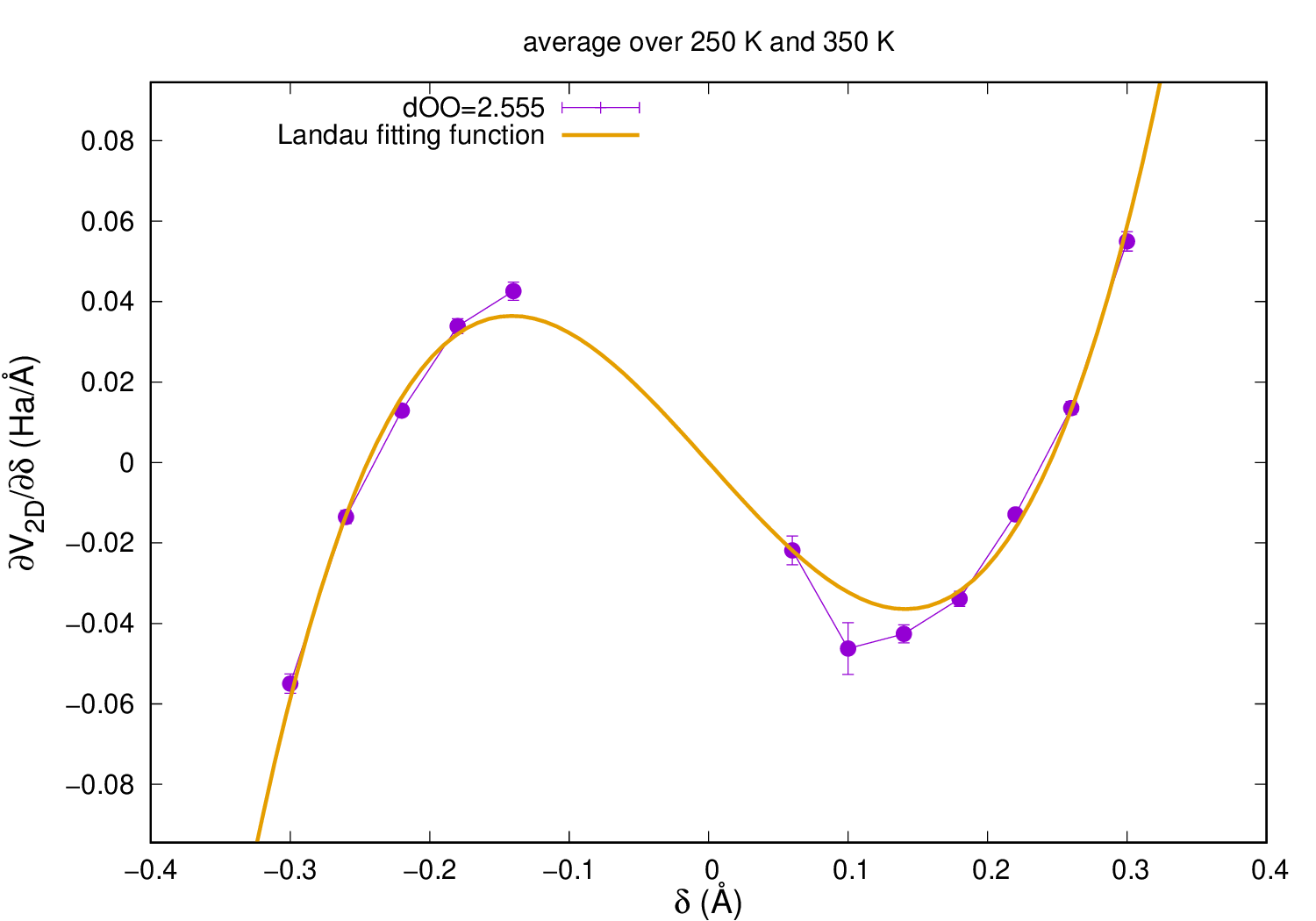}
  &
\includegraphics[width=0.32\columnwidth]{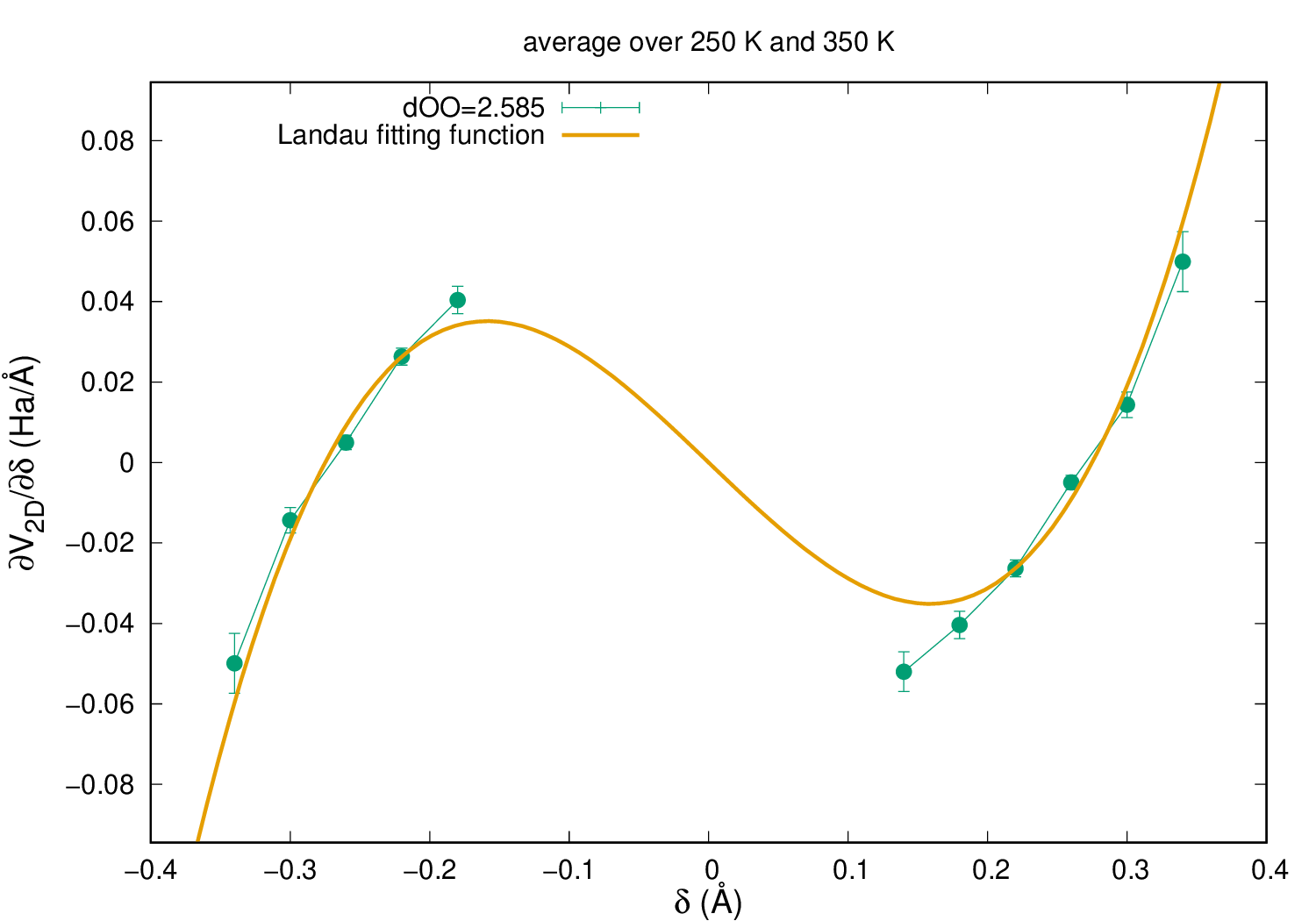}  
  \\
\end{tabular}
\caption{Fit of the QMC forces using $\partial V_{d_\textrm{O$_1$O$_2$}}(x)/\partial x$ as fitting
     function for different $d_\textrm{O$_1$O$_2$}$, where the potential $V_{d_\textrm{O$_1$O$_2$}}(x)$ is
    defined in Eq.~\ref{Landau_pot}. The points are calculated as
    average over the two temperatures of 250 K and 350 K. 
}
\label{Landau_fit_av}
\end{figure}

As a second step, we derive the $V_\textrm{2D}(d_\textrm{O$_1$O$_2$},\delta)$ potential. 
For every $d_\textrm{O$_1$O$_2$}$ slice, in Fig.~\ref{Landau_fit_av} we plot the estimated values of the  $\partial V_\textrm{2D}/\partial \delta$ derivative as a function of $\delta$, computed by averaging over the 250 K and 350 K classical MD samples, as we did for the $V_\textrm{1D}$ potential.

\newpage

\newpage

\begin{figure}[!h]
\centering 
\begin{tabular}{l r} 
  \includegraphics[width=0.5\columnwidth]{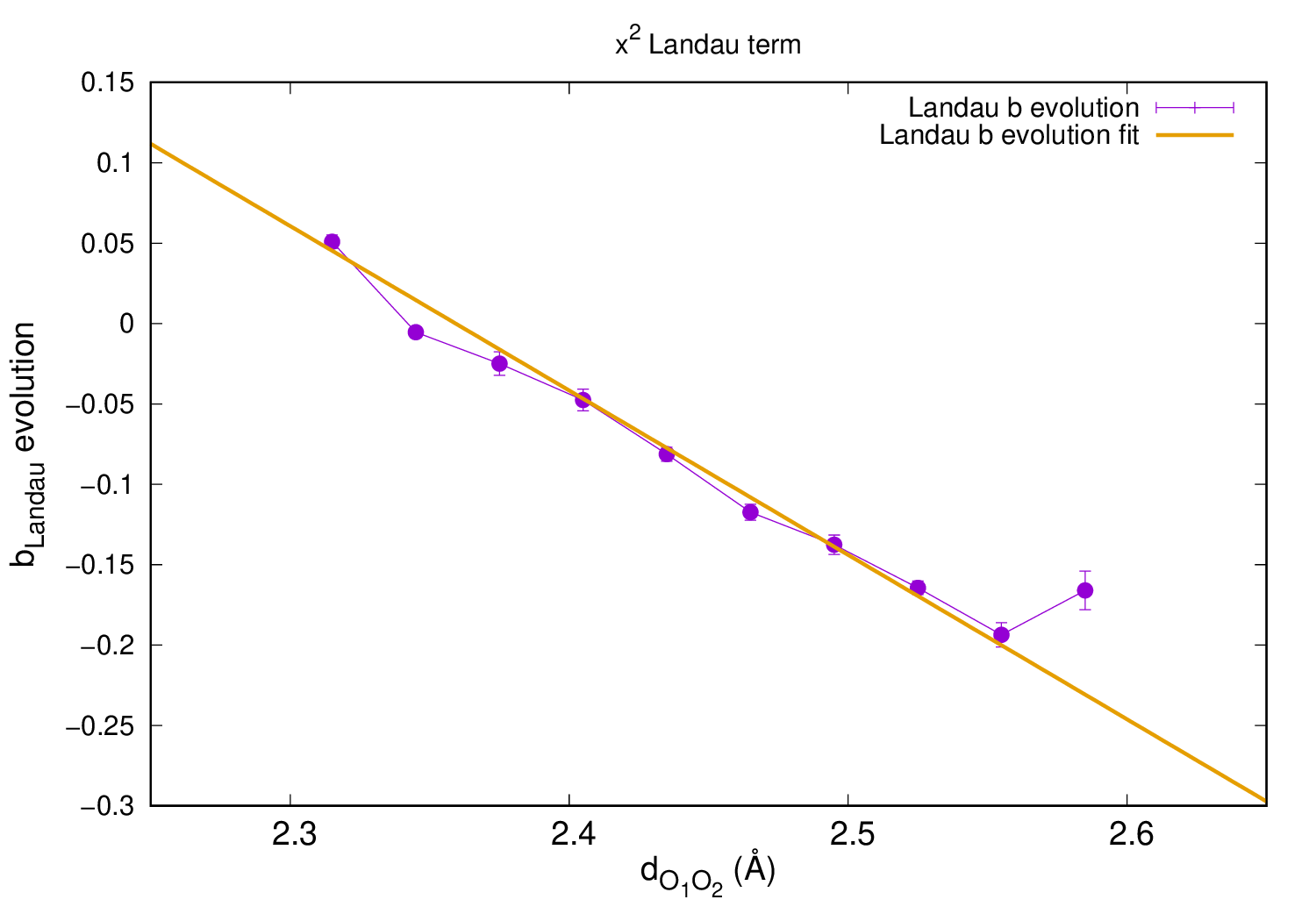} &
  \includegraphics[width=0.5\columnwidth]{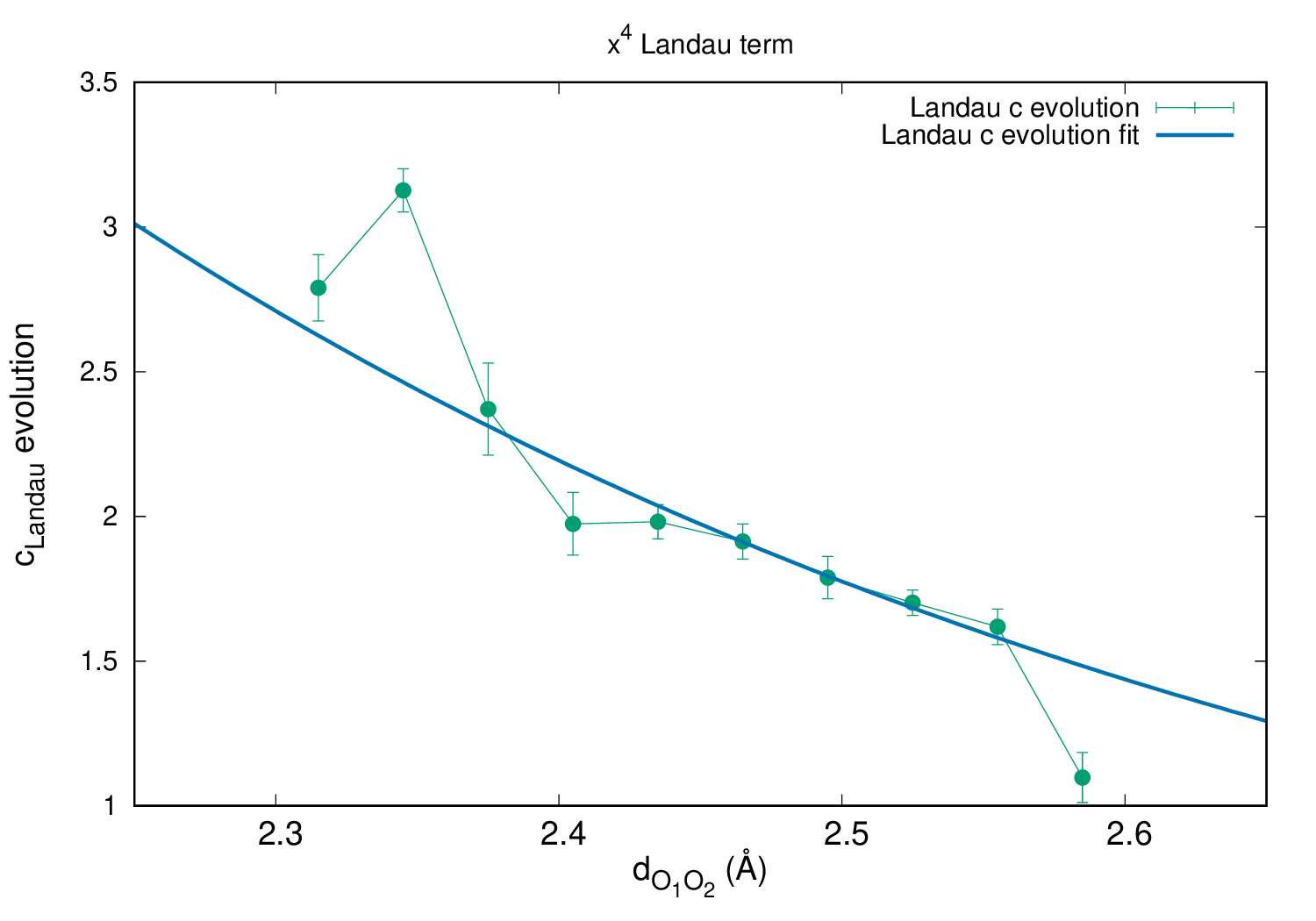}                  
  \\
\end{tabular}
\caption{Fit of the $b_\textrm{Landau}$ and
  $c_\textrm{Landau}$ dependence on the $d_\textrm{O$_1$O$_2$}$ distance, based
  on the functional forms in Eqs.~\ref{b_dep} and \ref{c_dep}. 
}
\label{Lparameters_vs_dOO_av}
\end{figure}

At every $d_\textrm{O$_1$O$_2$}$, we fit the derivative of the energy with respect to $\delta$, by using a symmetric quartic function, i.e. a Landau potential, as fitting model for the energy dependence:
\begin{equation}
  V_{d_\textrm{O$_1$O$_2$}} (x)=a_\textrm{Landau} + b_\textrm{Landau} x^2 + c_\textrm{Landau} x^4 ~~~~~\textrm{at fixed $d_\textrm{O$_1$O$_2$}$ distance.}
  \label{Landau_pot}
  \end{equation}
 The fits for selected $d_\textrm{O$_1$O$_2$}$ values are also reported in Fig.~\ref{Landau_fit_av}. The parameters $a_\textrm{Landau}$, $b_\textrm{Landau}$ and $c_\textrm{Landau}$ have thus
an implicit dependence on $d_\textrm{O$_1$O$_2$}$, which need to be
further included in the full $V_\textrm{2D}(d_\textrm{O$_1$O$_2$},\delta)$ function . We found that a good parametrisation for
$b_\textrm{Landau}$ is given by:
\begin{equation}
  b_\textrm{Landau}(d_\textrm{O$_1$O$_2$})=\alpha + \beta d_\textrm{O$_1$O$_2$},
  \label{b_dep}
 \end{equation}
  while for $c_\textrm{Landau}$ is:
 \begin{equation}
    c_\textrm{Landau}(d_\textrm{O$_1$O$_2$})=\epsilon \exp(-\gamma d_\textrm{O$_1$O$_2$}).
  \label{c_dep}
\end{equation}
Note that the $d_\textrm{O$_1$O$_2$}$-dependence in Eq.~\ref{c_dep} guarantees that the
potential in Eq.~\ref{Landau_pot} always binds for $\epsilon > 0$.
Fig.~\ref{Lparameters_vs_dOO_av} demonstrates how this
dependence, shown by the parameters evolution, is well
taken into account by the functional forms in
Eqs.~\ref{b_dep} and \ref{c_dep}.

From Eq.~\ref{b_dep} and its fitting parameters, the bifurcation point
turns out to be located at $d_\textrm{symm} \approx 2.37$\AA, in quite good agreement with the
relaxation of the ground state geometry.

The final 2D potential $V_\textrm{2D}(d_\textrm{O$_1$O$_2$},\delta)$ is thus
fully determined by the following function:
\begin{equation}
V_\textrm{2D}(d_\textrm{O$_1$O$_2$},\delta) = a_\textrm{Landau}(d_\textrm{O$_1$O$_2$})+b_\textrm{Landau}(d_\textrm{O$_1$O$_2$}) \delta^2 + c_\textrm{Landau}(d_\textrm{O$_1$O$_2$}) \delta^4,
\label{2D_pot}
\end{equation}
with $b_\textrm{Landau}(d_\textrm{O$_1$O$_2$})$ and
$c_\textrm{Landau}(d_\textrm{O$_1$O$_2$})$ already defined in
Eqs.~\ref{b_dep} and \ref{c_dep}, respectively, while
$a_\textrm{Landau}(d_\textrm{O$_1$O$_2$})$ is defined as follows:
\begin{equation}
  a_\textrm{Landau}(d_\textrm{O$_1$O$_2$})=V_\textrm{1D}(d_\textrm{O$_1$O$_2$})+\Delta(d_\textrm{O$_1$O$_2$}).
\label{a_dep}
\end{equation}
In the above Equation, $\Delta(d_\textrm{O$_1$O$_2$})$ is the proton barrier of
the 
Landau
potential $V_{d_\textrm{O$_1$O$_2$}}(x)$ in Eq.~\ref{Landau_pot}, such that the bottom of
$V_\textrm{2D}(d_\textrm{O$_1$O$_2$},\delta)$ at a given $d_\textrm{O$_1$O$_2$}$ distance
follows exactly the Morse potential $V_\textrm{1D}$ in Eq.~\ref{Morse_pot}. In particular,
$\Delta(d_\textrm{O$_1$O$_2$})$ reads:
\begin{equation}
\Delta(d_\textrm{O$_1$O$_2$})=
\begin{cases}
    0, & \text{if } d_\textrm{O$_1$O$_2$}  \leq d_\textrm{symm} \\
    \frac{b^2_\textrm{Landau}(d_\textrm{O$_1$O$_2$})}{4 c_\textrm{Landau}(d_\textrm{O$_1$O$_2$})},             & \text{otherwise}.
  \end{cases}
  \label{barrier}
\end{equation}

\begin{figure}[!hbt]
\centering 
\begin{tabular}{l r} 
  \includegraphics[width=0.5\columnwidth]{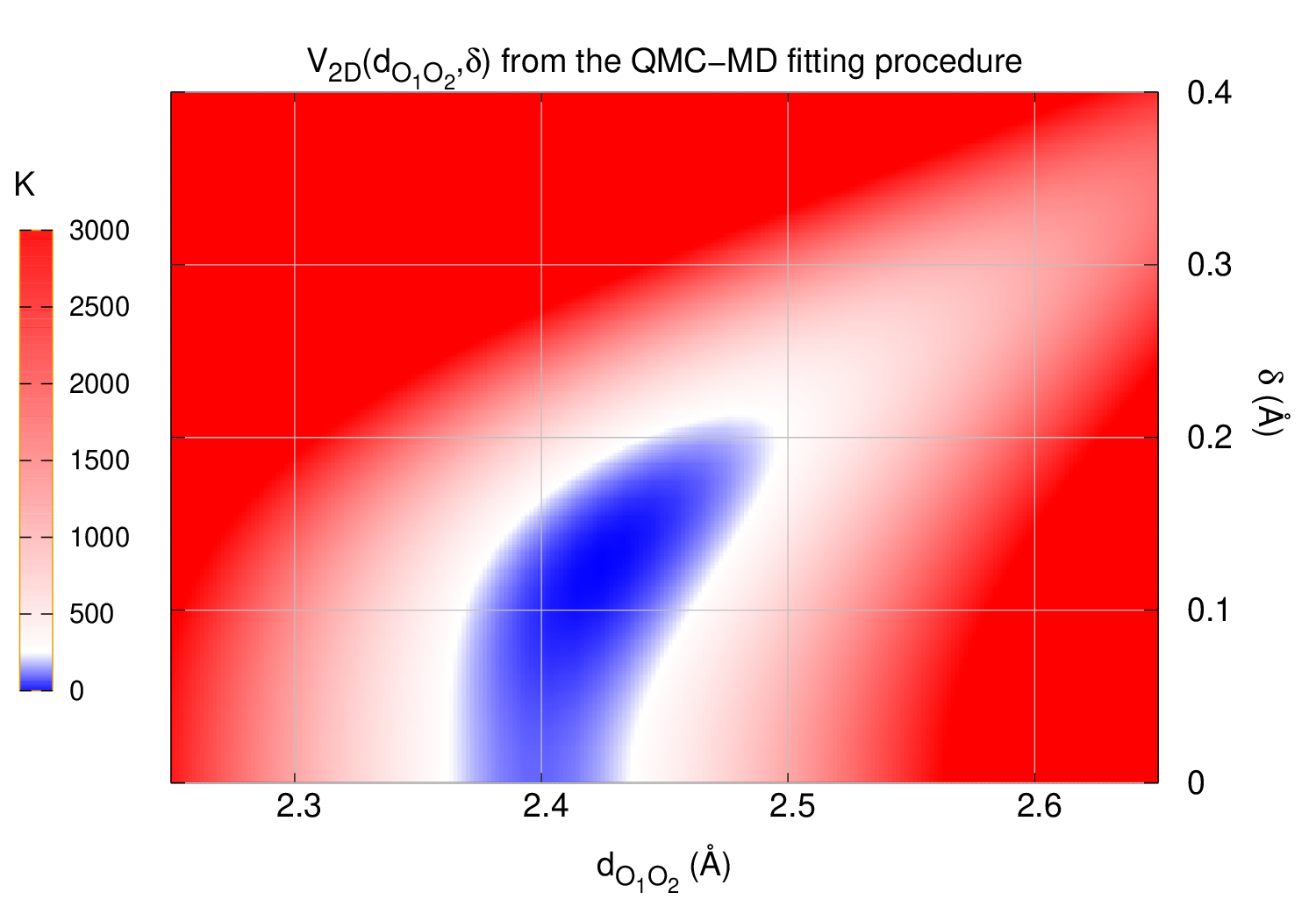}
  &
  \includegraphics[width=0.5\columnwidth]{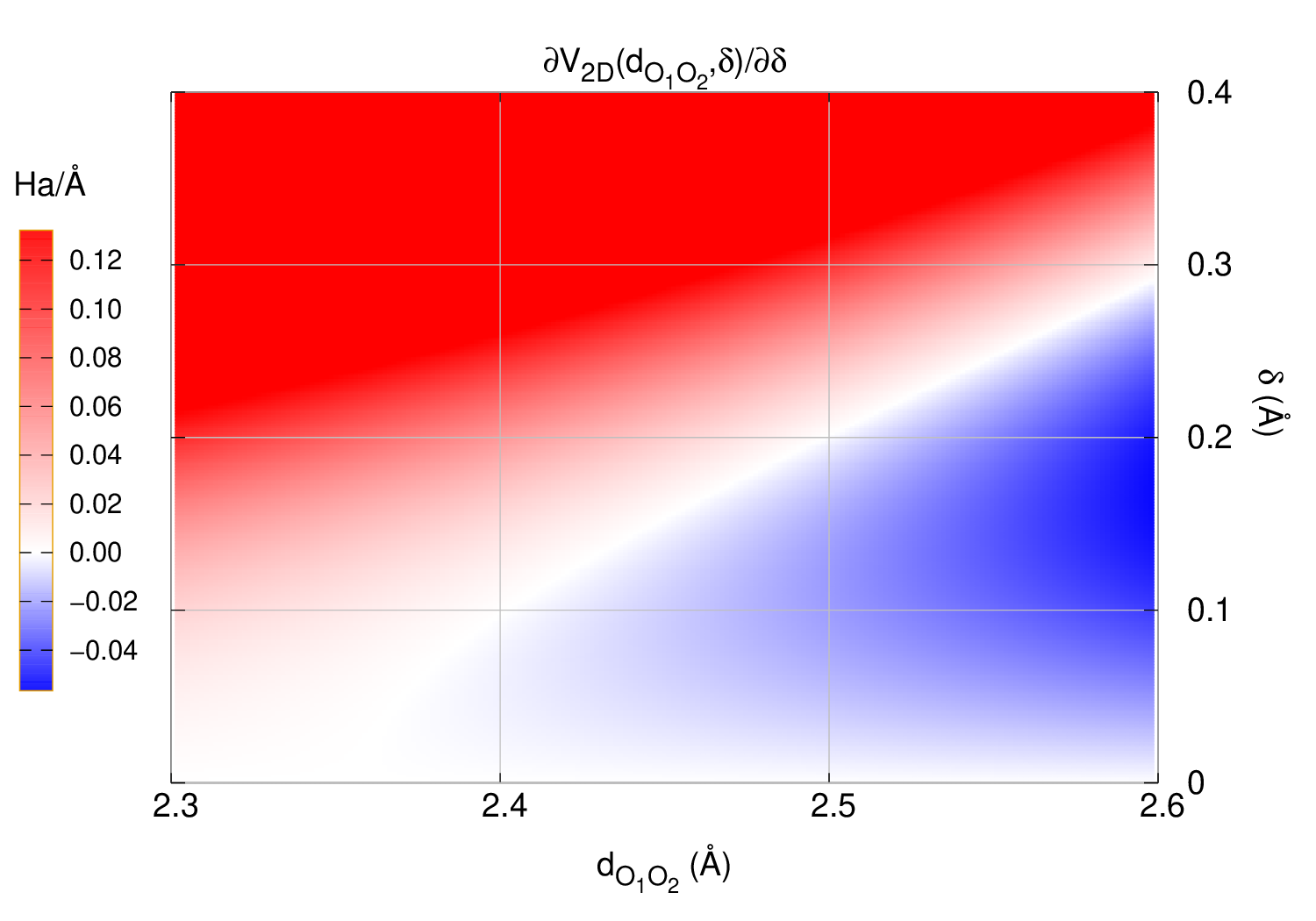}                  
  \\
\end{tabular}
\caption{Left panel: contour plot of the
  $V_\textrm{2D}(d_\textrm{O$_1$O$_2$},\delta)$ 2D model potential. Right panel:
  contour plot of the model-potential derivative
  $\frac{\partial}{\partial\delta} V_\textrm{2D}(d_\textrm{O$_1$O$_2$},\delta)$,
  to be compared with the contour plot of the mid- and bottom-left panels of Fig.~\ref{2DforceH_classical}, directly obtained from MD sampled datapoints.
  }
\label{fig_2D_pot_avOO_avOH}
\end{figure}

The resulting 2D potential $V_\textrm{2D}(d_\textrm{O$_1$O$_2$},\delta)$ and
its derivative with respect to $\delta$ are
drawn in the contour plot of
Fig.~\ref{fig_2D_pot_avOO_avOH}. 
$\frac{\partial}{\partial\delta} V_\textrm{2D}(d_\textrm{O$_1$O$_2$},\delta)$
compares very well with the same quantity directly evaluated by
QMC-driven classical MD at both 250 K and 350 K, as shown in
Fig.~\ref{2DforceH_classical}, mid- and bottom-left panels. This is an \emph{a posteriori} check of
the quality of our 2D-PES determination.

As we have seen, there is 
a residual
temperature dependence in the
determination of $V_\textrm{2D}(d_\textrm{O$_1$O$_2$},\delta)$, due to the projection scheme employed. 
Using a range
of temperatures $T \in [250, 350]$ K
guarantees an optimal sampling
of the configuration space during the MD, allowing for a more extended
determination of the 2D-PES model. In the main text of the paper, we used 
the function plotted in Fig.~\ref{fig_2D_pot_avOO_avOH}
to carry out an anharmonic vibrational analysis of the
shuttling mode. The range of temperatures at which the model potential $V_\textrm{2D}(d_\textrm{O$_1$O$_2$},\delta)$ has been derived is consistent with the temperatures where the PT shows a ``sweet spot'', supporting the outcome of our analysis.

\section{Proton transfer: adiabatic events versus quantum tunneling}
The picture unveiled in this work proves that the synergy of NQEs and thermal effects is fundamental to understand PT in an aqueous environment. Beside the ZPE, NQEs can contribute to the proton diffusion by means of instantaneous tunneling, which can further accelerate the PT dynamics.
Both adiabatic crossing boosted by the ZPE, and instantaneous tunneling are plausible and non-competing scenarios in our system. As we have seen, the former 
is certainly
favoured by the shortness of d$_{{\text{\scriptsize O}}_1{\text{\scriptsize O}}_2}$ in the short-Zundel configurations, where the barrier is absent along the shuttling trajectory $\delta$, while
the latter could sustain PT events even in case of high barriers, such as in Eigen-like configurations.

In order to assess the 
role played by
quantum tunneling,
we evaluate the localisation
level
of the excess proton during its dynamics, by computing the root-mean-square (RMS) displacement correlation functions\cite{Chandler1994}, $\mathcal{R}_{\textrm{d}_{\textrm{{\scriptsize OH}}^+}}(\tau) = \langle |\textrm{d}_{\textrm{\scriptsize OH}^+}(0) - \textrm{d}_{\textrm{{\scriptsize OH}}^+}(\tau)|^2 \rangle$, 
in imaginary time $\tau \in [0, \beta \hbar]$, with $\beta = 1/k_BT$ and $\textrm{d}_{\textrm{\scriptsize OH}^+} = |\mathbf{q}_\text{\scriptsize O}-\mathbf{q}_{\textrm{\scriptsize H}^+}|$. 
For localised states, trapped in potential energy minima, $\mathcal{R}_{\textrm{d}_{\textrm{\scriptsize OH}^+}}(\tau)$ is flat
in the intermediate $\tau$ range, while, for delocalised states, $\mathcal{R}_{\textrm{d}_{\textrm{\scriptsize OH}^+}}(\tau)$ is roughly parabolic \cite{Chandler1994}.
A quantum particles undergoing a tunneling event will look like a free particle with parabolic behavior, as unaffected by the tunneled potential energy barrier.

\begin{figure}[!htp]
\centering 
      \includegraphics[width=\columnwidth]{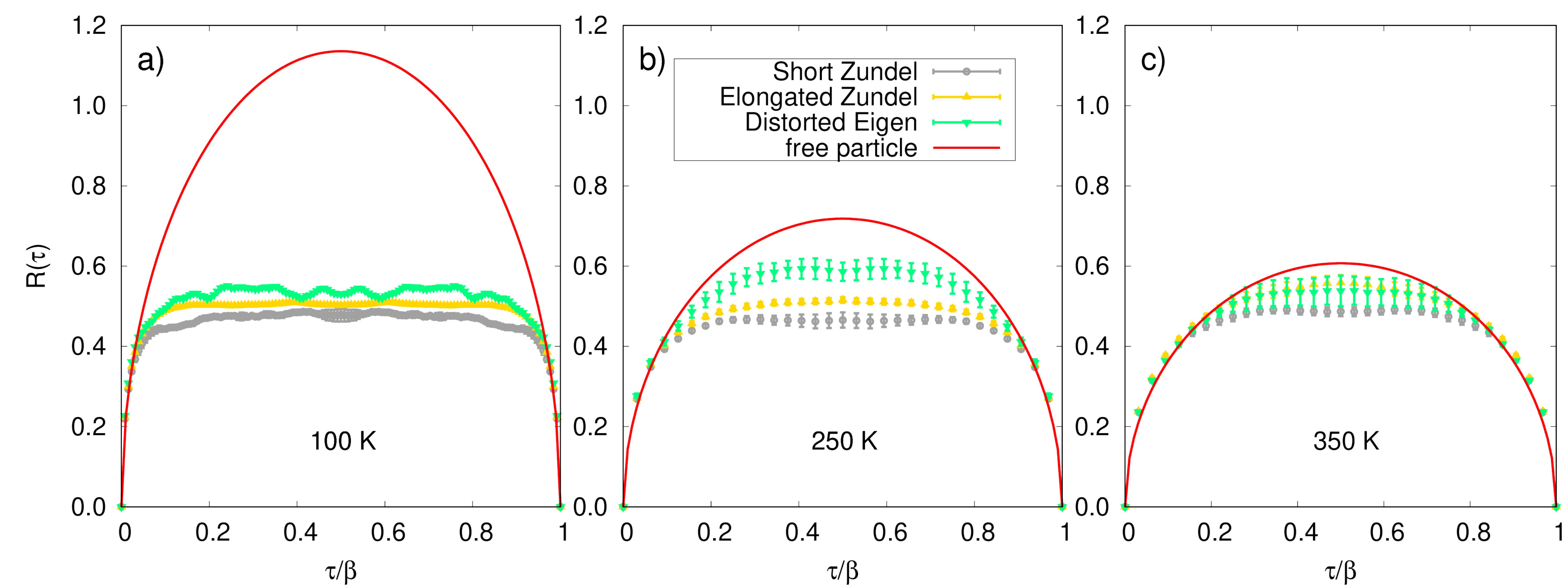}
\caption {\label{figure:correl} 
  Root mean square
  displacement $R(\tau)$ in quantum imaginary time $\tau^* \in [0,\beta \hbar/M] \equiv [0,1]$ for the proton displacement 
   with respect to the side
  oxygen atoms, $\textrm{d}_{\textrm{\scriptsize OH}^+} =
  |\mathbf{q}_\text{\scriptsize O}-\mathbf{q}_{\textrm{\scriptsize H}^+}|$, in the core of the H$_{13}$O$_6^+$ cation. Calculations are performed at 100 K, 250 K and 350 K, reported in panels a), b) and c), respectively.
  Samples are 
  taken from
  the instanton population, and averages are performed in a species-selected way: short-Zundel (gray points), elongated-Zundel (yellow points) and Eigen-like species (green points) are separated, according to the $\textrm{d}_{\text{\scriptsize O}_1\text{\scriptsize O}_2}$ distance at which they occur. As reference, the  root-mean-square displacement of the free proton at the given temperature is also reported in red solid lines.}    
\end{figure}
The evolution of the RMS displacement correlations functions $\mathcal{R}_{\textrm{d}_{\textrm{\scriptsize OH}^+}}(\tau)$ with the temperature is depicted in Fig.~\ref{figure:correl} for the instanton configurations, the most relevant for the PT description. Also in this case, we distinguish instantons belonging to the three different regimes. 
A striking difference is observed among the three temperatures investigated, i.e. 100 K, 250 K and 350 K.
For the lowest temperature (Fig.~\ref{figure:correl}(a)), the TS correlation function $\mathcal{R}_{\textrm{d}_{\textrm{\scriptsize OH}^+,\textrm{\scriptsize TS}}}(\tau)$ is flat, \emph{i.e.} more localised, in all regimes. At intermediate temperatures (Fig.~\ref{figure:correl}(b)), within the ``sweet spot'' range, the $\mathcal{R}_{\textrm{d}_{\textrm{\scriptsize OH}^+,\textrm{\scriptsize traj}}}(\tau)$ measured in the distorted-Eigen configurations is the closest to the free particle behavior, while the short-Zundel averages are clearly bound by the confining symmetric potential. This suggests that some PT events in the distorted Eigen, which need to 
overcome
taller barriers, could be enhanced by quantum tunneling, as additional PT channel, beside the very effective short-Zundel TS mechanism.
At the highest temperature (Fig.~\ref{figure:correl}(c), all $\mathcal{R}_{\textrm{d}_{\textrm{\scriptsize OH}^+,\textrm{\scriptsize traj}}}(\tau)$ are very close to the free particle. However, in this case, thermal effects, rather than quantum tunneling, make the instantons behave like free particles.

\setcounter{page}{1}
\renewcommand{\thepage}{R\arabic{page}} 

\bibliography{Bibliographie}

\bibliographystyle{sci-mathphys}

\end{document}